\documentclass[letterpaper,11pt]{article}

\usepackage{jheppub}
\usepackage{graphicx}
\usepackage[small]{subfigure}
\usepackage{amsmath}

\usepackage{xspace}

\newcommand{\eq}[1]{eq.~\eqref{eq:#1}}
\newcommand{\eqs}[2]{eqs.~\eqref{eq:#1} and \eqref{eq:#2}}
\renewcommand{\sec}[1]{section~\ref{sec:#1}}

\newcommand{\subsec}[1]{section~\ref{subsec:#1}}
\newcommand{\subsubsec}[1]{section~\ref{subsubsec:#1}}

\newcommand{\fig}[1]{figure~\ref{fig:#1}}
\newcommand{\figs}[2]{figures~\ref{fig:#1} and \ref{fig:#2}}
\newcommand{\tab}[1]{table~\ref{tab:#1}}

\newcommand{\abs}[1]{\lvert#1\rvert}
\newcommand{\ord}[1]{\mathcal{O}(#1)}

\newcommand{\df}{\mathrm{d}}

\newcommand{\Tau}{\mathcal{T}}

\newcommand{\cL}{\mathcal{L}}

\newcommand{\GeV}{\,\mathrm{GeV}}

\newcommand{\nn}{\nonumber}

\newcommand{\cut}{\mathrm{cut}}

\newcommand{\incl}{\mathrm{incl}}

\newcommand{\nons}{\mathrm{nons}}
\newcommand{\sing}{\mathrm{sing}}
\newcommand{\FO}{\mathrm{FO}}
\newcommand{\LO}{\mathrm{LO}}
\newcommand{\NLO}{\mathrm{NLO}}
\newcommand{\resum}{\mathrm{resum}}

\renewcommand{\max}{\mathrm{max}}

\newcommand{\lqcd}{\Lambda_\mathrm{QCD}}

\newcommand{\Ecm}{E_\mathrm{cm}}
\newcommand{\dsigMC}{\df\sigma^\textsc{mc}}

\newcommand{\geneva}{\textsc{Geneva}\xspace}
\newcommand{\mcatnlo}{\textsc{MC@NLO}\xspace}
\newcommand{\powheg}{\textsc{Powheg}\xspace}
\newcommand{\pythia}{\textsc{Pythia}\xspace}
\newcommand{\event}{\textsc{Event}2\xspace}
\newcommand{\Aleph}{ALEPH\xspace}
\newcommand{\Opal}{OPAL\xspace}
\newcommand{\sherpa}{\textsc{Sherpa}\xspace}

\allowdisplaybreaks[2]

\title{Combining Higher-Order Resummation with Multiple NLO Calculations and Parton Showers in GENEVA}

\author[a]{Simone Alioli,}
\author[a]{Christian W.~Bauer,}
\author[a]{Calvin Berggren,}
\author[b]{Andrew Hornig,}
\author[c]{Frank J.~Tackmann,}
\author[a]{Christopher K.~Vermilion,}
\author[a]{Jonathan R.~Walsh,}
\author[a]{Saba Zuberi}

\affiliation[a]{Ernest Orlando Lawrence Berkeley National Laboratory, University of California, Berkeley, CA 94720, U.S.A.}
\affiliation[b]{Department of Physics, University of Washington, Seattle, WA 98195, U.S.A.}
\affiliation[c]{Theory Group, Deutsches Elektronen-Synchrotron (DESY), D-22607 Hamburg, Germany}

\emailAdd{salioli@lbl.gov}
\emailAdd{cwbauer@lbl.gov}
\emailAdd{cjberggren@lbl.gov}
\emailAdd{ahornig@uw.edu}
\emailAdd{frank.tackmann@desy.de}
\emailAdd{ckvermilion@lbl.gov}
\emailAdd{jwalsh@lbl.gov}
\emailAdd{szuberi@lbl.gov}

\abstract{
We extend the lowest-order matching of tree-level matrix elements with parton showers to give a complete description at the next higher perturbative accuracy in $\alpha_s$ at both small and large jet resolutions, which has not been achieved so far. This requires the combination of the higher-order resummation of large Sudakov logarithms at small values of the jet resolution variable with the full next-to-leading-order (NLO) matrix-element corrections at large values. As a by-product, this combination naturally leads to a smooth connection of the NLO calculations for different jet multiplicities.
In this paper, we focus on the general construction of our method and discuss its application to $e^+e^-$ and $pp$ collisions.
We present first results of the implementation in the \geneva Monte Carlo framework.
We employ $N$-jettiness as the jet resolution variable, combining its next-to-next-to-leading logarithmic resummation with fully exclusive
NLO matrix elements, and \pythia8 as the backend for further parton showering and hadronization.
For hadronic collisions, we take Drell-Yan production as an example to apply our construction.
For $e^+e^-\to$ jets, taking $\alpha_s(m_Z) = 0.1135$ from fits to LEP thrust data, together with the \pythia8 hadronization model, we obtain good agreement with LEP data for a variety of $2$-jet observables.
}

\keywords{QCD, Monte Carlo, NLO Computations, Resummation, Jets, Collider Physics\vspace{-2ex}}

\begin{document}

{\flushright DESY 12-221\\November 29, 2012\\[-7ex]}

\addtolength{\textheight}{2cm}
\addtolength{\footskip}{-2cm}
\maketitle
\addtolength{\textheight}{-2cm}
\clearpage
\addtolength{\footskip}{2cm}

\section{Introduction}
\label{sec:intro}

Accurate and reliable theoretical predictions for measurements at collider experiments require the inclusion of QCD effects beyond the lowest perturbative accuracy in the strong coupling $\alpha_s$. This is especially important in the complex environment of the LHC, which requires precise predictions for a broad spectrum of observables. Higher-order corrections in $\alpha_s$ are important to predict total cross sections and other inclusive observables. Exclusive jet observables, such as jet-vetoed cross sections, require the all-orders resummation of logarithmically enhanced contributions. For many observables, an accurate description across phase space demands a combination of both types of corrections.  For experimental analyses to benefit from these advances, it is crucial to provide the best possible theoretical predictions in the context of fully exclusive Monte Carlo event generators.

The goal of modern Monte Carlo programs is to provide a proper
description of the physics at every jet resolution scale. This is the
motivation for the by-now standard combination of matrix elements with
parton showers (ME/PS).~\cite{Catani:2001cc, Lonnblad:2001iq} Here, the parton shower provides the correct
lowest-order description at small jet resolution scales, where the
resummation of large Sudakov logarithms is needed, while at large
jet resolution scales the exact tree-level matrix elements are needed
to provide the correct lowest-order description. Hence, the ME/PS
merging provides theoretical predictions at the formally leading $\ord{1}$
accuracy relative to the lowest meaningful perturbative order. Once
one has a consistent matching between these two limits of phase space,
the possibility to include exact tree-level matrix elements for
several jet multiplicities follows almost automatically by iteration.

Given the necessity of higher-order perturbative corrections to make
accurate predictions, it is important to extend the perturbative
accuracy of the Monte Carlo description to formal $\ord{\alpha_s}$ accuracy
relative to the lowest order. This requires including the formally next
higher-order corrections that are relevant at each scale.  At small
scales, i.e., small values of the jet resolution variable, this
requires improving the leading-logarithm (LL) parton shower resummation with higher-order
logarithmic resummation, while at large scales this requires including
the fully differential next-to-leading-order (NLO) matrix elements. It
is important to realize that typically a large part of phase space,
often including the experimentally relevant region, is characterized
by intermediate scales, i.e., by a transition from small to large
scales. In the end, providing an accurate description of this
transition region requires a careful combination of both types
of corrections.

Such a Monte Carlo description at relative $\ord{\alpha_s}$ accuracy
across phase space has never been achieved and is the subject
of our paper.  (We briefly summarize the existing efforts to combine
NLO corrections with parton showers in \subsec{previous} below.)
The crucial starting point in our approach is that all
perturbative inputs to the Monte Carlo are formulated in terms of
well-defined physical jet cross sections~\cite{Bauer:2008qh, Bauer:2008qj}.
This allows us to systematically increase the
perturbative accuracy by incorporating results for the relevant ingredients
to the desired order in fixed-order and resummed perturbation theory.

An essential aspect of any higher-order prediction is a reliable
estimate of its perturbative uncertainty due to neglected
higher-order corrections. To the extent that parton shower Monte
Carlos provide perturbative predictions, they should be held to the
same standards. An important benefit in our approach is that we have
explicit control of the perturbative uncertainties and are able to
estimate reliable fixed-order and resummation uncertainties. As a
result, in \geneva each event comes with an estimate of its
perturbative uncertainty; i.e., \geneva provides event-by-event theory
uncertainties.\footnote{Further uncertainties, e.g.
due to nonperturpative effects such as hadronization, must
be evaluated as well for a complete uncertainty analysis.}

In our approach, the Monte Carlo not only benefits from the
resummation, but in turn also provides important
benefits to analytic resummed predictions. For one, it greatly
facilitates the comparison with experimental data, as it allows easy
application of arbitrary kinematic selection cuts, which can often be
tedious to take into account in analytic predictions. More
importantly, resummed predictions require nonperturbative inputs which
can be treated as power corrections at intermediate scales but become
$\ord{1}$ corrections at very small scales. Here, these are provided
``on-the-fly'' by the nonperturbative hadronization model. In essence,
we are able to combine the precision and theoretical control offered
by higher-order resummed predictions with the versatility and
flexibility offered by fully exclusive Monte Carlos.

In this paper, we focus on the theoretical construction. We leave a discussion of the implementation details of the \geneva Monte Carlo framework to a dedicated publication.%
\footnote{The current \geneva framework and implementation is new and independent of the earlier work in refs.~\cite{Bauer:2008qh, Bauer:2008qj}.}
We will however highlight some of the main technical issues we had to overcome and discuss some implementation details in the application sections. In the remainder of this section, we briefly summarize the existing efforts to include NLO corrections in parton shower Monte Carlos and give a short overview of our basic construction. In \sec{master}, we discuss in detail the requirements to obtain full $\alpha_s$ accuracy as well as our method to achieve it. In \sec{ee}, we discuss the application to $e^+e^-\to $ jets, where we combine next-to-next-to-leading logarithmic (NNLL) resummation with NLO matrix elements, and present results from the implementation in \geneva together with a comparison to LEP measurements. In \sec{pp}, we discuss the application to hadronic collisions and show first results for Drell-Yan production, $pp\to Z/\gamma^* \to \ell^+ \ell^- +$ jets, obtained with \geneva. We conclude in \sec{conclusions}.

\subsection{Previous Approaches Combining NLO Corrections with Parton Showers}
\label{subsec:previous}

Over the past decade, many steps have been taken to include NLO corrections into Monte Carlo programs~\cite{Collins:2000gd, Collins:2000qd, Potter:2001ej, Dobbs:2001dq, Frixione:2002ik, Frixione:2003ei, Kramer:2003jk, Soper:2003ya, Nagy:2005aa, Kramer:2005hw, Nason:2004rx, Frixione:2007vw, Alioli:2010xd, Torrielli:2010aw, Hoche:2010pf, Frixione:2010ra, Frederix:2011zi, Platzer:2011bc}. By now, the \mcatnlo~\cite{Frixione:2002ik, Frixione:2003ei} and \powheg~\cite{Nason:2004rx, Frixione:2007vw, Alioli:2010xd} methods are routinely able to consistently combine the fixed NLO calculation of an inclusive jet cross section for a given jet multiplicity with additional parton showering. These methods have also been extended to include the full tree-level matrix elements for additional jet multiplicities~\cite{Bauer:2008qh, Hamilton:2010wh, Hoche:2010kg, Giele:2011cb}.

Recently, efforts have been made to extend these approaches in order to combine NLO matrix elements for several jet multiplicities with parton showers~\cite{Lavesson:2008ah, Alioli:2011nr, Hoeche:2012yf, Gehrmann:2012yg, Frederix:2012ps, Platzer:2012bs, Lonnblad:2012ix}. We will discuss some issues faced by some of these approaches in \subsubsec{existing}. Here, we would like to stress that including several NLO matrix elements by itself does not provide a full extension of the lowest-order ME/PS matching to relative $\ord{\alpha_s}$ perturbative accuracy, since the fixed NLO corrections only suffice to increase the perturbative accuracy in the region of large jet resolution scales. To the same extent that the inclusion of the LL Sudakov factors in the ME/PS merging are needed to get meaningful results at intermediate and small jet scales, higher-order resummation is necessary to improve the perturbative accuracy in this region.

In our approach, the full information from NLO matrix elements for several jet multiplicities is automatically included as follows: For a given Born process with $N$ partons, a small jet scale corresponds to the exclusive $N$-jet region, and here the $N$-parton virtual NLO corrections are incorporated in conjunction with the higher-order resummation; in fact, they are a natural ingredient of it. On the other hand, a large jet scale corresponds to the inclusive $(N+1)$-jet region with additional hard emissions. Here, the $(N+1)$-parton virtual NLO corrections are included in the usual way by the fixed NLO calculation for $N+1$ jets.

\subsection{Brief Overview of Our Construction}

The starting point of our approach is the separation of the inclusive $N$-jet cross section into an exclusive $N$-jet region and an inclusive $(N+1)$-jet region,
\begin{equation}\label{eq:inclxsec}
\sigma_{\geq N}
= \int\!\df\Phi_N\, \frac{\df\sigma}{\df\Phi_N}(\Tau^\cut)
+ \int\!\df\Phi_{N+1}\,\frac{\df\sigma}{\df\Phi_{N+1}}(\Tau) \,\theta(\Tau > \Tau^\cut)
\,.\end{equation}
Here $\Tau \equiv \Tau(\Phi_{N+1})$ is a suitable resolution variable,
which is a function of $\Phi_{N+1}$, and $\df\sigma/\df\Phi_{N+1}(\Tau)$ denotes the
fully differential cross section for a given $\Tau$. In ME/PS merging, this role is
played by the variable that determines the merging scale.
However, in our case the parameter $\Tau^\cut$ is not a jet-merging cut but
instead serves as an infrared cutoff for the calculation of
$\df\sigma/\df\Phi_{N+1}(\Tau)$ and ideally is taken as small as possible.

In the $N$-jet region at small $\Tau$ (both above and below $\Tau^\cut$), logarithms of $\Tau$ become large and must be resummed to maintain consistent perturbative accuracy to some order in $\alpha_s$. On the other hand, in the $(N+1)$-jet region at large $\Tau$, a fixed-order expansion in $\alpha_s$ will suffice. To consistently match the resummed and fixed-order calculations, we use the following prescription for the jet cross sections entering in \eq{inclxsec}:
\begin{align}\label{eq:introcumspec} 
\frac{\df\sigma}{\df\Phi_{N}}(\Tau^\cut)
&= \frac{\df\sigma^\resum}{\df\Phi_N}(\Tau^\cut)
+ \biggl[\frac{\df\sigma^\FO}{\df\Phi_{N}}(\Tau^\cut)
- \frac{\df\sigma^\resum}{\df\Phi_N}(\Tau^\cut)\bigg\vert_\FO \biggr]
\,, \nn \\[1ex]
\frac{\df\sigma}{\df\Phi_{N+1}}(\Tau)
&= \frac{\df\sigma^\FO}{\df\Phi_{N+1}}(\Tau)
\biggl[\frac{\df\sigma^\resum}{\df\Phi_N \df\Tau}\bigg/\frac{\df\sigma^\resum}{\df\Phi_N\,\df\Tau}\bigg\vert_\FO \biggr] 
\,.\end{align}
The superscript ``resum'' indicates an analytically resummed calculation and ``FO'' indicates a fixed-order calculation or expansion.  This construction properly reproduces the fixed-order calculation at large $\Tau$, the resummed calculation at small $\Tau$, and smoothly interpolates between them.

It is straightforward to extend our formulation to combine higher jet multiplicities at NLO with higher-order resummation, as we will show. This is done by replacing $\df \sigma^\FO/\df \Phi_{N+1}$ in \eq{introcumspec} with an inclusive $(N+1)$-jet cross section separated into the exclusive $(N+1)$-jet and inclusive $(N+2)$-jet cross sections and iteratively applying \eq{introcumspec}.

The key ingredients in our approach are the higher-order resummation of the jet resolution variable, the fully differential fixed-order calculation, and the parton shower and hadronization. While each of these components is known, there is a sensitive interplay of constraints between them that must be satisfied to achieve a consistent combination. This is precisely what is accomplished in the \geneva framework and is the focus of this paper.

\section{General Construction}
\label{sec:master}

In this section, we derive our theoretical construction in a process-independent manner.
We start in \subsec{singlediff} with a slightly simplified setup, considering the singly differential
spectrum in the jet resolution variable. We use this to discuss in detail the perturbative
structure and the accuracy in the different phase space regions. In \subsec{fullydiff}, we discuss the
extension to the fully differential case and how to combine the fixed-order expansion and resummation
in this situation. In \subsec{morejets}, we further generalize these results to include several jet multiplicities
by iteration. Finally, in \subsec{partonshower}, we discuss the Monte Carlo implementation
and how to attach parton showering and hadronization.

\subsection{What Resummation Can Do for Monte Carlo}
\label{subsec:singlediff}

\subsubsection{Basic Setup}

The basic idea of Monte Carlo integration is to randomly generate points in phase space (``events'') that are distributed according to some differential (probability) distribution. By summing over all points that satisfy certain selection criteria, we are able to perform arbitrary integrals of the distribution. In our case, that distribution is the fully differential cross section, allowing one to compute arbitrary observables. For simplicity, we will first focus on the singly differential cross section in some phase space resolution (or jet resolution) variable $\Tau$ of dimension one. The precise definition of $\Tau$ is not important at the moment, so we keep it generic for now. We use the convention that the limit $\Tau \to 0$ corresponds to Born kinematics, i.e., the tree-level cross section is $\sim\delta(\Tau)$. We also require that $\Tau$ is an IR-safe observable, such that the differential cross section $\df\sigma/\df\Tau$ can in principle be well defined to all orders in perturbation theory and for $\Tau > 0$ contains no IR divergences.

To give an example, for our application to $e^+e^- \to 2/3$ jets in \sec{ee}, we will use 2-jettiness $\Tau_2 = \Ecm(1 - T)$, where $T$ is the usual thrust~\cite{Farhi:1977sg}. Alternatives include other $2$-jet event shapes. For Drell-Yan in \sec{pp}, we will use beam thrust~\cite{Stewart:2009yx}. An alternative would be the $p_T$ of the leading jet. If the Born cross section we are interested in has $N$ signal jets,%
\footnote{As usual, we assume that the Born cross section is defined with appropriate cuts on the $N$ signal jets, so that it does not contain any IR divergences by itself.} then $\Tau$ could be $N$-jettiness or the largest $p_T$ of any additional jet.  The important point is that we can think of $\Tau$ as a resolution variable which determines the scale of additional emissions in the $\Phi_{\geq N+1}$ phase space, such that for $\Tau \leq \Tau^\cut$ there are no emissions above the scale $\Tau^\cut$. For later convenience, we also define the dimensionless equivalent of $\Tau$ as
\begin{equation}
\tau = \frac{\Tau}{Q}
\,.\end{equation}
Here, $Q$ is the relevant hard-interaction scale in the Born process, e.g., $Q\equiv\Ecm$ for $e^+e^-\to$ jets or $Q\equiv m_{\ell^+\ell^-}$ for Drell-Yan $pp\to Z/\gamma^* \to \ell^+ \ell^-$. In terms of $\tau$, the limit $\tau\ll 1$ corresponds to the exclusive limit close to Born kinematics. For $\tau \sim 1$, there are additional emissions at the hard scale $\Tau\sim Q$, which means we are far away from Born kinematics and we should switch the description to consider the corresponding Born process with one additional hard jet.

To describe the differential $\Tau$ spectrum, we want the Monte Carlo to generate events at specific values of $\Tau$, which are distributed according to the differential cross section $\df\sigma/\df\Tau$. The total cross section is then simply given by summing over all events, 
\begin{equation}
\sigma = \int \!\df \Tau \,\frac{\df \sigma}{\df \Tau}
\,.\end{equation}
The essential problem every Monte Carlo generator faces is that in perturbation theory the differential cross section $\df\sigma/\df\Tau$ contains IR divergences from real emissions for $\Tau\to 0$, which only cancel against the corresponding virtual IR divergences upon integration over the $\Tau\to 0$ region. As a result, the perturbative spectrum for $\Tau\to 0$ can only be defined in a distributional sense in terms of plus and delta distributions [see \eq{dsigmadtausing} below]. To deal with this, we have to introduce a small cutoff $\Tau^\cut$ and define the cumulant of the spectrum as 
\begin{equation}
\sigma(\Tau^\cut) = \int \!\df \Tau \,\frac{\df \sigma}{\df \Tau}\,  \theta(\Tau < \Tau^\cut)
\,.\end{equation}
In the Monte Carlo, the total cross section is then obtained by combining the cumulant and spectrum as
\begin{equation} \label{eq:MCsinglediff}
\sigma = \sigma(\Tau^\cut) + \int \!\df \Tau \,\frac{\df \sigma}{\df \Tau}\,  \theta(\Tau > \Tau^\cut)
\,.\end{equation}
In practice, this is implemented by generating two distinct types of events: (i) events that have $\Tau = 0$ and relative weights given by $\sigma(\Tau^\cut)$, and (ii) events that have nonzero values $\Tau > \Tau^\cut$ and relative weights given by $\df\sigma/\df\Tau$. The first type of events have Born kinematics and represents the tree-level and virtual corrections together with the corresponding real emissions integrated below $\Tau^\cut$. The second type of events contains one or more partons in the final state, since the real-emission corrections determine the shape of the spectrum for nonzero $\Tau$. We now have two basic conditions:
\begin{enumerate}
\item From a numerical point of view, we want the value of $\Tau^\cut$ to be as small as possible, so as to describe as much differential information as possible. In practice, our ability to reliably compute the cumulant $\sigma(\Tau^\cut)$ in perturbation theory sets a lower limit on the possible value of $\Tau^\cut \gtrsim$ few times $\lqcd$.
\item Since $\Tau^\cut$ is an unphysical parameter, we want the dependence on it to drop out (to the order we are working at). In practice, this is guaranteed by including the corresponding dominant higher-order corrections in the cumulant and spectrum.
\end{enumerate}

\subsubsection{Perturbative Expansion and Order Counting}

In perturbation theory, the differential cross section in $\tau$ and the cumulant in $\tau^\cut$ have the general form
\begin{equation} \label{eq:dsigmadtau}
\frac{\df\sigma}{\df\tau}
= \frac{\df\sigma^\sing}{\df\tau} + \frac{\df\sigma^\nons}{\df\tau}
\,,\qquad
\sigma(\tau^\cut)
= \int_0^{\tau^\cut}\!\!\df\tau\,\frac{\df\sigma}{\df\tau}
= \sigma^\sing(\tau^\cut) + \sigma^\nons(\tau^\cut)
\,,\end{equation}
where we have distinguished ``singular'' and ``nonsingular'' contributions. For $\tau\to 0$, the singular terms in $\df\sigma^\sing/\df\tau$ scale like $1/\tau$, while the nonsingular terms in $\df\sigma^\nons/\df\tau$ contain at most integrable singularities. For the cumulant, this means that $\sigma^\sing(\tau^\cut)$ contains all terms in $\sigma(\tau^\cut)$ enhanced by logarithms $\ln^k(\tau^\cut)$, while $\sigma^\nons(\tau^\cut = 0) = 0$.

The singular part of the spectrum is given by
\begin{equation} \label{eq:dsigmadtausing}
\frac{\df\sigma^\sing}{\df\tau} = \sigma_B \Bigl[ C_{-1}(\alpha_s)\,\delta(\tau) + \sum_{n\geq 0} C_n(\alpha_s)\, \cL_n(\tau) \Bigr]
\,,\end{equation}
where $\sigma_B$ denotes the Born cross section, and we denote the usual plus distributions as
\begin{equation}
\cL_n(x) = \biggl[\frac{\theta(x)\ln^n(x)}{x}\biggr]_+
\,,\qquad
\int_0^{x^\cut}\!\df x\, \cL_n(x) = \frac{\ln^{n+1}(x^\cut)}{n+1}
\,.\end{equation}
They encode the cancellation between real and virtual IR divergences. The corresponding singular contribution to the cumulant cross section integrated up to $\tau \leq \tau^\cut$ is
\begin{align}
\sigma^\sing(\tau^\cut) &= \sigma_B \biggl[ C_{-1}(\alpha_s) + \sum_{n\geq 0} C_n(\alpha_s)\, \frac{\ln^{n+1}(\tau^\cut)}{n+1} \biggr]
\,.\end{align}
At $\ord{\alpha_s^k}$, only the coefficients $C_n(\alpha_s)$ with $n \leq 2k-1$ contribute, so $\df\sigma/\df\tau$ has logarithms up to $\alpha_s^n L^{2n-1}/\tau$, while $\sigma(\tau^\cut)$ has logarithms up to $\alpha_s^n L_\cut^{2n}$, where we use the abbreviations
\begin{equation}
L \equiv \ln(\tau)
\,,\qquad
L_\cut \equiv \ln(\tau^\cut)
\,.\end{equation}

The $\alpha_s$ expansion of the coefficients $C_{-1}(\alpha_s)$ and $C_n(\alpha_s)$ in the singular contributions can be written as
\begin{equation} \label{eq:Cndef}
C_{-1}(\alpha_s) = 1 + \sum_{k\geq 1} c_{k,-1}\, \alpha_s^k
\,,\qquad
C_n(\alpha_s) = \sum_{2k \geq n+1} c_{kn}\, \alpha_s^k
\,.\end{equation}
Similarly, the $\alpha_s$ expansion of the nonsingular contributions can be written as
\begin{equation} \label{eq:fnonsexpansion}
\frac{\df\sigma^\nons}{\df\tau} = \sigma_B\, \sum_{k\geq 1} f_k^\nons(\tau)\,\alpha_s^k
\,,\qquad
F_k^\nons(\tau^\cut) = \int_0^{\tau^\cut}\!\df\tau\, f_k^\nons(\tau)
\,.\end{equation}
Using \eqs{Cndef}{fnonsexpansion}, the spectrum and cumulant up to $\ord{\alpha_s^2}$ are given by
\begin{align} \label{eq:fixedas2}
\frac{1}{\sigma_B} \frac{\df\sigma}{\df\tau} \bigg\vert_{\tau > 0}
&=\quad \frac{\alpha_s}{\tau} \Bigl[ c_{11} L \,\,+ c_{10} \mspace{120mu} + \tau f_1^\nons(\tau) \Bigr]
\nn\\ & \quad
+  \frac{\alpha_s^2}{\tau} \Bigl[c_{23} L^3 + c_{22} L^2 + c_{21} L + c_{20} + \tau f_2^\nons(\tau) \Bigr]
+ \ord{\alpha_s^3}\,,
\\[1ex]\nn
\frac{1}{\sigma_B} \sigma(\tau^\cut)
&= 1 + \alpha_s \Bigl[ \frac{c_{11}}{2} L_\cut^2 \,+ c_{10} L_\cut \,\,+ c_{1,-1} \mspace{155mu} + F_1^\nons(\tau^\cut) \Bigr]
\nn\\\nn & \qquad
+  \alpha_s^2 \Bigl[ \frac{c_{23}}{4} L_\cut^4 + \frac{c_{22}}{3} L_\cut^3 + \frac{c_{21}}{2} L_\cut^2 + c_{20} L_\cut + c_{2,-1} + F_2^\nons(\tau^\cut) \Bigr]
+ \ord{\alpha_s^3}
\,.\end{align}
Note that the $c_{k,-1}$ constant term in the singular corrections, which contains the finite virtual corrections to the Born process, only appears in the cumulant.

\begin{figure*}[t!]
\begin{center}
\includegraphics[scale=0.5]{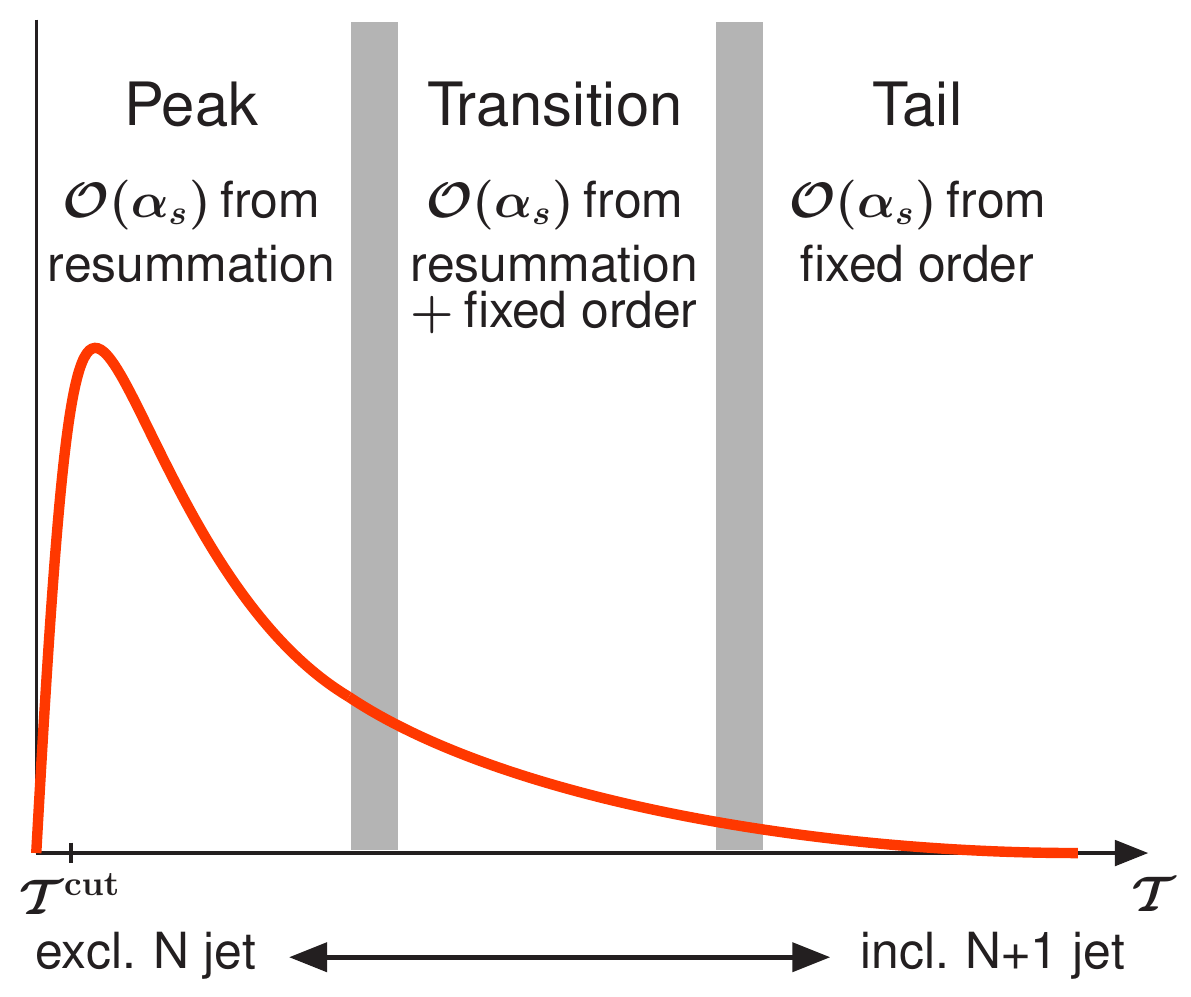}%
\end{center}
\vspace{-0.5ex}
\caption{ Illustration of the different parametric regions in the jet resolution.}
\label{fig:regions}
\end{figure*}

We now distinguish three parametrically different regions in $\tau$, which are illustrated in \fig{regions}:
\begin{itemize}
\item \emph{Resummation (``peak'') region $\tau \ll 1$}:
In this limit, the logarithms in the singular contributions are large, such that parametrically one has to count%
\footnote{In analytic resummation, the counting and resummation of logarithms is performed in the exponent of the cross section, where one counts $\alpha_s L \sim 1$. For the purpose of our argument in this section, it is sufficient to adopt the weaker scaling in \eq{logcounting} and only count logarithms in the cross section. In our results, we always perform the full resummation in the exponent, as discussed in \subsubsec{eeResum}.}
\begin{equation} \label{eq:logcounting}
\alpha_s L^2 \sim 1
\,,\qquad
\alpha_s L_\cut^2 \sim 1
\,.\end{equation}
This means one has to resum the towers of logarithms $(\alpha_s L^2)^n$ in the spectrum and $(\alpha_s L_\cut^2)^n$ in the cumulant in \eq{fixedas2} to all orders in $\alpha_s$ to obtain a meaningful perturbative approximation at some order.
At the same time, the nonsingular corrections can be regarded as power suppressed, since they are of relative $\ord{\tau}$.
\item \emph{Fixed-order (``tail'') region $\tau \sim 1$}:
In this limit, the logarithms are not enhanced, and a fixed-order expansion in $\alpha_s$ is applicable. The singular and nonsingular contributions are equally important and both must be included at the same order in $\alpha_s$. In particular, there are typically large cancellations between these for $\tau\sim 1$, so it is actually crucial not to resum the singular contributions in this region, since otherwise this cancellation would be spoiled.
\item \emph{Transition region}: The transition between the resummation and fixed-order regions.
\end{itemize}
There are of course no strict boundaries between the different regions. This is why it is important to have a proper description not just in the two limits but also in the transition region, which connects the resummation and fixed-order regions. In fact, in practice the experimentally relevant region is often somewhere in the transition region, where both types of perturbative corrections can be important.

\subsubsection{Lowest Perturbative Accuracy}
\label{subsubsec:lo}

For the Monte Carlo to provide a proper description at all values of $\Tau$, it has to include at least the lowest-order terms relevant for each region. Keeping only these, and dropping all other terms, the spectrum and the cumulant are given by
\begin{align} \label{eq:lowestorder}
\frac{1}{\sigma_B}\frac{\df\sigma}{\df\tau} \bigg\vert_{\tau > 0}
&= \frac{\alpha_s}{\tau} \Bigl[ L\, f_0(\alpha_s L^2) + f_1(\alpha_s L^2) + \tau f_1^\nons(\tau) \Bigr]
\,,\nn\\[1ex]
\frac{1}{\sigma_B} \sigma(\tau^\cut)
&= 1 + \alpha_s \Bigl[ L_\cut^2\, F_0(\alpha_s L_\cut^2) + L_\cut\, F_1(\alpha_s L_\cut^2) \Bigr]
\,.\end{align}
where the functions $f_{0,1}$ and $F_{0,1}$ are given in terms of the coefficients $c_{ij}$ in \eq{Cndef} as
\begin{align}
\mathrm{LL}_\sigma: && f_0(\alpha_s L^2) &= \sum_{n \geq 0} c_{n+1,2n+1} (\alpha_s L^2)^n
\,,&
F_0(\alpha_s L^2) &= \sum_{n\geq 0}\frac{c_{n+1,2n+1}}{2(n+1)} (\alpha_s L_\cut^2)^n
\,,\nn\\
\mathrm{NLL}_\sigma: && f_1(\alpha_s L^2) &= \sum_{n\geq 0} c_{n+1,2n} (\alpha_s L^2)^n
\,,&
F_1(\alpha_s L^2) &= \sum_{n\geq 0} \frac{c_{n+1,2n}}{2n+1} (\alpha_s L_\cut^2)^n
\,.\end{align}
The $f_0$ and $F_0$ resum the leading-logarithmic series in the cross section, which we denote as LL$_\sigma$. The functions $f_1$ and $F_1$ resum the next-to-leading-logarithmic series in the cross section, which we denote as NLL$_\sigma$.

In the resummation region at $\tau \ll 1$, the LL$_\sigma$ terms in the spectrum scale as $L \sim 1/\sqrt{\alpha_s}$ (relative to the overall $\alpha_s/\tau$ scaling) and provide the lowest level of approximation. The NLL$_\sigma$ terms scale as $\sim 1$, and one can argue about whether they are needed as well in order to get a meaningful lowest-order prediction. Formally, they are necessary to obtain the spectrum at $\sim\alpha_s/\tau$, which one might consider the natural leading-order scaling of the spectrum (or equivalently if one does not want to rely on the $\sim 1/\sqrt{\alpha_s}$ enhancement of the LL series). Experience shows that the NLL terms are indeed numerically important. For example, in analytic resummations, one rarely gets a sensible prediction without going at least to NLL. Similarly, to obtain sensible predictions from a parton shower, it is almost mandatory to include important physical effects such as momentum conservation in the parton splitting and the choice of $\alpha_s$ scale~\cite{Catani:1990rr}. In the cumulant, the LL$_\sigma$ series in $F_0$ scales as $\sim 1$ and must be included. The NLL$_\sigma$ series in $F_1$ scales as $\sim\sqrt{\alpha_s}$ and, for consistency, should be included in the cumulant if it is included in the spectrum.

In the fixed-order region at $\tau\sim 1$, the lowest meaningful order in the spectrum is given by the complete $\ord{\alpha_s}$ terms, requiring one to include the $c_{11}$ and $c_{10}$ terms, which are part of the $f_0$ and $f_1$ functions, as well as the nonsingular corrections $f_1^\nons(\tau)$. Since we take $\tau^\cut$ to be small, the cumulant is always in the resummation region. Hence, its nonsingular corrections $F_1^\nons(\tau)$ [see \eq{fixedas2}] are suppressed by $\ord{\alpha_s \tau^\cut}$ and can be safely neglected.

The leading level of accuracy in \eq{lowestorder} closely corresponds to what is achieved in the standard ME/PS matching. In this case, the LL resummation is provided by the parton shower Sudakov factors (either generated by the shower or multiplied by hand), where the jet resolution variable corresponds to the shower evolution variable, since that is the variable for which the shower directly resums the correct LL$_\sigma$ series. The LL$_\sigma$ series has a well-known and simple exponential structure,
\begin{equation}
c_{n+1,2n+1} = \frac{c_{11}^{n+1}}{2^n n!}
\qquad\Rightarrow\qquad
f_0(\alpha_s L^2) = \exp\Bigl[\frac{c_{11}}{2}\, \alpha_s L^2\Bigr]
\,,\end{equation}
such that
\begin{align} \label{eq:PS}
\frac{1}{\sigma_B}\frac{\df\sigma}{\df\tau} \bigg\vert_{\tau > 0}
= c_{11}\,\alpha_s\frac{L}{\tau}\, \exp\Bigl[\frac{c_{11}}{2}\, \alpha_s L^2\Bigr]
\,,\qquad
\frac{1}{\sigma_B} \sigma(\tau^\cut)
= \exp\Bigl[\frac{c_{11}}{2}\, \alpha_s L_\cut^2\Bigr]
\,.\end{align}
The resummation exponent at LL$_\sigma$ is given by the integral over the leading $c_{11}\, \alpha_s \ln(\tau)/\tau$ term in the spectrum. This is precisely what the standard parton shower veto algorithm exploits to generate the resummation exponent. The analogous structure does not hold at NLL$_\sigma$, which is why the parton shower cannot resum the NLL$_\sigma$ series by exponentiating the integral of the $c_{10}\,\alpha_s/\tau$ term. As already mentioned, in practice, parton showers include important partial NLL effects, so practically this provides a numerically close approximation to the correct NLL$_\sigma$ series. The nonsingular corrections in the spectrum, $f_1^\nons(\tau)$, are obtained by including the full tree-level matrix element for one additional emission. Since the full matrix element also includes the $c_{11}$ and $c_{10}$ terms, this requires a proper matching procedure to avoid double counting these terms. At LL$_\sigma$, a simple way to do this is to multiply the full fixed-order result from the matrix element with the shower's LL$_\sigma$ resummation exponent,
\begin{equation} \label{eq:CKKWL}
\frac{1}{\sigma_B}\frac{\df\sigma}{\df\tau} \bigg\vert_{\tau > 0}
= \frac{\alpha_s}{\tau} \Bigl[ c_{11} L + c_{10} + \tau f_1^\nons(\tau) \Bigr] \exp\Bigl[\frac{c_{11}}{2}\, \alpha_s L^2\Bigr]
\,,\end{equation}
which corresponds to the CKKW-L~\cite{Catani:2001cc, Lonnblad:2001iq, Krauss:2002up, Lavesson:2005xu} procedure. The reason this gives the spectrum correctly at LL$_\sigma$ is the simple structure in \eq{PS}, where the LL$_\sigma$ exponent multiplies the $c_{11}$ term in the spectrum.\footnote{As before, since this simple LL$_\sigma$ structure does not hold in general at NLL$_\sigma$, this procedure does not yield the resummed spectrum at NLL$_\sigma$, even if one were to multiply the spectrum with the NLL$_\sigma$ resummation exponent}. At large $\tau\sim1$, the exponent in \eq{CKKWL} can be expanded as $1 + \ord{\alpha_s}$, so \eq{CKKWL} gives the correct leading fixed-order result.

Compared to \eq{lowestorder}, the NLO matching performed in \mcatnlo and \powheg amounts to adding to the cumulant the $c_{1,-1}$ singular constant, containing the $\ord{\alpha_s}$ virtual corrections, as well as the nonsingular contributions $F_1^\nons(\tau)$. Assuming the same set of NLL terms are included in the cumulant and spectrum, this achieves that inclusive quantities that are integrated over a large range of $\tau$, such as the total cross section, are correctly reproduced at fixed NLO, which provides them with $\ord{\alpha_s}$ accuracy. In these approaches, the goal is not to improve the perturbative accuracy of the spectrum (or the cumulant at small $\tau^\cut$), which has the same leading accuracy as in \eq{CKKWL}.

\subsubsection{Next-To-Lowest Perturbative Accuracy}
\label{subsubsec:nlo}

We now want to improve the Monte Carlo description in \eq{MCsinglediff} from the lowest-order accuracy, given by \eq{lowestorder}, to the next-to-lowest perturbative accuracy in $\alpha_s$. This requires us to include the appropriate higher-order corrections in each region, which gives
\begin{align} \label{eq:nexttolowestorder}
\frac{1}{\sigma_B}\frac{\df\sigma}{\df\tau} \bigg\vert_{\tau > 0}
&= \frac{\alpha_s}{\tau} \Bigl[ L\, f_0(\alpha_s L^2) + f_1(\alpha_s L^2) + \tau f_1^\nons(\tau) \Bigr]
\nn\\ & \quad
+  \frac{\alpha_s^2}{\tau} \Bigl[L\, f_2(\alpha_s L^2) + f_3(\alpha_s L^2) + \tau f_2^\nons(\tau) \Bigr]
\,,\nn\\[1ex]
\frac{1}{\sigma_B} \sigma(\tau^\cut)
&= 1 + \alpha_s \Bigl[ L_\cut^2\, F_0(\alpha_s L_\cut^2) + L_\cut\, F_1(\alpha_s L_\cut^2)
+ c_{1,-1} + F_1^\nons(\tau^\cut) \Bigr]
\nn\\ & \quad
+  \alpha_s^2 \Bigl[ L_\cut^2\, F_2(\alpha_s L_\cut^2) + L_\cut\, F_3(\alpha_s L_\cut^2)
\Bigr]
\,,\end{align}
where we denote the series of logarithms resummed by the functions $f_2$ and $F_2$ by NLL$'_\sigma$ and the series resummed by $f_3$ and $F_3$ by NNLL$_\sigma$. They can again be written in terms of the $c_{ij}$ coefficients in \eq{Cndef} as
\begin{align}
\mathrm{NLL}'_\sigma: && f_2(\alpha_s L^2) &= \sum_{n \geq 0} c_{n+2,2n+1} (\alpha_s L^2)^n
\,,&
F_2(\alpha_s L^2) &= \sum_{n\geq 0}\frac{c_{n+2,2n+1}}{2(n+1)} (\alpha_s L_\cut^2)^n
\,,\nn\\
\mathrm{NNLL}_\sigma: && f_3(\alpha_s L^2) &= \sum_{n\geq 0} c_{n+2,2n} (\alpha_s L^2)^n
\,,&
F_3(\alpha_s L^2) &= \sum_{n\geq 0} \frac{c_{n+2,2n}}{2n+1} (\alpha_s L_\cut^2)^n
\,.\end{align}
In the resummation region, the NLL$'_\sigma$ series in the spectrum scales as $\sim\alpha_s^{3/2}$ and thus provides the $\sim\alpha_s$ correction to the LL$_\sigma$ series in $f_0$. Similarly, the NNLL$_\sigma$ series scales as $\sim\alpha_s^2$ providing the $\sim\alpha_s$ correction to the NLL$_\sigma$ series in $f_1$.
They can again be obtained by performing the standard resummation in the exponent of the cross section to NLL$'$ and NNLL respectively. (Here, NLL$'$ refers to those parts of the full NNLL resummation that arise from the combination of the one-loop matching corrections with the NLL resummation, see \subsubsec{eeResum} and \tab{expcounting}.)

In the fixed-order region, increasing the perturbative accuracy by $\sim\alpha_s$ requires the complete $\ord{\alpha_s^2}$ corrections, including the $f_2^\nons(\tau)$ nonsingular corrections. Similarly, for the cumulant, $F_2$ and $F_3$ resum the NLL$'_\sigma$ and NNLL$_\sigma$ series of logarithms, which scale as $\sim\alpha_s$ and $\sim\alpha_s^{3/2}$, respectively, and provide the $\sim\alpha_s$ improvement over the LL$_\sigma$ and NLL$_\sigma$ series in $F_0$ and $F_1$. In addition, going to the next higher order in the cumulant requires including the full singular constant $c_{1,-1}$,%
\footnote{Formally, $c_{1,-1}$ belongs to the NLL$'_\sigma$ series in the cumulant, but for the sake of discussion, we keep it explicit.}
as well as the nonsingular corrections $F_1^\nons(\tau)$, which both scale as $\sim\alpha_s$.

It is instructive to see where the information from the virtual NLO matrix elements enters in \eq{nexttolowestorder}. As already mentioned, the virtual NLO corrections to the Born process are given by $c_{1,-1}$. In addition, by multiplying the LL series it contributes part of $f_2$ and $F_2$. Hence, consistently combining the virtual corrections with the resummation requires one to go to at least NLL$'$. The virtual NLO corrections with one extra emission (plus the integral over the two-emission tree-level matrix element) yield the full $\ord{\alpha_s^2}$ corrections in the spectrum, i.e., both the singular $c_{2k}$ terms as well as the nonsingular $f_2^\nons$ terms in \eq{fixedas2}. Adding these corrections again requires one to avoid double counting the singular $c_{2k}$ terms that are already included in the resummation. In analytic resummation, it is well known how to do this, namely by simply adding the nonsingular corrections. These are obtained by taking the difference of the full NLO corrections and the singular NLO corrections, where the latter are given by expanding the resummed result to fixed order. Since this construction involves the virtual contribution to both the Born process and the process with one extra emission, we see that going consistently to higher order in both the resummation and fixed-order regions naturally leads to a combination of the information from two successive NLO matrix elements.

\subsubsection{Merging NLO Matrix Elements with Parton Shower Resummation Only}
\label{subsubsec:existing}

We stress that, for a description at the next-higher perturbative accuracy across the whole range in $\tau$, it is not sufficient to include the fixed NLO corrections to the spectrum and take care of the double counting with the parton shower resummation. This only provides the proper NLO description in the fixed-order region at large $\tau$. In the transition and resummation regions, a proper higher-order description necessitates higher-order resummation. Of course, this is not a problem if the only goal is to improve the fixed-order region at large $\tau$, as is the case for example in a recent \mcatnlo publication~\cite{Frederix:2012ps}.

However, including the fixed NLO corrections outside the fixed-order region, as is done in \sherpa's recent NLO merging~\cite{Hoeche:2012yf, Gehrmann:2012yg}, can actually make things worse in two respects: 
First, numerically this will typically force the spectrum to shift toward the fixed-order result and away from the resummed one. Since this can shift the spectrum in the wrong direction, it can potentially make the result \emph{less} accurate.%
\footnote{One can see this for example in the case of $2$-jettiness in \fig{TauVal} in \sec{ee}. Here, the NLL$'$+LO$_3$ result is much closer to the slightly higher NNLL$'$+NLO$_3$ best prediction than the fixed NLO$_3$ result. We have checked that in this case, adding the NLO$_3$ to the NLL$'$+LO$_3$ by expanding it to $\ord{\alpha_s^2}$ forces the result to move in the wrong direction toward the lower NLO$_3$.}
At the same time, the perturbative uncertainties from fixed-order scale variation decrease, which only aggravates this problem. Multiplying the NLO corrections to the spectrum with LL parton shower Sudakov factors (see, e.g., ref.~\cite{Hamilton:2012np}) can mitigate this to some extent but does not solve the problem.
The only consistent way to include the fixed NLO corrections to the spectrum outside the fixed-order region, and in particular obtain reliable perturbative uncertainties, is to properly combine them with a higher-order resummation.

Second, this explicitly spoils the formal $\ord{\alpha_s}$ accuracy of the inclusive cross section. To see this, consider adding the fixed NLO corrections to the lowest-order spectrum and cumulant in \eq{lowestorder}, properly taking care of the double counting at $\ord{\alpha_s^2}$, which gives
\begin{align}
\frac{1}{\sigma_B}\frac{\df\sigma}{\df\tau} \bigg\vert_{\tau > 0}
&= \frac{\alpha_s}{\tau} \Bigl[ L\, f_0(\alpha_s L^2) + f_1(\alpha_s L^2) + \tau f_1^\nons(\tau) \Bigr]
+  \frac{\alpha_s^2}{\tau} \Bigl[c_{21} L + c_{20} + \tau f_2^\nons(\tau) \Bigr]
\,,\nn\\[1ex]
\frac{1}{\sigma_B} \sigma(\tau^\cut)
&= 1 + \alpha_s \Bigl[ L_\cut^2\, F_0(\alpha_s L_\cut^2) + L_\cut\, F_1(\alpha_s L_\cut^2)
+ c_{1,-1} + F_1^\nons(\tau^\cut) \Bigr]
\,.\end{align}
Using these expressions yields for the inclusive cross section
\begin{align} \label{eq:naiveNLO}
\frac{1}{\sigma_B} \sigma
&=  \frac{1}{\sigma_B} \sigma(\tau^\cut) + \int_{\tau^\cut}^1 \!\df\tau \frac{1}{\sigma_B}\frac{\df\sigma}{\df\tau}
\nn\\
&= 1 + \alpha_s \Bigl[ c_{1,-1} + F_1^\nons(1) \Bigr] - \alpha_s^2 \Bigl[ \frac{c_{21}}{2} L_\cut^2 + c_{20} L_\cut \Bigr]
\,.\end{align}
While the first two terms give the correct NLO inclusive cross section, the $\ord{\alpha_s^2}$ terms induced by the fixed NLO corrections in the spectrum formally scale as $\alpha_s$ and $\alpha_s^{3/2}$ and therefore spoil the formal $\ord{\alpha_s}$ perturbative accuracy for the inclusive cross section and in fact for any inclusive observable. This directly contradicts the claim in refs.~\cite{Hoeche:2012yf, Gehrmann:2012yg} that this description maintains the higher-order accuracy of the underlying matrix elements in their respective phase space range. It only preserves the fixed $\ord{\alpha_s}$ terms, which in the context of combining fixed-order corrections with a logarithmic resummation is necessary but not sufficient to preserve the higher perturbative accuracy.

This problem cannot be avoided by multiplying the $\alpha_s^2$ corrections in the spectrum with the LL parton shower Sudakov factors, since this does not provide the proper NLL$'_\sigma$ and NNLL$_\sigma$ series. Note also that we have already assumed in \eq{naiveNLO} that the full NLL$_\sigma$ series is included in the spectrum and cumulant. In general, the parton shower cannot provide this, which means there will be even $\alpha_s^2 L_\cut^3 \sim \sqrt{\alpha_s}$ terms induced in \eq{naiveNLO}.

Pragmatically, the inclusive cross section can be restored to formal $\ord{\alpha_s}$ accuracy by either explicitly including the corresponding $\alpha_s^2$ corrections in the cumulant to cancel these terms, where numerical methods to do so have been described very recently in refs.~\cite{Lonnblad:2012ng, Platzer:2012bs, Lonnblad:2012ix}, or alternatively by explicitly restricting the fixed NLO corrections in the spectrum to the fixed-order region at large $\tau$, such that the induced $\ord{\alpha_s^2}$ terms in the total cross section are not logarithmically enhanced and are formally $\ord{\alpha_s^2}$. This is essentially the approach taken in ref.~\cite{Frederix:2012ps}. However, neither of these approaches improves the perturbative accuracy in the spectrum outside the fixed-order region.

\subsection{What Monte Carlo Can Do for Resummation}
\label{subsec:fullydiff}

For $\Tau$ being the resolution variable between $N$ and more than $N$ jets, we showed in the previous subsection that combining the NLO matrix-element corrections for $N$ and $N+1$ partons at the level of the singly differential $\Tau$ spectrum is equivalent to combining the NNLL resummation of the singular contributions with the higher-order nonsingular contributions. Our goal now is to extend this singly differential description to the fully differential case, in order to use the full $N$-parton and $(N+1)$-parton information of the matrix elements. We will use the notation (N)LO$_N$ or (N)LO$_{N+1}$ to indicate up to which fixed order in $\alpha_s$ the $N$-parton or $(N+1)$-parton matrix elements are included.

To start with, it is straightforward to generalize the jet resolution spectrum $\df\sigma/\df\Tau$ and its cumulant $\sigma(\Tau^\cut)$ to include the full dependence on the $N$-body Born phase space,
\begin{align}
\frac{\df\sigma}{\df\Tau} &\quad\to\quad \frac{\df\sigma}{\df\Phi_N\df\Tau}
\,,\nn\\
\sigma(\Tau^\cut) &\quad\to\quad \frac{\df \sigma}{\df\Phi_N}(\Tau^\cut)
= \int\!\df \Tau\, \frac{\df\sigma}{\df\Phi_N \df \Tau}\, \theta(\Tau < \Tau^\cut)
\,,\end{align}
such that \eq{MCsinglediff} becomes
\begin{equation} \label{eq:MCPhiNdiff}
\frac{\df\sigma_\incl}{\df\Phi_N}
= \frac{\df\sigma}{\df\Phi_N}(\Tau^\cut) + \int \!\df \Tau \,\frac{\df \sigma}{\df\Phi_N\df \Tau}\,  \theta(\Tau > \Tau^\cut)
\,.\end{equation}
Here, $\df\sigma_\incl/\df\Phi_N$ is the \emph{inclusive} $N$-jet cross section.
The discussion in \subsec{singlediff} can be precisely repeated in this case, since the perturbative structure of the differential spectrum $\df\sigma/\df\Phi_N\df\Tau$ with respect to $\Tau$ is precisely the same as in \eqs{dsigmadtau}{dsigmadtausing}.  Namely, we can write it as the sum of
singular and nonsingular contributions,
\begin{equation} \label{eq:dsigmadPhiNdtau}
\frac{\df\sigma}{\df\Phi_N\df\Tau}
= \frac{\df\sigma^\sing}{\df\Phi_N\df\Tau} + \frac{\df\sigma^\nons}{\df\Phi_N\df\Tau}
\,.\end{equation}
The nonsingular contributions are general functions of $\Phi_N$ and $\Tau$, but as before are integrable in $\Tau$ for $\Tau \to 0$. The singular contributions have the structure
\begin{align}
\frac{\df\sigma^\sing}{\df\Phi_N\df\Tau}
&= \frac{\df\sigma_B}{\df\Phi_N} \biggl[ C_{-1}(\Phi_N, \alpha_s)\,\delta(\Tau)
+ \sum_{n\geq 0} C_n(\Phi_N, \alpha_s)\, \frac{1}{Q}\cL_n\Bigl(\frac{\Tau}{Q}\Bigr) \biggr]
\,,\end{align}
where $\df\sigma_B/\df\Phi_N$ is now the fully differential Born cross section. Since the singular contributions arise from the cancellation of virtual and real IR singularities, which only know about $\Phi_N$, their $\Tau$ dependence naturally factorizes from the $\Phi_N$ kinematics of the underlying hard process. This is what allows the resummation of the singular terms to higher orders for a given point in $\Phi_N$. At LL, the entire $\Phi_N$ dependence is that of the Born cross section. At higher logarithmic orders, this is not the case anymore, since the coefficients $C_n$ can have nontrivial $\Phi_N$ dependence. In addition, the precise definition of $\Tau$ also becomes important. Depending on its definition, the higher-order singular coefficients can depend on clustering effects or other types of nonglobal logarithms~\cite{Dasgupta:2001sh, Dasgupta:2002bw, Delenda:2006nf, Hornig:2011tg, Kelley:2012kj, Kelley:2012zs}, which can be difficult to resum to high enough order with currently available methods. Therefore, it is important to choose a resolution variable with simple resummation properties. An example is $N$-jettiness, for which the complete NNLL resummation for arbitrary $N$ is known~\cite{Stewart:2010tn, Jouttenus:2011wh}. For the purpose of our discussion below, we will assume that a resummed result for the spectrum and its cumulant in \eq{MCPhiNdiff} at sufficiently high order is available to us.

We can think of the cumulant $\df \sigma/\df\Phi_N(\Tau_\cut)$ in \eq{MCPhiNdiff} as the \emph{exclusive} $N$-jet cross section with no additional emissions (jets) above the scale $\Tau^\cut$, while the spectrum $\df\sigma/\df\Phi_N\df\Tau$ for $\Tau > \Tau^\cut$ is the corresponding \emph{inclusive} $(N+1)$-jet cross section. While the cumulant $\df\sigma/\df\Phi_N(\Tau^\cut)$ is differential in $\df\Phi_N$ and thus already as differential as it can be, the spectrum contains a projection from the full $\df\Phi_{\geq N+1}$ phase space down to $\df\Phi_N\df\Tau$. To also be fully differential in the $(N+1)$-jet phase space, we can generalize \eq{MCPhiNdiff} to
\begin{align} \label{eq:MCfullydiff}
\frac{\df\sigma_\incl}{\df\Phi_N}
&= \frac{\df \sigma}{\df\Phi_N}(\Tau^\cut)
+ \int\!\frac{\df\Phi_{N+1}}{\df\Phi_N}\, \frac{\df\sigma}{\df\Phi_{N+1}}(\Tau)\, \theta(\Tau > \Tau^\cut)
\,,\end{align}
where $\df\sigma/\df\Phi_{N+1}(\Tau)$ denotes the fully differential spectrum for a given $\Tau \equiv \Tau(\Phi_{N+1})$. We explicitly denote the dependence on $\Tau$ and $\Tau^\cut$ to clearly distinguish the spectrum from the cumulant. We have also used the shorthand notation
\begin{equation} \label{eq:PhiNprojection}
\frac{\df\Phi_{N+1}}{\df\Phi_N} \equiv \df\Phi_{N+1} \delta(\Phi_N-\Phi_N(\Phi_{N+1}))
\,,\end{equation}
where $\Phi_N(\Phi_{N+1})$ denotes a projection from an $(N+1)$-body phase space point to an $N$-body phase space point. This projection defines what we mean by $N$ jets at higher orders in perturbation theory.
Note that beyond LO, both the cumulant $\df\sigma/\df\Phi_N(\Tau^\cut)$ and spectrum $\df\sigma/\df\Phi_{N+1}(\Tau)$ must be well-defined jet cross sections; i.e., they require a specific IR-safe projection from $\Phi_{\geq k+1}$ to $\Phi_k$ for both $k = N$ and $k = N+1$. We will see below where this definition enters. Using \eq{MCfullydiff} at the next-higher perturbative accuracy requires us to combine the higher-order resummation in $\Tau$ for the cumulant and spectrum with the fully exclusive $N$-jet and $(N+1)$-jet fixed-order calculations at NLO$_N$ and NLO$_{N+1}$. To achieve this, we have to construct appropriate expressions for the cumulant $\df \sigma/\df\Phi_N(\Tau^\cut)$ and the spectrum $\df\sigma/\df\Phi_{N+1}(\Tau)$, which  we do in the next two subsections.

\subsubsection{Matched Cumulant}
\label{subsubsec:cumulant}

We start by discussing the cumulant in \eq{MCfullydiff}. Since the resummation is naturally differential in the $\df\Phi_N$ of the underlying Born process, we can combine the resummed result with the fixed-order one by adding the fixed-order nonsingular contributions to it,
\begin{equation} \label{eq:sigmaNTaucut}
\frac{\df\sigma}{\df\Phi_{N}}(\Tau^\cut)
= \frac{\df\sigma^\resum}{\df\Phi_N}(\Tau^\cut)
+ \biggl[\frac{\df\sigma^\FO}{\df\Phi_{N}}(\Tau^\cut)
- \frac{\df\sigma^\resum}{\df\Phi_N}(\Tau^\cut)\bigg\vert_\FO \biggr]
\,.\end{equation}
The first term contains the resummed contributions, while the difference of the two terms in square brackets provides the remaining nonsingular corrections that have not already been included in the resummation. The NLO$_N$ fixed-order result is given by
\begin{equation} \label{eq:sigmaNLOPhiN}
\frac{\df\sigma^\NLO}{\df\Phi_{N}}(\Tau^\cut)
= B_N(\Phi_N) + V_N(\Phi_N) + \int\!\df\Tau \theta(\Tau < \Tau^\cut) \int\!\frac{\df\Phi_{N+1}}{\df\Phi_N\df\Tau}\, B_{N+1}(\Phi_{N+1})
\,,\end{equation}
where $B_N$ and $B_{N+1}$ are the $N$-parton and $(N+1)$-parton tree-level (Born) contributions, $V_N$ is the $N$-parton one-loop virtual correction, and we abbreviated
\begin{equation} \label{eq:PhiNTauprojection}
\frac{\df\Phi_{N+1}}{\df\Phi_N\df\Tau}
\equiv \df\Phi_{N+1}\,\delta[\Tau - \Tau(\Phi_{N+1})]\,\delta[\Phi_N - \Phi_N(\Phi_{N+1})]
\,.\end{equation}
Here, $\Tau(\Phi_{N+1})$ implements the definition of $\Tau$. The NLO$_N$ result also depends on the projection from $\Phi_{N+1}$ to $\Phi_N$, i.e., the precise NLO definition of $\Phi_N$. However, this dependence only appears in the nonsingular corrections. For a given definition of $\Tau$, the singular NLO corrections do not depend on how the remaining $\Phi_{N+1}$ phase space is projected onto $\Phi_N$, since they arise from the IR limit in which all (IR-safe) definitions agree. In \eq{sigmaNTaucut}, the singular contributions inside the full fixed-order cumulant, $\df\sigma^\FO/\df\Phi_N(\Tau^\cut)$ are canceled by the NLO expansion of the resummed result at NLL$'_\sigma$ or higher, leaving only the nonsingular fixed-order contributions in square brackets.

\subsubsection{Matched Spectrum}
\label{subsubsec:spectrum}

To properly combine the higher-order resummation in $\Tau$ with the fully differential $(N+1)$-jet fixed-order calculation, the inclusive $(N+1)$-jet spectrum $\df\sigma/\df\Phi_{N+1}(\Tau)$ in \eq{MCfullydiff} has to fulfill two basic matching conditions,
\begin{align} \label{eq:mastercond1}
\text{Condition 1:} && \int\! \frac{\df\Phi_{N+1}}{\df\Phi_N\df\Tau}\,
\frac{\df\sigma}{\df\Phi_{N+1}}(\Tau)
&= \frac{\df\sigma}{\df\Phi_N\, \df \Tau}
\,,\\[2ex]
\label{eq:mastercond2}
\text{Condition 2:} && \frac{\df\sigma}{\df\Phi_{N+1}}(\Tau) \bigg\vert_\FO
&= \frac{\df\sigma^\FO}{\df\Phi_{N+1}}
\,.\end{align}
The first condition states that integrating the fully differential spectrum over the additional radiative phase space has to reproduce the correct spectrum in $\Tau$ including the desired resummation and fixed-order nonsingular corrections, such that \eq{MCfullydiff} reproduces \eq{MCPhiNdiff}. The second condition states that the fixed-order expansion of the fully differential spectrum has to reproduce the full $(N+1)$-jet fixed-order calculation, where at NLO$_{N+1}$,
\begin{equation} \label{eq:sigmaNLOPhiNplus1}
\frac{\df\sigma^\NLO}{\df\Phi_{N+1}}
= B_{N+1}(\Phi_{N+1}) + V_{N+1}(\Phi_{N+1}) + \int\!\frac{\df\Phi_{N+2}}{\df\Phi_{N+1}}\, B_{N+2}(\Phi_{N+2})
\,.\end{equation}
Here, $B_{N+1}$ and $B_{N+2}$ are the $(N+1)$-parton and $(N+2)$-parton tree-level (Born) contributions, and $V_{N+1}$ is the $(N+1)$-parton one-loop virtual correction. Integrating over $\df\Phi_{N+2}$ in the last term now requires a projection from $\Phi_{N+2}$ to $\Phi_{N+1}$,
\begin{equation} \label{eq:PhiNp1projection}
\frac{\df\Phi_{N+2}}{\df\Phi_{N+1}} \equiv \df\Phi_{N+2}\,\delta[\Phi_{N+1} - \Phi_{N+1}(\Phi_{N+2})]
\,,\end{equation}
analogous to \eq{PhiNprojection}, which now defines precisely what we mean by $N+1$ jets at NLO.

In principle, there is some freedom to construct an expression for $\df\sigma/\df\Phi_{N+1}(\Tau)$ that satisfies both conditions to the order one is working. Our master formula to combine the resummed spectrum $\df\sigma^\resum/\df\Phi_N\df\Tau$ with the fully differential $\df\sigma^\FO/\df\Phi_{N+1}$ is given by
\begin{equation} \label{eq:mastermult}
\frac{\df\sigma}{\df\Phi_{N+1}}(\Tau)
= \frac{\df\sigma^\FO}{\df\Phi_{N+1}}
\biggl[ \frac{\df\sigma^\resum}{\df\Phi_N \df\Tau} \bigg/ \frac{\df\sigma^\resum}{\df\Phi_N\,\df\Tau}\bigg\vert_\FO \biggr]
\,.\end{equation}
Expanding the right-hand side to a given fixed order, we can see immediately that \hyperref[eq:mastercond2]{Condition 2} is satisfied by construction. Imposing \hyperref[eq:mastercond1]{Condition 1} yields the consistency (or ``matching'') condition
\begin{equation} \label{eq:masterconsistency}
\frac{\df\sigma}{\df\Phi_N\df\Tau}
= \biggl[\ \frac{\df\sigma^\FO}{\df\Phi_N\df\Tau}\bigg/\frac{\df\sigma^\resum}{\df\Phi_N \df\Tau}\bigg\vert_\FO \biggr]\ \frac{\df\sigma^\resum}{\df\Phi_N \df\Tau}
\,.\end{equation}
If the resummed result already has the nonsingular contributions at the desired fixed order added in, then the term in brackets is by construction equal to unity for any value of $\Tau$. Otherwise, the expansion of the resummed result reproduces the singular terms of the full fixed-order result, leaving the nonsingular fixed-order contributions, such that we get
\begin{equation}
\frac{\df\sigma}{\df\Phi_N\df\Tau}
=
\frac{\df\sigma^{\sing,\resum}}{\df\Phi_N \df\Tau} +
\frac{\df\sigma^\nons}{\df\Phi_N\df\Tau}\
\biggl[\
\frac{\df\sigma^{\sing,\resum}}{\df\Phi_N \df\Tau}\bigg/\frac{\df\sigma^\sing}{\df\Phi_N \df\Tau}\ \biggr]
\,.\end{equation}
Here, $\df\sigma^{\sing,\resum}$ denotes the pure resummed result only containing the resummation of the singular contributions. Hence, \eq{mastermult} not only multiplies in the additional dependence on $\Phi_{N+1}/\Phi_N$ at fixed order, but if needed also adds the nonsingular corrections to the spectrum multiplied by the higher-order resummation factor. (Note that for the expansion of the resummed result to indeed reproduce all the singular terms at the desired fixed order, the resummation has to be carried out to sufficiently high order, which we have already seen in \subsec{singlediff}.)

To apply \hyperref[eq:mastercond1]{Condition 1}, we have to integrate \eq{sigmaNLOPhiNplus1} using the projection onto $\Phi_N$ and $\Tau$ in \eq{PhiNTauprojection}. Therefore, to get the correct $\Tau$ spectrum at NLO$_{N+1}$, the projection in \eq{PhiNp1projection} has to satisfy
\begin{equation} \label{eq:Taucond}
\Tau[\Phi_{N+1}(\Phi_{N+2})] = \Tau(\Phi_{N+2})
\,;\end{equation}
i.e., it has to preserve the value of $\Tau$ when constructing the projected $\Phi_{N+1}$ point. Usually, the simplest way to handle this would be to use the left-hand side to define $\Tau(\Phi_{N+2})$. However, in our case, \eq{Taucond} provides a very nontrivial condition on the projection since $\Tau(\Phi_{N+2})$ is already defined by our choice of jet resolution variable, which in particular has to be resummable. This turns out to be a nontrivial technical challenge one has to overcome to be able to satisfy \hyperref[eq:mastercond1]{Condition 1}. We will see where this enters in \subsubsec{eeFixedOrder} and \subsubsec{ppFixedOrder}.

Note that to ensure that the resummation factor in square brackets in \eq{mastermult} is well behaved in the fixed-order region at large $\Tau$, it is important to turn off the resummation such that the ratio of the resummed spectrum and its expansion becomes $\ord{1}$ up to higher fixed-order corrections. In principle, the fixed-order result in the denominator can also become negative at very small values of $\Tau$. This is not a problem in practice, since this region is explicitly avoided by imposing the cut $\Tau >\Tau^\cut$.

\subsubsection{Perturbative Accuracy and Order Counting}
\label{subsubsec:ordercounting}

The appropriate order counting in the resummation and fixed-order regions is precisely the same as in \subsec{singlediff}, so there is no need to repeat it here. Applying \eq{mastermult} at the very lowest order, namely LL$_\sigma$ resummation with LO$_{N+1}$ fixed-order corrections, we get
\begin{equation} \label{eq:meps}
\frac{\df\sigma_{\geq N+1}}{\df\Phi_{N+1}} \bigg\vert_{\Tau > 0}
= B_{N+1}(\Phi_{N+1})\, \exp\Bigl[\frac{c_{11}}{2}\, \alpha_s L^2\Bigr]
\,,\end{equation}
where $B_{N+1}(\Phi_{N+1})$ scales as $\alpha_s/\Tau$ relative to $B_N(\Phi_N)$ at small $\Tau$, and we used that at LL$_\sigma$ the ratio in brackets in \eq{mastermult} is just the resummation exponent. This directly corresponds to the CKKW-L procedure~\cite{Catani:2001cc, Lonnblad:2001iq, Krauss:2002up, Lavesson:2005xu}, which multiplies the tree-level matrix elements with the shower Sudakov factors. Hence, we can think of our master formula \eq{mastermult} as a consistent extension of this to higher orders.

\begin{table}
\centering
\begin{tabular}{c||cl|cl|cl}
\hline\hline
& \multicolumn{2}{c|}{inclusive $N$-jet} & \multicolumn{2}{c|}{exclusive $N$-jet} & \multicolumn{2}{c}{inclusive $(N+1)$-jet}
\\
notation & fixed order & accuracy & log. order & accuracy  & fixed order & accuracy
\\\hline
LL$_\Tau$+LO$_{N+1}$ & LO$_N$ & $\sim 1$ & LL & $\sim \alpha_s^{-1/2} $ & LO$_{N+1}$ & $\sim 1$
\\\hline
NLL$_\Tau$ & LO$_N$ & $\sim 1$ & NLL & $\sim 1 $ & - & -
\\
NLL$_\Tau$+LO$_{N+1}$ & LO$_N$ & $\sim 1$ & NLL & $\sim 1 $ & LO$_{N+1}$ & $\sim 1$
\\\hline
NLL$'_\Tau$+LO$_{N+1}$ & NLO$_N$ & $\sim \alpha_s$ & NLL$'$ & $\sim\alpha_s^{1/2}$ & LO$_{N+1}$ & $\sim 1$
\\
NNLL$_\Tau$+NLO$_{N+1}$ & NLO$_N$ & $\sim \alpha_s$ & NNLL & $\sim\alpha_s$ &  NLO$_{N+1}$ & $\sim\alpha_s$
\\
NNLL$'_\Tau$+NLO$_{N+1}$ & NLO$_N$ & $\sim \alpha_s$ & NNLL$'$ & $\sim\alpha_s^{3/2}$ &  NLO$_{N+1}$ & $\sim\alpha_s$
\\\hline\hline
\end{tabular}
\caption{Fixed and resummation orders and their achieved accuracy in $\alpha_s$.}
\label{tab:orders}
\end{table}

As demonstrated in \subsec{singlediff}, going to the next higher perturbative accuracy in all phase space regions requires the NLL$'_\sigma$ and NNLL$_\sigma$ series of logarithms. We obtain these by performing the full NLL$'$ and NNLL resummation in the exponent, as well as the fixed NLO$_N$ and NLO$_{N+1}$ corrections in the cumulant and spectrum, respectively. The resummation naturally connects both jet multiplicities, since the NLO$_N$ corrections are included in the cumulant and are part of the resummation for the spectrum starting at NLL$'$, where they effectively predict the singular NLO$_{N+1}$ contributions, and the full NLO$_{N+1}$ corrections are obtained by adding the nonsingular corrections to the spectrum. In the following, we will use the notation (N)NLL$'_\Tau$+(N)LO$_{N+1}$ to indicate the resummation order for the employed jet resolution variable together with the $(N+1)$-jet fixed order. For simplicity, we do not explicitly denote the $N$-jet fixed order and keep it implicit in the resummation order, i.e., LO$_N$ at (N)LL and NLO$_N$ at NLL$'$ and above. This is summarized in Table~\ref{tab:orders}.

An immediate and important question to ask is to what accuracy resummed spectra for jet resolution variables other than $\Tau$ are predicted in our approach. A detailed theoretical investigation of the formal resummation order one attains for other variables would be very interesting but is beyond the scope of the present work.  What is certainly clear is that other variables will not be resummed at the same formal level as the primary jet resolution variable $\Tau$ itself. However, we know that other variables are correct to NLO$_{N+1}$, while at the same time, the inclusive cross section is not changed, as it is independent of which variable one integrates over. This implies that the NLO$_{N+1}$ corrections for other variables do not induce uncanceled, higher-order logarithmic terms as in \eq{naiveNLO}, and hence, some higher-order resummation must be partially retained for other observables as well. Numerically, the higher-order resummation in $\Tau$ provides an improved weighting of the IR region of phase space, from which other variables are expected to benefit as well. We can validate to what accuracy other variables are obtained by comparing predictions from our highest order to the analytically resummed results for other observables, which we do in \subsec{eeEvShapes}.

\subsection{Extension to More Jet Multiplicities}
\label{subsec:morejets}

The method proposed in this paper is completely general and can be extended to more jet multiplicities essentially by iterating the procedure discussed in \subsec{fullydiff}. We start by introducing separate jet resolution variables $\Tau_N$ to distinguish $N$ from $N+1$ jets, $\Tau_{N+1}$ to distinguish $N+1$ from $N+2$ jets, and so on. One can choose any IR-safe observable that goes to zero in the limit of $N$ pencil-like jets.  For each $N$, the inclusive $N$-jet cross section is obtained by combining the cumulant and spectrum for $\Tau_N$ as in \eq{MCfullydiff},
\begin{align} \label{eq:dsigmainclgen}
\frac{\df\sigma_\incl}{\df\Phi_N}
&= \frac{\df \sigma}{\df\Phi_N}(\Tau_N^\cut)
+ \int\!\frac{\df\Phi_{N+1}}{\df\Phi_N}\, \frac{\df\sigma}{\df\Phi_{N+1}}(\Tau_N)\, \theta(\Tau_N > \Tau_N^\cut)
\,,\nn\\
\frac{\df\sigma_\incl}{\df\Phi_{N+1}}
&= \frac{\df\sigma}{\df\Phi_{N+1}}(\Tau_{N+1}^\cut)
+ \int\!\frac{\df\Phi_{N+2}}{\df\Phi_{N+1}}\, \frac{\df\sigma}{\df\Phi_{N+2}}(\Tau_{N+1})\, \theta(\Tau_{N+1} > \Tau_{N+1}^\cut)
\,,\nn\\
&\quad\vdots
\nn\\
\frac{\df\sigma_\incl}{\df\Phi_{N_\max}}
&= \frac{\df\sigma}{\df\Phi_{N_\max}}(\Tau_{N_\max}^\cut \to \infty)
\,.\end{align}
The exception is the highest jet multiplicity, $N_\max$, for which $\Tau_{N_\max}^\cut = \infty$, corresponding to the fact that no additional jets are resolved.

For the cumulants in \eq{dsigmainclgen}, the discussion in \subsubsec{cumulant} applies separately for each $N$, so the cumulants matched to higher resummed and fixed order are given, as in \eq{sigmaNTaucut}, by
\begin{equation}
\frac{\df\sigma}{\df\Phi_{N}}(\Tau_N^\cut)
= \frac{\df\sigma^\resum}{\df\Phi_N}(\Tau_N^\cut)
+ \biggl[\frac{\df\sigma^\FO}{\df\Phi_{N}}(\Tau_N^\cut)
- \frac{\df\sigma^\resum}{\df\Phi_N}(\Tau_N^\cut)\bigg\vert_\FO \biggr]
\,.\end{equation}

The fully differential $\Tau_N$ spectra $\df\sigma/\df\Phi_{N+1}(\Tau_N)$ are now obtained recursively as follows. We start with the highest jet multiplicity, $N_\max$, for which no resummation is needed since $\Tau_{N_\max}^\cut$ is essentially removed. Furthermore, the highest jet multiplicity is, by construction, only required at leading order, where the result is simply given by the Born contribution,
\begin{equation}
\label{eq:highestmult}
\frac{\df\sigma}{\df\Phi_{N_\max}}(\Tau_{N_\max}^\cut \to \infty) = \frac{\df\sigma^\LO}{\df\Phi_{N_\max}} = B_{N_\max}(\Phi_{N_\max})
\,.\end{equation}
For each $N < N_\max$, we apply the discussion in \subsubsec{spectrum}. To combine the resummation in $\Tau_N$ with the $(N+1)$-jet fixed-order calculation, the fully differential $\Tau_N$ spectrum $\df\sigma/\df\Phi_{N+1}(\Tau_N)$ must satisfy the matching conditions as in \eqs{mastercond1}{mastercond2},
\begin{align} \label{eq:mastercondgen1}
\int\! \frac{\df\Phi_{N+1}}{\df\Phi_N\df\Tau_N}\,
\frac{\df\sigma}{\df\Phi_{N+1}}(\Tau_N)
&= \frac{\df\sigma}{\df\Phi_N \df \Tau_N}
\,,\\[1ex]
\label{eq:mastercondgen2}
\frac{\df\sigma}{\df\Phi_{N+1}}(\Tau_N) \bigg\vert_\FO
&= \frac{\df\sigma^\FO}{\df\Phi_{N+1}}
\,.\end{align}
These can be satisfied by a straightforward generalization of \eq{mastermult},
\begin{align} \label{eq:mastermultgen}
\frac{\df\sigma}{\df\Phi_{N+1}}(\Tau_N)
&= \frac{\df\sigma_\incl}{\df\Phi_{N+1}}\
\biggl[\ \frac{\df\sigma^\resum}{\df\Phi_N \df\Tau_N}\bigg/\frac{\df\sigma^\resum}{\df\Phi_N \df\Tau_N}\bigg\vert_\FO \biggr]
\,.\end{align}
The prefactor on the right-hand side is now the inclusive $(N+1)$-jet cross section from \eq{dsigmainclgen}. This is what ties together the different jet multiplicities. The condition in \eq{mastercondgen2} now leads to the consistency condition
\begin{equation} \label{eq:mastergenFOconsistency}
\frac{\df\sigma_\incl}{\df\Phi_{N+1}} \bigg\vert_\FO = \frac{\df\sigma^\FO}{\df\Phi_{N+1}}
\,,\end{equation}
which states that for each $N$, the cumulant and spectrum in $\Tau_{N+1}$ must be included to sufficiently high order so as to reproduce the $(N+1)$-jet fixed order that is required by the $\Tau_N$ spectrum. Imposing the condition in \eq{mastercondgen1} yields the consistency condition for the $\Tau_N$ spectrum,
\begin{align} \label{eq:mastergenresumconsistency}
\frac{\df\sigma}{\df\Phi_N\df\Tau_N}
&= \biggl[\ \int\! \frac{\df\Phi_{N+1}}{\df\Phi_N\df\Tau_N}\,
\frac{\df\sigma_\incl}{\df\Phi_{N+1}}
\bigg/\frac{\df\sigma^\resum}{\df\Phi_N \df\Tau_N}\bigg\vert_\FO \biggr]\ \frac{\df\sigma^\resum}{\df\Phi_N \df\Tau_N}
\,,\end{align}
which is the generalization of \eq{masterconsistency}. To satisfy \eq{mastergenFOconsistency} at NLO$_{N+1}$, it requires that
\begin{equation}
\Tau_N[\Phi_{N+1}(\Phi_{N+2})] = \Tau_N(\Phi_{N+2})
\,,\end{equation}
as in \eq{Taucond}. That is, for each $N$, the projection from $\Phi_{N+2}$ to $\Phi_{N+1}$ which defines the $(N+1)$-jet cross section at NLO has to preserve the value of $\Tau_N$. In addition, \eq{mastergenresumconsistency} requires that, upon integration, the $\Tau_{N+1}$ resummation contained in $\df\sigma_\incl/\df\Phi_{N+1}$ does not interfere with the $\Tau_N$ resummation, e.g., by inducing higher-order logarithms in $\Tau_N$.
Since \eq{dsigmainclgen} relates the $\df \sigma^\incl / \df \Phi_N$ to $\df\sigma/\df\Phi_{N+1}(\Tau_N)$, the relationship in \eq{mastermultgen} gives rise to a recursive definition, which when combined with the result for the highest jet multiplicity in \eq{highestmult} determines $\df\sigma/\df\Phi_{N+1}(\Tau_N)$ for all $N$.

In the Monte Carlo implementation, the phase space is split up recursively as
\begin{align} \label{eq:dsigmaMCgen}
\frac{\dsigMC_{\geq N}}{\df\Phi_N}
&= \frac{\dsigMC_N}{\df\Phi_N}(\Tau_N^\cut)
+ \int\!\frac{\df\Phi_{N+1}}{\df\Phi_N}\, \frac{\dsigMC_{\geq N+1}}{\df\Phi_{N+1}}\, \theta(\Tau_N > \Tau_N^\cut)
\,,\nn\\
\frac{\dsigMC_{\geq {N+1}}}{\df\Phi_{N+1}}
&= \frac{\dsigMC_{N+1}}{\df\Phi_{N+1}}(\Tau_{N+1}^\cut)
+ \int\!\frac{\df\Phi_{N+2}}{\df\Phi_{N+1}}\, \frac{\dsigMC_{\geq N+2}}{\df\Phi_{N+2}}\, \theta(\Tau_{N+1} > \Tau_{N+1}^\cut)
\,,\\\nn
& \quad\vdots
\,,\end{align}
where in each step, the total cross section for $N$ or more jets is separated into an exclusive $N$-jet cross section, which is assigned to partonic events with $N$ final-state partons, and the integral over the remaining cross section for $N+1$ or more jets. For the highest multiplicity, $N_\max$, the remaining cross section for $N_\max$ or more jets is represented by events with $N_\max$ final-state partons.

Note that the structure of \eq{dsigmaMCgen} is very similar to \eq{dsigmainclgen}. The crucial difference is that in \eq{dsigmaMCgen}, each inclusive cross section on the left-hand side is the same that appears under the integral on the right-hand side in the line above. By comparing \eq{dsigmaMCgen} with \eq{dsigmainclgen} and repeatedly inserting \eq{mastermultgen}, we obtain the higher-order, ``fully resummed,'' exclusive $N$-jet cross sections that serve as inputs to the Monte Carlo. Abbreviating the resummation factor in \eq{mastermultgen} as
\begin{equation}
U_N(\Phi_N, \Tau_N) = \frac{\df\sigma^\resum}{\df\Phi_N \df\Tau_N}\bigg/\frac{\df\sigma^\resum}{\df\Phi_N \df\Tau_N}\bigg\vert_\FO
\,,\end{equation}
we obtain
\begin{align} \label{eq:sigmaNMCfinal}
\frac{\dsigMC_N}{\df\Phi_N}(\Tau_N^\cut)
&= \frac{\df\sigma}{\df\Phi_N}(\Tau_N^\cut)
\,,\nn\\
\frac{\dsigMC_{N+1}}{\df\Phi_{N+1}}(\Tau_{N+1}^\cut)
&= \frac{\df\sigma}{\df\Phi_{N+1}}(\Tau_{N+1}^\cut)\, U_N(\Phi_N, \Tau_N)
\,,\nn\\
& \quad\vdots
\nn\\
\frac{\dsigMC_{\geq N_\max}}{\df\Phi_{N_\max}}
&= \frac{\df\sigma}{\df\Phi_{N_\max}}(\Tau_{N_\max}^\cut\to\infty)\,
U_N(\Phi_N, \Tau_N)\, U_{N+1}(\Phi_{N+1}, \Tau_{N+1})
\nn\\ & \quad \times \dotsb \times U_{N_\max-1}(\Phi_{N_\max-1}, \Tau_{N_\max - 1})
\,.\end{align}
The careful reader will have noticed that the above is in one-to-one correspondence to the structure generated by a parton shower with up to $N_\max$ emissions. The crucial difference is that, in our case, all ingredients are well-defined physical jet cross sections defined in terms of a global jet resolution variable. This allows us to systematically increase the perturbative accuracy by computing the relevant ingredients to higher order in resummed and fixed-order perturbation theory as well as to systematically estimate the perturbative uncertainties.  The analogous parton-shower-like structure underlies the CKKW-L ME/PS merging, which replaces the splitting functions in the shower with the full tree-level matrix elements. Restricting \eq{sigmaNMCfinal} to the lowest order as in \eq{meps}, it reduces to the ME/PS merging as a special case.

In principle, the above construction allows us to go to even higher fixed and resummation order, as long as the fixed-order ingredients are available and the resummation is known to a correspondingly high enough order. It also lets us combine as many jet multiplicities as we like at the order they are available.
In particular, it is straightforward to add additional multiplicities at the lowest accuracy in a CKKW-L-like fashion.

\subsection{Attaching Parton Showering and Hadronization}
\label{subsec:partonshower}

In the Monte Carlo, a point in $\Phi_N$ is represented by $N$ (massless) four-vectors together with the appropriate flavor information. We then generate events with $N$ to $N_\max$ partons and assign the $N$-parton events the weight $\dsigMC_N/\df\Phi_N(\Tau_N^\cut)$, the $(N+1)$-parton events the weight $\dsigMC_{N+1}/\df\Phi_{N+1}(\Tau_{N+1}^\cut)$, and so on. The events with $N_\max$ partons are assigned the weight $\dsigMC_{\geq N_\max}/\df\Phi_{N_\max} = B_{N_\max} (\Phi_{N_\max})$. The $\theta(\Tau_N > \Tau_N^\cut)$ functions in \eq{dsigmaMCgen} are included in the weight, which means that all events with $\geq N+1$ partons that have $\Tau_N < \Tau_N^\cut$ get zero weight.%
\footnote{Technically, the split up of phase space is usually flavor-aware. This means that an event with $\Tau_N < \Tau_N^\cut$ is only set to zero if the closest two partons produce a QCD singularity.} In this way, by summing up the weights of all events, we can integrate up the cross sections in \eq{dsigmaMCgen}, including arbitrary kinematic cuts in $\Phi_{N}$, $\Phi_{N+1}$, etc. What is important is that, although the events contain massless partons, they represent the exclusive jet cross sections of \eq{sigmaNMCfinal}. (From the resummation point of view, the massless partons represent the kinematics of the hard function.)

In the next step, the events are given as a starting point to a parton shower, whose purpose it is to fill up the jets with additional emissions inside the jets without changing the weight of the event. Formally, this means that the shower should not be allowed to change the underlying distribution in the jet resolution variable, since this has already been computed at the higher perturbative accuracy. For example, starting from an event with $N+1$ partons with kinematics $\Phi_{N+1}$ and weight $\dsigMC_{N+1}/\df\Phi_{N+1}(\Tau_{N+1}^\cut)$, the fully showered event should have the same jet kinematics $\Phi_{N+1}$ as the unshowered event from which it originated. Most importantly, the showered event should have the same value of $\Tau_N(\Phi_{N+1})$ and should have $\Tau_{N+1} < \Tau_{N+1}^\cut$ so it still has the correct weight $\dsigMC_{N+1}/\df\Phi_{N+1}(\Tau_{N+1}^\cut)$. In the cumulant $N$-jet bin, the shower is allowed to fill out the phase space from $\Tau_N = 0$ to $\Tau_N^\cut$. Since for the highest jet multiplicity, $\Tau_{N_\mathrm{max}}^\cut\to\infty$, the shower fills out the remaining phase space.
In practice, these are quite nontrivial constraints on the shower. The easiest way to enforce them is to repeatedly run the shower on the same event until it produces an acceptable showered event, where we allow the value of $\Tau_N$ to be changed at most by a numerically small amount consistent with a power correction.
This method is of course computationally intensive (though it is not computationally prohibitive), since one may have to rerun the shower many times, and it would be interesting to develop a more efficient way of constraining the shower for this purpose. Notice that in this procedure no events are discarded, so the cross section is not changed.

In the final step, the showered event is passed to the hadronization routine. In this case, there are no constraints on the kinematics of the hadronized event; i.e., the hadronization is allowed to smear out the $\Tau_N$ spectrum. The reason is that our perturbative calculation does not take into account nonperturbative effects, which are instead supplied by the hadronization. This is discussed in more detail in \subsubsec{eePythia}.

\section{Application to $e^+ e^-$ Collisions}
\label{sec:ee}

In this section, we apply the framework described in \sec{master} to $e^+ e^- \to 2/3$ jets, implemented in the \geneva Monte Carlo.
The higher-order resummation for $2$-jet event shapes in $e^+e^-$ collisions is very well understood and many precise measurements from LEP exist, which are used, for example, for precise determinations of the strong coupling constant $\alpha_s$~\cite{Dissertori:2007xa, Becher:2008cf, Dissertori:2009ik, Chien:2010kc, Abbate:2010xh, Bethke:2011tr, Abbate:2012jh}.

In this context,  one important aspect is the interplay between both resummed and fixed-order perturbative contributions with the nonperturbative corrections. Here, the \geneva framework provides an important development by being able to combine the perturbative higher-order resummation with the nonperturbative information provided by \pythia's hadronization model~\cite{Sjostrand:2006za,Sjostrand:2007gs}. For example, this allows us to use a common theoretical framework to make predictions for different phase space regions and different observables.

The $e^+e^-$ implementation also provides an important and powerful validation of our approach and its practical feasibility, while avoiding the additional complications arising for hadronic collisions, such as initial-state radiation and parton distribution functions (PDFs). The implementation and first results for $pp$ collisions are presented in \sec{pp}.

In our $e^+e^-$ implementation, we use 2-jettiness, $\Tau_2$, as the $2$-jet resolution variable, which is defined as~\cite{Stewart:2010tn}
\begin{equation}\label{eq:defTau2}
\Tau_2 =\Ecm \biggl(1 - \max_{\hat n} \frac{\sum_k \abs{\hat n\cdot \vec p_k}}{\sum_k \abs{\vec p_k} }\biggr)
\,,\end{equation}
and is simply related to thrust $T$~\cite{Farhi:1977sg} by $\Tau_2 = \Ecm (1-T)$. Its kinematic limits are $0\leq\Tau_2\leq\Ecm/2$. In the limit $\Tau_2 \to 0$, there are precisely $2$ pencil-like jets in the final state, while for $\Tau_2 \sim \Ecm$, there are $3$ or more jets. We perform the resummation in $\Tau_2$ to NNLL$'$ and include the full NLO$_2$, NLO$_3$, and LO$_4$ fixed-order matrix elements, i.e., we obtain NNLL$'_\Tau$+NLO$_3$ predictions.

The default running parameters for our $e^+ e^-$ studies are $\Ecm = 91.2\GeV$, $\alpha_s(m_Z) = 0.1135$, and \pythia 8.170 with $e^+e^-$ tune 1.%
\footnote{The $\alpha_s$ value used inside \pythia's parton shower is not changed from the value set in the tune. This is not inconsistent, since here the strong coupling functions as a phenomenological parameter, regulating the amount of showering.}
Using this value of $\alpha_s(m_Z)$ is motivated by the fact that it was obtained from fits to the thrust spectrum using N$^3$LL$'$ resummation. These fits were performed in a region (corresponding to $6\GeV \leq \Tau_2 \leq 30\GeV$ for our $\Ecm$) where the nonperturbative corrections due to hadronization are power suppressed and can be described by a single nonperturbative parameter, which leads to a shift in the spectrum and is included in the fit in ref.~\cite{Abbate:2010xh}. We find that this value of $\alpha_s(m_Z)$, in conjunction with \pythia's tune 1, provides overall the best description of the data, including the peak region below $\Tau_2 \leq 6\GeV$ and other 2-jet event shapes. For comparison, we show results using the world average $\alpha_s(m_Z) = 0.1184$~\cite{Beringer:1900zz} as well as from using \pythia tune 3.

In the next subsection, we summarize the various ingredients that go into the master formula, with the intention of giving a concise and informative overview, while leaving a detailed discussion of our implementation to a separate publication. In \subsec{eeTau2}, we discuss the $\Tau_2$ spectrum, validating our implementation using analytic predictions as well as comparing our results to LEP data. In \subsec{eeEvShapes}, we present our results for other $2$-jet variables, namely $C$-parameter, heavy jet mass, and jet broadening, comparing \geneva's predictions at NNLL$'_\Tau$+NLO$_3$ to the analytic higher-order resummation for each variable as well as to the experimental measurements. In all cases, we find good consistency and agreement with the data.

\subsection{Ingredients}
\label{subsec:eeIngredients}

The master formula is given by
\begin{align} \label{eq:MCfullydiff23}
\frac{\df\sigma_{\incl}}{\df\Phi_2}
&= \frac{\df \sigma}{\df\Phi_2}(\Tau_2^\cut)
+ \int\!\frac{\df\Phi_{3}}{\df\Phi_2}\, \frac{\df\sigma}{\df\Phi_{3}}(\Tau_2)\, \theta(\Tau_2 > \Tau_2^\cut)
\,,\end{align}
where
\begin{align} \label{eq:dsigma23}
\frac{\df\sigma}{\df\Phi_{2}}(\Tau_2^\cut)
&= \frac{\df\sigma^\resum}{\df\Phi_2}(\Tau_2^\cut)
\,, \nn \\[1ex]
\frac{\df\sigma}{\df\Phi_{3}}(\Tau_2)
&= \frac{\df\sigma_\incl}{\df\Phi_{3}}
\biggl(\frac{\df\sigma^\resum}{\df\Phi_2 \df\Tau_2}\bigg/\frac{\df\sigma^\resum}{\df\Phi_2\,\df\Tau_2}\bigg\vert_\FO \biggr) 
\,.\end{align}
Its three key ingredients are the higher-order resummation of 2-jettiness, which we include at NNLL$'_\Tau$+LO$_3$, the full fixed-order matrix elements at NLO$_2$, NLO$_3$, and LO$_4$, and the interface to parton showering and hadronization, for which we use \pythia 8.

Following the construction in \subsec{morejets} with $N_{\rm max} = 4$, the inclusive 3-jet cross section is separated into 3 and 4 or more jet contributions using 3-jettiness, $\Tau_3$, as our 3-jet resolution variable,
\begin{align}\label{eq:ee34master}
\frac{\df\sigma_\incl}{\df\Phi_{3}}
&= \frac{\df \sigma}{\df\Phi_{3}}(\Tau_{3}^\cut)
+\int\!\frac{\df\Phi_{4}}{\df\Phi_{3}}\, \frac{\df \sigma}{\df\Phi_{4}}(\Tau_3)\, \theta(\Tau_{3} > \Tau_{3}^\cut)
\,.\end{align}
For $e^+e^-$ collisions, $N$-jettiness is defined by~\cite{Stewart:2010tn}
\begin{equation}
\label{eq:eeNjettiness}
\Tau_N = \sum_k \min_{i} \bigl(E_k - \hat{n}_i \cdot \vec{p}_k \bigr)
\,,\end{equation}
where $i=1,\cdots,N$ and $\hat{n}_i$ is a unit vector along the direction of the $i$th jet, where the jet directions can be determined by a jet algorithm or by directly minimizing $\Tau_N$.\footnote{This definition agrees with $\Tau_2$ in \eq{defTau2} for massless final-state particles, which is the limit in which resummation is carried out. It does affect the nonperturbative corrections when including hadron masses~\cite{Salam:2001bd, Mateu:2012nk}. We use the definition of $\Tau_2$ in \eq{defTau2} to be able to directly compare to the experimental data for thrust.} There are $N$ pencil-like jets in the limit $\Tau_N \to 0$ and $N$ or more jets in the limit $\Tau_N \sim\Ecm$.

As discussed in \subsec{morejets}, the master formula naturally incorporates the resummation of the 3-jet resolution variable in \eq{ee34master} and extends to higher jet multiplicities, i.e., $N_\max>4$. However, since our current focus is on the main conceptual development of combining the higher-order resummation with the fixed NLO matrix elements for 2 and 3 jets, we leave these extensions to future work. As we will not be interested in the $\Tau_3$ spectrum or other exclusive $3$-jet observables, it is sufficient for our purposes to calculate the two terms on the right-hand side of \eq{ee34master} at fixed order
(i.e., we do not include resummation for $\Tau_3$). Thus, we use
\begin{align} \label{eq:FO34terms}
\frac{\df \sigma}{\df\Phi_{3}}(\Tau_{3}^\cut) =  \frac{\df \sigma^\FO}{\df\Phi_{3}}(\Tau_{3}^\cut) \,,
\qquad \frac{\df \sigma}{\df\Phi_{4}} = B_4(\Phi_4)
\,.\end{align}
In the results that follow, we use $\Tau_2^\cut$ value between $0.5-1\GeV$, which is selected randomly from a flat distribution. This smoothing out of $\Tau_2^\cut$ avoids small numerical discontinuities that can arise with a sharp cutoff.
For $\Tau_3^\cut$, we use $\Tau_3^\cut=2\GeV$. This value is chosen small enough that the NLO$_3$ calculation is fully exclusive and our results are insensitive to scales below $\Tau_3^\cut$. Changing $\Tau_3^\cut$ by a factor of two up and down, the results
remain unchanged, with any variations well within our perturbative uncertainties.

\subsubsection{Resummation}
\label{subsubsec:eeResum}

\begin{table}[t]
  \centering
  \begin{tabular}{l | c c c c c}
  \hline \hline
  & \multicolumn{2}{c}{Fixed-order corrections} & \multicolumn{3}{c}{Resummation input} \\
  & singular     & nonsingular & $\gamma_x$ & $\Gamma_\mathrm{cusp}$ & $\beta$ \\ \hline
  LL             & LO$_2$ & - & - & $1$-loop & $1$-loop \\
  NLL            & LO$_2$  & -      & $1$-loop & $2$-loop & $2$-loop \\
  NLL$'$         & NLO$_2$ & - & $1$-loop & $2$-loop & $2$-loop \\
  NLL$'$+LO$_3$  & NLO$_2$ & LO$_3$ & $1$-loop & $2$-loop & $2$-loop \\
  NNLL+LO$_3$    & NLO$_2$ & LO$_3$ & $2$-loop & $3$-loop & $3$-loop \\ 
  NNLL$'$        & NNLO$_2$ & - & $2$-loop & $3$-loop & $3$-loop \\
  NNLL$'$+NLO$_3$ & NNLO$_2$ & NLO$_3$ & $2$-loop & $3$-loop & $3$-loop \\
  \hline\hline
  \end{tabular}
  \caption{Perturbative inputs included at a given order in resummed and fixed-order perturbation theory. The columns in the resummation input refer to the noncusp anomalous dimension ($\gamma_x$), the cusp anomalous dimension ($\Gamma_\mathrm{cusp}$), and the QCD beta function ($\beta$).}
\label{tab:expcounting}
\end{table}

Our jet resolution variable, $\Tau_2$, has the important property that it can be factorized. The factorization theorem for the $\Tau_2$ spectrum provides the resummed prediction that is one of the primary inputs to our master formula in \eq{dsigma23}. It is obtained by using the framework of Soft Collinear Effective Theory (SCET)~\cite{Bauer:2000ew, Bauer:2000yr, Bauer:2001ct, Bauer:2001yt} and allows the resummation to be systematically carried out to higher orders and combined with the nonsingular fixed-order result. Our highest-order resummed input to the master formula has NNLL$'$ resummation. We use the standard resummation formalism, where the large logarithms are resummed in the exponent of the cross section, with the corresponding resummation orders summarized in \tab{expcounting}.

We write the jet resolution distribution in $\Tau_2$ as
\begin{equation}\label{eq:tau2singnsing}
\frac{\df\sigma_2^\resum}{\df\Phi_2\,\df\Tau_2}
= \frac{\df\sigma_2^\sing}{\df \Omega_2\,\df\Tau_2} + \frac{\df\sigma_2^\nons}{\df\Omega_2\,\df\Tau_2}
\,,\end{equation}
where the separation into singular and nonsingular contributions was discussed in \subsec{fullydiff} [see \eq{dsigmadPhiNdtau}].
The singular contribution is given by~\cite{Fleming:2007qr, Schwartz:2007ib}
\begin{align}\label{eq:tau2fact}
\frac{\df\sigma_2^\sing}{\df\Omega_2\,\df\Tau_2}
&= \frac{\df\sigma_B}{\df\Omega_2}\, H_2(\Ecm^2, \mu)
\int\!\df s_1 \df s_2\, J_1(s_1, \mu)\, J_2(s_2, \mu)\, S_2\Bigl(\Tau_2 - \frac{s_1}{\Ecm} - \frac{s_2}{\Ecm}, \mu \Bigr) 
\,.\end{align}
Here, $\df \Phi_2 = \df\Omega_2 = \df \cos\theta \df\phi$ is the angular phase space for the orientation of the thrust axis with respect to the beam, and $\df\sigma_B/\df\Omega_2$ is the tree-level 2-parton cross section. Note that the overall dependence on $\Omega_2$ here is that of the Born cross section, which is correct in the limit $\Tau_2 \to 0$ in which \eq{tau2fact} is obtained. The hard function $H_2$ in \eq{tau2fact} contains the fixed-order $2$-parton matrix elements, which describe the short-distance corrections at the scale $\Ecm$. The jet functions $J_{1}$ and $J_2$ describe the back-to-back collinear final-state radiation along the thrust axis, and the soft function $S_2$ describes the soft radiation between the jets. The soft function contains perturbative and nonperturbative components, which can be separated as~\cite{Korchemsky:1999kt, Hoang:2007vb, Ligeti:2008ac}
\begin{align}\label{eq:S2nonpert}
S_2(\Tau_2,\mu) = \int\!\df k \, S^\mathrm{pert}_2(\Tau_2-k,\mu) \, f(k, \mu)
\,,\end{align}
where $S^\mathrm{pert}_2(\Tau_2-k,\mu)$ is the perturbative soft function, while the shape function $f(k,\mu)$ describes the nonperturbative hadronization corrections. For $\Tau_2 \sim \lqcd$, the shape function gives an $\ord{1}$ contribution to the cross section, while for $\Tau_2 \gg \lqcd$ it can be expanded, and only the leading $\ord{\lqcd/\Tau_2}$ nonperturbative power correction is relevant. For further discussion and the derivation of the factorization theorem, see refs.~\cite{Fleming:2007qr, Schwartz:2007ib}. The resummed prediction used in \geneva only includes the perturbative soft function, while the nonperturbative corrections are provided by the hadronization in \pythia.

The nonsingular contribution in \eq{tau2singnsing} is given by the spectrum at fixed order with the singular terms subtracted. It includes all $\ord{\Tau_2/\Ecm}$ corrections to the singular distribution to a given order in $\alpha_s$.  The $\ord{\alpha_s}$ nonsingular corrections in $\Tau_2$ are known analytically and can be taken from ref.~\cite{Abbate:2010xh}, so we include them in our resummed result.
Each function in \eq{tau2fact} depends on the renormalization scale $\mu$ and the characteristic scale of the physics it describes. These are $\mu_H \sim \Ecm$, $\mu_J \sim \sqrt{\Tau_2 \Ecm}$, and $\mu_S\sim \Tau_2$ for the hard, jet, and soft functions, respectively. Renormalization group evolution (RGE) between the soft, collinear, and hard scales resums the logarithms of the form $\ln \mu_S/\mu_H \sim \ln \Tau_2/\Ecm$ and $\ln \mu_J^2/\mu_H^2\sim\ln\Tau_2/\Ecm$ in the factorized singular distribution in \eq{tau2fact}. The anomalous dimensions and singular fixed-order corrections required at a given resummation order are summarized in \tab{expcounting}.

The resummed cumulant in \eq{dsigma23} is obtained in an analogous way to the resummed $\Tau_2$ distribution. It is given by a singular and nonsingular component,
\begin{align}
\frac{\df\sigma_2^\resum}{\df\Phi_2} (\Tau_2^\cut)
= \frac{\df\sigma_2^\sing}{\df \Omega_2}(\Tau_2^\cut) + \frac{\df\sigma_2^\nons}{\df\Omega_2}(\Tau_2^\cut)
\,,\end{align}
where the singular contribution is obtained by integrating \eq{tau2fact} over $\Tau_2$ from $0$ to $\Tau_2^\cut$. The nonsingular contribution to the cumulant is given by the difference between the fixed-order result and the resummed singular terms expanded to fixed order.

The perturbative uncertainties in the resummed spectrum are estimated by scale variation and receive a contribution from two distinct sources, the fixed-order corrections and the higher-order logarithmic resummation. The fixed-order uncertainties are estimated by a correlated overall variation of all scales by factors of two. The resummation uncertainties are instead estimated by varying the lower scales $\mu_J(\Tau_2)$ and $\mu_S(\Tau_2)$, which are functions of $\Tau_2$, and are referred to as profile scales~\cite{Ligeti:2008ac, Abbate:2010xh, Berger:2010xi}. The profile scales satisfy the criteria that, in the resummation region, $\mu_{J,S}(\Tau_2)$ have their canonical scaling (given above) and in the fixed-order region, $\mu_{J,S}(\Tau_2) \sim \mu_H$, which turns off the resummation. In the transition region, the profile scales provide a smooth interpolation between the resummation and fixed-order regions. These three regions are determined based on where the fixed-order singular contributions dominate over the nonsingular ones. The variations in the profile scales subject to the above constraints determine the resummation uncertainty, where we take the largest absolute variation from the central scale. The resummation uncertainties are combined in quadrature with the fixed-order uncertainties to generate our theory uncertainty estimate. For a given partonic event in \geneva, each profile scale variation gives rise to a different event weight, which is computed analytically. Hence, we can provide each event with its own perturbative uncertainty estimate by assigning it several weights from the profile scale variation in addition to its central weight.

\subsubsection{Fixed Order}
\label{subsubsec:eeFixedOrder}

As we can see from \eqs{ee34master}{FO34terms}, we need the 3-jet cumulant $\df \sigma / \df \Phi_3 (\Tau_3^\cut)$ as well as the Born 4-parton cross section $B_4(\Phi_4)$. The Born $4$-parton cross section is trivial and requires no further discussion.
To calculate the 3-jet cumulant at NLO$_3$, we use the generic formula given in \eq{sigmaNLOPhiNplus1},
\begin{equation} \label{eq:dsigma3NLO}
\frac{\df \sigma}{\df \Phi_3} = B_3(\Phi_3) + V_3(\Phi_3)
+ \int\!\frac{\df \Phi_4}{\df \Phi_3}\, B_4(\Phi_4)\, \theta(\Tau_3 < \Tau_3^\cut)
\,,\end{equation}
where
\begin{equation}
\frac{\df\Phi_4}{\df\Phi_3} \equiv \df\Phi_4\,\delta[\Phi_3 - \Phi^\Tau_3(\Phi_4)]
\,.\end{equation}
The projection $\Phi^\Tau_3(\Phi_4)$ defines what we mean by $\Phi_3$ at NLO$_3$. It implicitly depends on our choice of resolution variable since \eq{Taucond} requires it to satisfy
\begin{equation} \label{eq:Phi3Taucond}
\Tau_2[\Phi_3^\Tau(\Phi_4)] = \Tau_2(\Phi_4)
\,.\end{equation}

To deal with the IR singularities that are present both in $V_3$ and
in the integral of $B_4$ over $\Phi_4$, we use the
FKS subtraction method~\cite{Frixione:1995ms}. We introduce a set of projecting
functions, $\theta^{\Tau}_m(\Phi_4)$, that partition the phase space
into nonoverlapping regions, such that $\sum_m \theta_m^\Tau(\Phi_4) = 1$.
In our case, this partition is effectively determined by the resolution
variable, which is indicated by the superscript.
The resulting  partition must be such that each region $m$ contains at
most one collinear and one soft singularity.  Then, we can write
\begin{equation}
\df \Phi_4 = \sum_m \df \Phi_3\, \df \Phi^m_{\rm rad}\, \theta^{\Tau}_m(\Phi_3,\Phi^m_{\rm rad})
\,,\end{equation}
where $\Phi^m_{\rm rad}$ denotes the radiative phase space describing a $1 \to 2$ splitting in each region.

For each region $m$, we define a mapping that identifies which
particle in $\Phi_3$ is undergoing the $1 \to 2$ splitting, which
generates the
\begin{equation} 
\Phi_4^{m} \equiv \Phi_4^{m}(\Phi_3,\Phi^m_{\rm rad})
\end{equation}
phase space point. It also unequivocally defines how the recoil is shared
amongst the remaining particles in the event, which is needed to enforce total momentum
conservation.  Notice that our definition of the $\theta^{\Tau}_m$-functions
leaves us the freedom to include phase space regions in the partition which do not contain any IR singularity. This freedom is in fact essential to be able to satisfy \hyperref[eq:mastercond1]{Condition 1} in eq.~(\ref{eq:mastercond1}), namely, to ensure that in each phase space region the correct functional form for the resolution variable $\Tau_2(\Phi_4)$ is used. For example, in $e^+ e^- \to q \bar{q} g $ production, the region in which the quark and antiquark are closest to each other does not contain
any QCD singularity. Nevertheless, it must be treated as a separate region in the phase space partition, since in this region the invariant mass between the quark and antiquark determines the value of $\Tau_2$.

With this notation, one can write
\begin{equation}
\frac{\df \sigma}{\df \Phi_3} = B_3(\Phi_3) + V_3(\Phi_3) + \sum_m  \int \df \Phi^m_{\rm rad} B_4(\Phi_4^m) \theta^\Tau_m(\Phi_3,\Phi^m_{\rm rad})\theta(\Tau_3 < \Tau_3^\cut)\,.
\end{equation}
If the region  $m$ contains an IR divergence, the FKS subtraction requires one to define the soft, collinear, and soft-collinear limits of $\Phi^m_{\rm rad}$, which we denote as  $\Phi_{\rm rad}^{m,{\rm s}}$,  $\Phi_{\rm rad}^{m,{\rm c}}$, and  $\Phi_{\rm rad}^{m,{\rm cs}}$, respectively, together with the resulting points in the 4-body phase space $\Phi_4^{m,{\rm s}}$, $\Phi_4^{m,{\rm c}}$, and $\Phi_4^{m,{\rm cs}}$. We can then write
\begin{align}
\label{eq:fixedspectrum}
\frac{\df \sigma}{\df \Phi_3} = & B_3(\Phi_3) + V_3(\Phi_3) + I(\Phi_3) + \sum_m  \int \df \Phi^m_{\rm rad} \bigg[ B_4(\Phi_4^{m}) \theta^\Tau_m(\Phi_3,\Phi^m_{\rm rad}) \theta(\Tau_3 < \Tau_3^\cut)
\nn\\
&
- \frac{\df \Phi_{\rm rad}^{m,{\rm s}}}{\df \Phi^m_{\rm rad}} B_4(\Phi_4^{m,{\rm s}}) \theta^{\rm s}_m(\Phi_3,\Phi_{\rm rad}^{m,{\rm s}}) 
 - \frac{\df \Phi_{\rm rad}^{m,{\rm c}}}{\df \Phi^m_{\rm rad}} B_4(\Phi_4^{m,{\rm c}}) 
  + \frac{\df \Phi_{\rm rad}^{m,{\rm cs}}}{\df \Phi^m_{\rm rad}} B_4(\Phi_4^{m,{\rm cs}})  \bigg]
\,,
\end{align}
where  $\theta^{\rm s}_m$ encodes the soft limit of the $\theta^\Tau_m$-functions  and we have used the fact that in the collinear and soft-collinear limits the $\theta^\Tau_m$-functions are trivially satisfied. Also, since in each of these limits $\Tau_3 \equiv 0$, the $\theta(\Tau_3 < \Tau_3^\cut)$ functions  are satisfied by construction.

If $m$ is not singular, in principle, no such subtraction is needed, and one could simply evaluate the 4-parton tree-level matrix element  $B_4(\Phi_4^{m})$. However, given that the integral of the subtraction counterterms  over the whole phase space is known analytically for both massless and massive partons~\cite{Frixione:1995ms,Frederix:2009yq,Alioli:2010xd},
\begin{align}
I(\Phi_3) = & \sum_m \Big[ \int \df \Phi_{\rm rad}^{m,{\rm s}} B_4(\Phi_4^{m,{\rm s}}) \theta^{\rm s}_m(\Phi_3,\Phi_{\rm rad}^{m,{\rm s}}) +  \int \df \Phi_{\rm rad}^{m,{\rm c}} B_4(\Phi_4^{m,{\rm c}}) \nn \\& \quad - \int \df \Phi_{\rm rad}^{m,{\rm cs}} B_4(\Phi_4^{m,{\rm cs}})\Big]\,,
\end{align}
we found it easier not to restrict the integration of the subtraction counterterms only in the singular regions of phase space but to extend it across all of phase space.\footnote{These integrals can also easily be defined by restricting the integration of the  FKS variable $\xi$ up to some $\xi_{\rm cut}$ value. These are, however,  not in direct correspondence to the partition of phase space we are considering. } This ensures the complete cancellation of and the independence of the final results from the subtraction terms. The procedure outlined above takes care of all IR divergences, making  the integrand in the square brackets of \eq{fixedspectrum} as well as the sum of $V(\Phi_3) + I(\Phi_3)$ IR finite.

The crucial point, discussed in \subsubsec{spectrum}, is that our construction requires the phase space map that generates $\Phi_4^{m}(\Phi_3,\Phi^m_{\rm rad})$ to preserve the value of $\Tau_2$; i.e.,
\begin{equation}
\Tau_2[\Phi_4^{m}(\Phi_3,\Phi^m_{\rm rad})] = \Tau_2(\Phi_3)\,.
\end{equation}
Comparing this to \eq{Phi3Taucond}, we see that the map $\Phi_4^m(\Phi_3, \Phi_{\rm rad}^m)$ must be precisely the inverse of $\Phi_3^\Tau(\Phi_4)$ in the region $m$.
In principle, this condition can be relaxed to only hold up to power corrections.  Additionally, the map can fail to preserve $\Tau_2$ in a region of phase space that gives a power-suppressed contribution to the cross section.  The phase space maps used in the standard FKS implementations~\cite{Frixione:1995ms,Frixione:2007vw} were not designed to preserve the value of $\Tau_2$, and thus, they change its value by an $\ord{\Tau_3 / \Tau_2}$ amount over a large region of phase space.\footnote{Generically, an emission that takes a 3-parton event to a 4-parton event will change $\Tau_2$ by the scaling $\Tau_2 (\Phi_4) - \Tau_2(\Phi_3) \sim \Tau_3$.}  Since 4-parton events with $\Tau_3 < \Tau_3^\cut$ and $\Tau_2 > \Tau_2^\cut$ are the only real emission contributions included in the NLO calculation for the 3-jet cumulant, \eq{dsigma3NLO}, one can impose the restriction $\Tau_3^\cut \ll \Tau_2^\cut$ and use the standard FKS phase space maps.  However, this hierarchy strongly restricts $\Tau_3^\cut$, and it is preferable to define a map that is specifically designed for our goals.  We have constructed such a map, which preserves the exact value of $\Tau_2$ up to power corrections, except in a region of phase space whose contribution to the cross section scales as $\ord{\Tau_2 / \Ecm}$.  In this region, the value of $\Tau_2$ is altered by an $\ord{\Tau_3/\Tau_2}$ amount, meaning the net correction scales as $\ord{\Tau_3 / \Ecm}$.  Therefore, enforcing the much looser constraint $\Tau_3^\cut \ll \Ecm$ is sufficient to achieve our purposes. We postpone the detailed discussion of this map to a dedicated publication describing the implementation of \geneva.

\subsubsection{Parton Shower and Hadronization}
\label{subsubsec:eePythia}

The phase space points $\Phi_2$ and $\Phi_3$ represent jet kinematics, which are defined by the jet resolution variable 2-jettiness.  As discussed in \subsec{partonshower}, we require that the parton shower does not change the underlying hard jet distribution so that the higher-order weights we calculate, $\df \sigma/\df \Phi_2 (\Tau_2^\cut)$ and $\df \sigma/\df \Phi_3 (\Tau_2)$, are correctly assigned. Without any constraints the parton shower will not preserve the value of $\Tau_2$. We address this problem in our current implementation in a physically motivated way.  For small $\Tau_2$, the resummed singular jet resolution spectrum dominates and is determined up to power corrections of order $\lambda \sim \Tau_2/\Ecm$. We require that for $\Phi_{\geq 3}$ events, the change in $\Tau_2$ due to showering, $\Delta \Tau_2$, satisfies $\Delta \Tau_2/\Tau_2<\lambda$.
This represents a power correction to the 2-jettiness spectrum, which scales as $1/\Tau_2$ for small $\Tau_2$.  For the 2-jet cumulant bin, we require that $\Phi_2$ events, which have $\Tau_2 = 0$ when unshowered, remain in the 2-jet bin after showering up to a power suppressed correction, with $\Tau_2< \Tau_2^\cut (1+\lambda')$. Here, the effect of a small nonzero $\lambda'$ induces a change to the shape of the distribution generated by \pythia that scales as a power correction and does not affect the formal accuracy of the spectrum.
Formally, we work in the limit where $\lambda$ and $\lambda'$ are effectively taken to zero. For $\lambda = 0$, the shower would be required to exactly preserve $\Tau_2$, making it maximally inefficient. Therefore, for the results shown in this section, we use small nonzero values $\lambda=2\, \lambda'=0.05$. We have checked that these are small enough to be in the asymptotic region where the results become independent of the precise value.

Furthermore, the shower must also be restricted to not change the NLO$_3$ result. This requires that, for 3-parton events, we effectively only allow showers to start from $\Tau_3^{\cut}$. Likewise, we limit the showering of 4-partons events down from their $\Tau_3(\Phi_4)$ value. This can be seen as a proxy for what would be the correct approach in a $\Tau_N$-ordered shower.

We use \pythia 8.170 with $e^+ e^-$ tune 1 for showering and hadronization. The choice of tune for $e^+ e^-$ data in \pythia affects both the time-like showering and hadronization model. However, since in our implementation we restrict the shower from changing the $\Tau_2$ spectrum, the effect of changing the tune in \pythia primarily reflects the uncertainty from hadronization in \geneva. We have checked that this is also the case for observables other than $\Tau_2$ by verifying that the effect of the tune on the showered \geneva predictions is very small compared to the change due to hadronization. The uncertainty from hadronization is associated with the nonperturbative contribution to the soft function in \eq{S2nonpert} in our framework and is not included in our event-by-event perturbative uncertainties, which are derived from the analytical resummation and fixed-order matching.  A complete uncertainty analysis should also include uncertainties due to hadronization as well as due to the remaining amount of parton showering.  As an indication of the size of the uncertainty from hadronization, we also show \geneva hadronized results using $e^+ e^-$ tune 3. The shift from the partonic to the showered results could be taken as a conservative upper limit on the remaining showering uncertainty.

It is important to note that we use the standard tunes in \pythia, without changing any internal parameters. Since in our approach the shower evolution in standalone \pythia8  is substituted with higher-order resummation above the 2-jet resolution scale, we advocate that a separate tuning of \geneva + \pythia8 should be employed to obtain the best results. This would also allow one to obtain meaningful estimates of hadronization and remaining showering uncertainties.

\subsection{Validation Using the Jet Resolution Spectrum}
\label{subsec:eeTau2}

Before comparing the \geneva prediction for various $e^+ e^-$ spectra to analytic predictions and LEP data, we first validate the implementation of our procedure to combine higher-order resummation and full NLO matrix-element corrections by using the jet resolution spectrum. At the level of the singly differential $\Tau_2$ spectrum only, the standard approach to resummation achieves the same matching between resummation and fixed order by adding the nonsingular contribution to the resummed result. This provides a nontrivial crosscheck of the master formula and in particular validates the event-by-event theory uncertainties generated by \geneva.
For each comparison in this section, we show the peak, transition, and tail regions, described in \subsec{singlediff}, at the LEP center-of-mass energy $\Ecm=91.2 \GeV$.
In all cases, the error bars or bands on the \geneva histograms are built from its event-by-event perturbative uncertainties. The statistical uncertainties from Monte Carlo integration are much smaller and are not shown.

\subsubsection{Partonic Results}
\label{subsubsec:partonicresultsee}

\begin{figure*}[t!]
\subfigure[\hspace{1ex} Peak Region]{%
\parbox{0.5\columnwidth}{\includegraphics[scale=0.5]{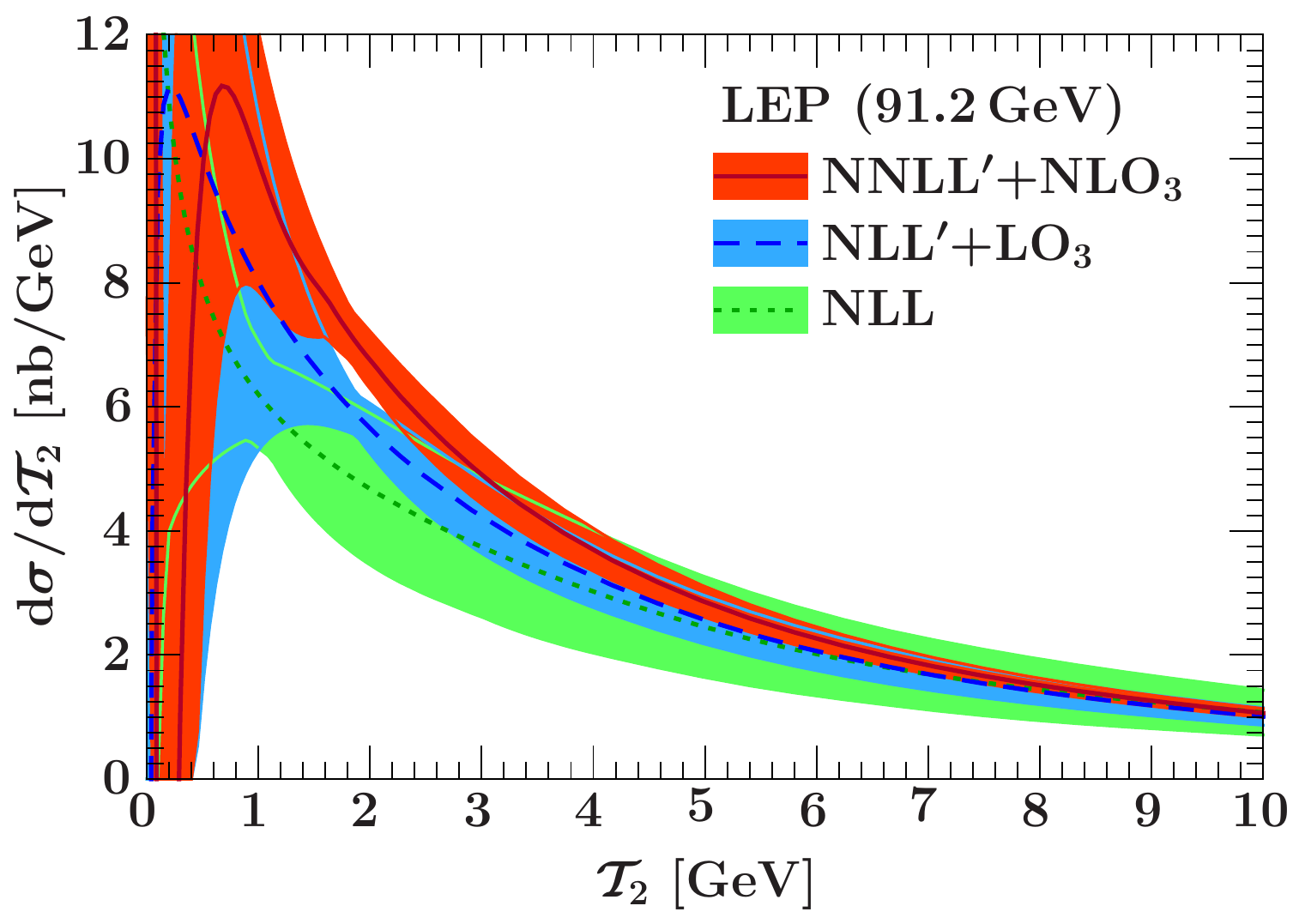}}%
\label{fig:TauConvPeak}}%
\hfill%
\subfigure[\hspace{1ex} Transition Region]{%
\parbox{0.5\columnwidth}{\includegraphics[scale=0.5]{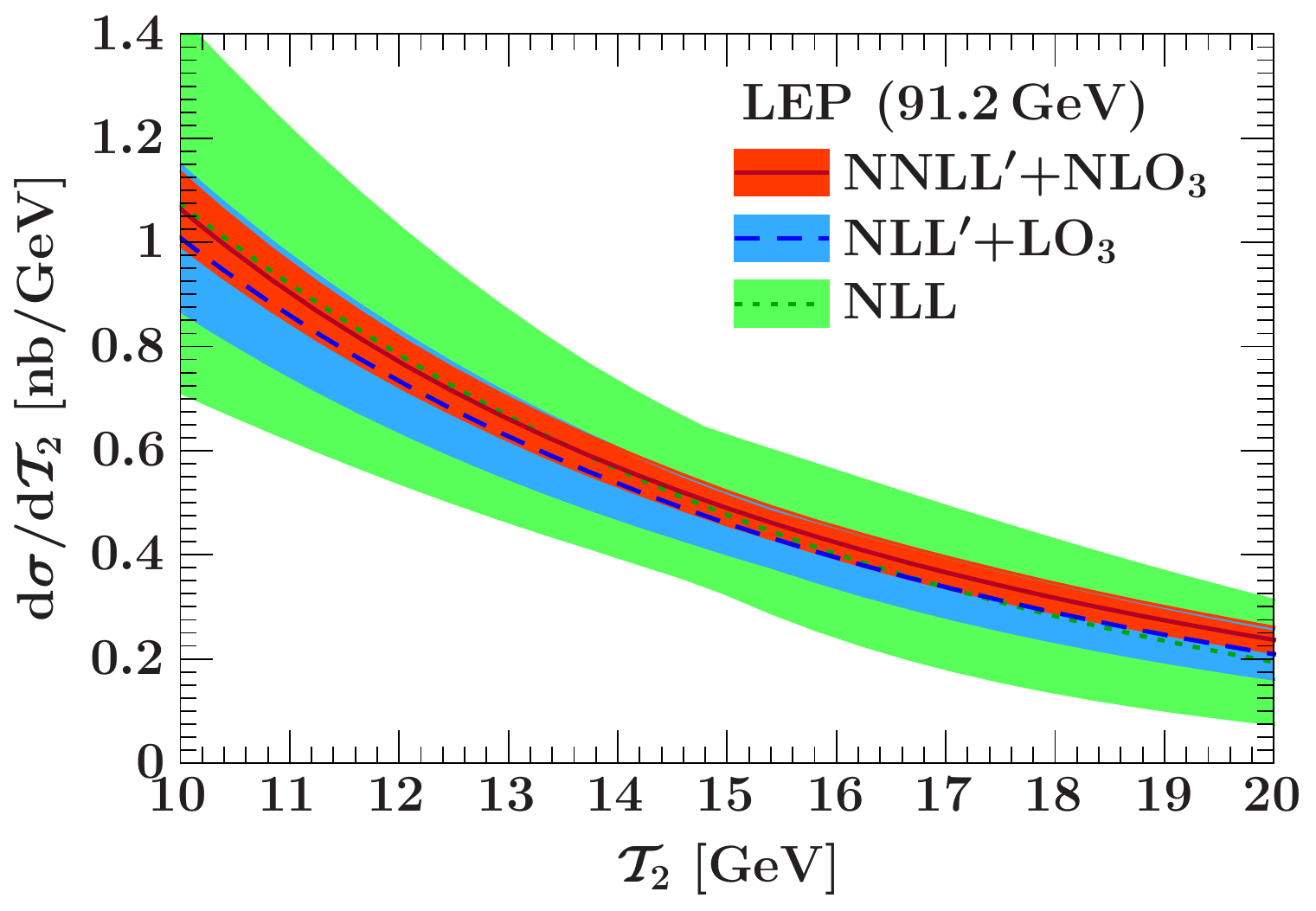}}%
\label{fig:TauConvTrans}}%
\\[-0.5ex]%
\begin{center}
\subfigure[\hspace{1ex} Tail Region]{%
\parbox{0.5\columnwidth}{\includegraphics[scale=0.5]{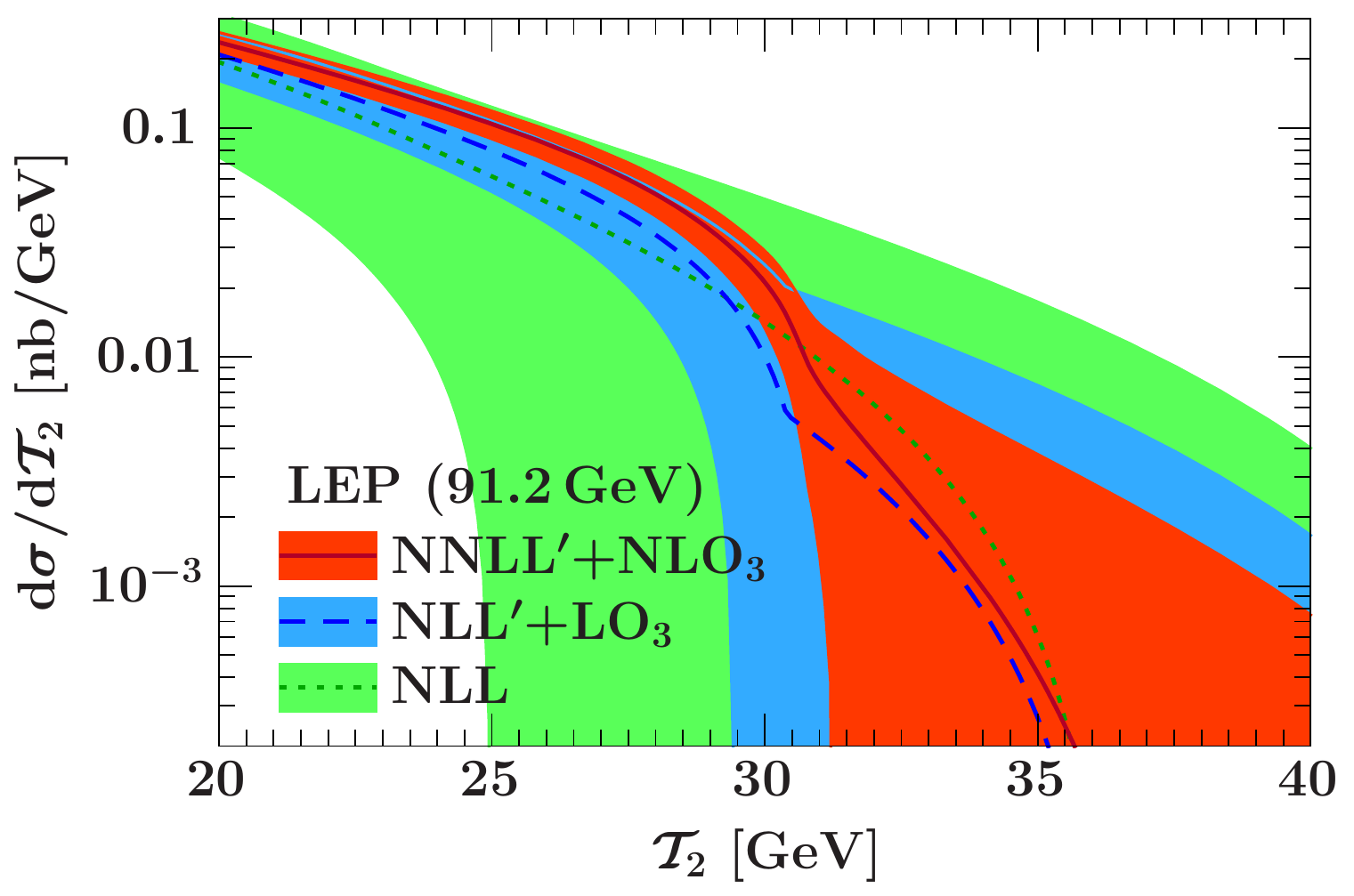}}%
\label{fig:TauConvTail}}%
\end{center}
\vspace{-2ex}
\caption{Analytic resummation of $\Tau_2$ matched to fixed order. The central value is shown along with the band from scale uncertainties, as discussed in \subsubsec{eeResum},  at NLL, NLL$'$+LO$_3$, and NNLL$'$+NLO$_3$. }
\label{fig:TauConv}
\end{figure*}

The analytic resummed $\Tau_2$ spectrum is shown in \fig{TauConv} at successively higher orders: NLL, NLL$'$+LO$_3$, and NNLL$'$+NLO$_3$ (see \tab{expcounting} for the order-counting definitions). The perturbative uncertainties are generated by using the same profile scale variations employed in \geneva and discussed in \subsubsec{eeResum}. The theory uncertainties decrease at increasing order and demonstrate excellent convergence at all values of $\Tau_2$. Below $\Tau_2 < 0.5\GeV$, we enter the purely nonperturbative region, and the scale uncertainties diverge since even resummed perturbation theory breaks down. In the far tail, the scale uncertainties also grow rapidly, which reflects missing higher fixed-order corrections. The uncertainties in the NNLL$'$+NLO$_3$ prediction diverge past the 3-parton endpoint at $\Ecm/3$, where the fixed-order prediction is only correct at leading order for 4 partons. In the transition region, there is a smooth interpolation between the resummation and fixed-order regions.

\begin{figure*}[t!]
\subfigure[\hspace{1ex}  Peak Region]{%
\parbox{0.333\columnwidth}{\includegraphics[scale=0.333]{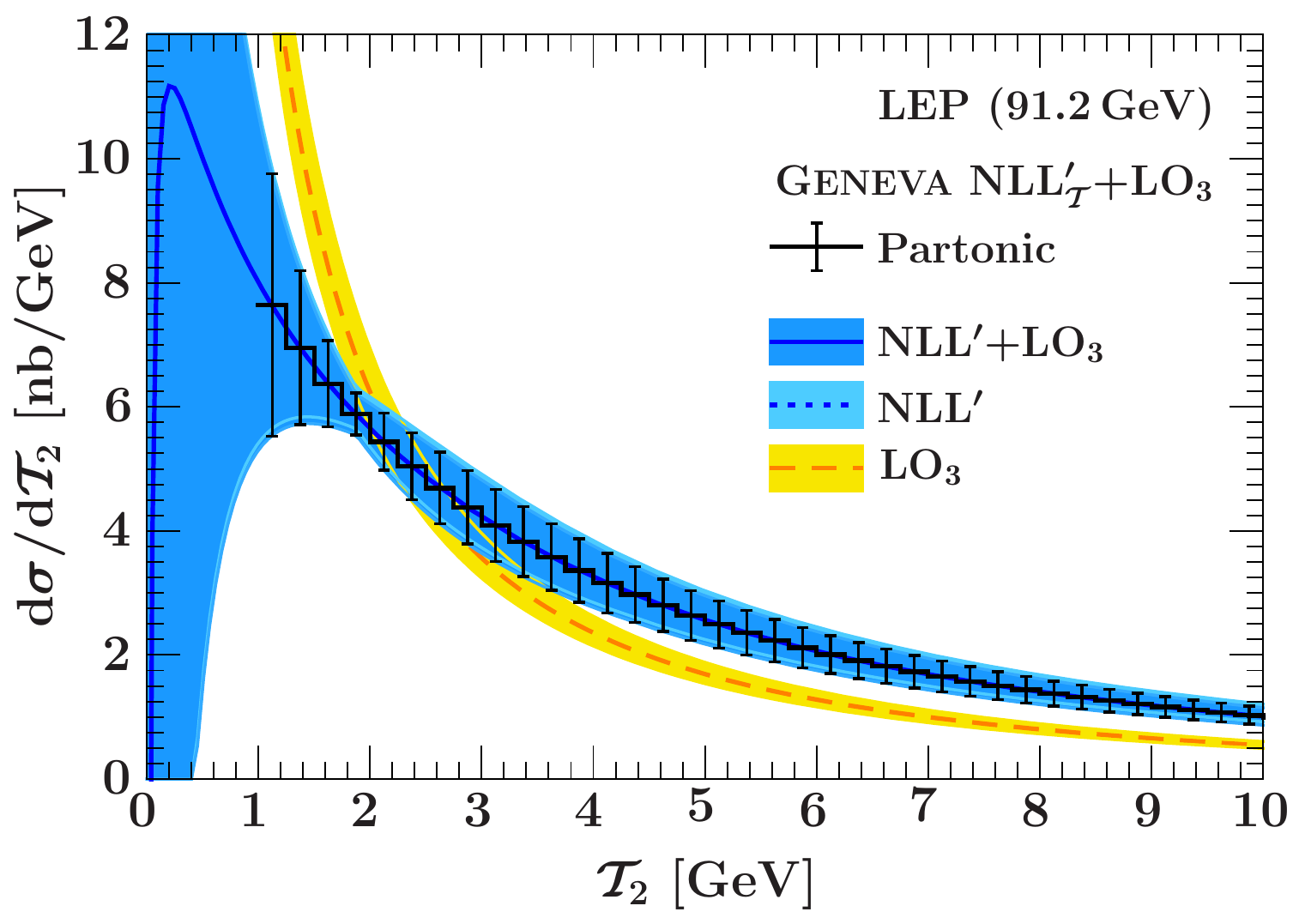}}%
\label{fig:TauNLLpPeak}}%
\hfill%
\subfigure[\hspace{1ex} Transition Region]{%
\parbox{0.333\columnwidth}{\includegraphics[scale=0.333]{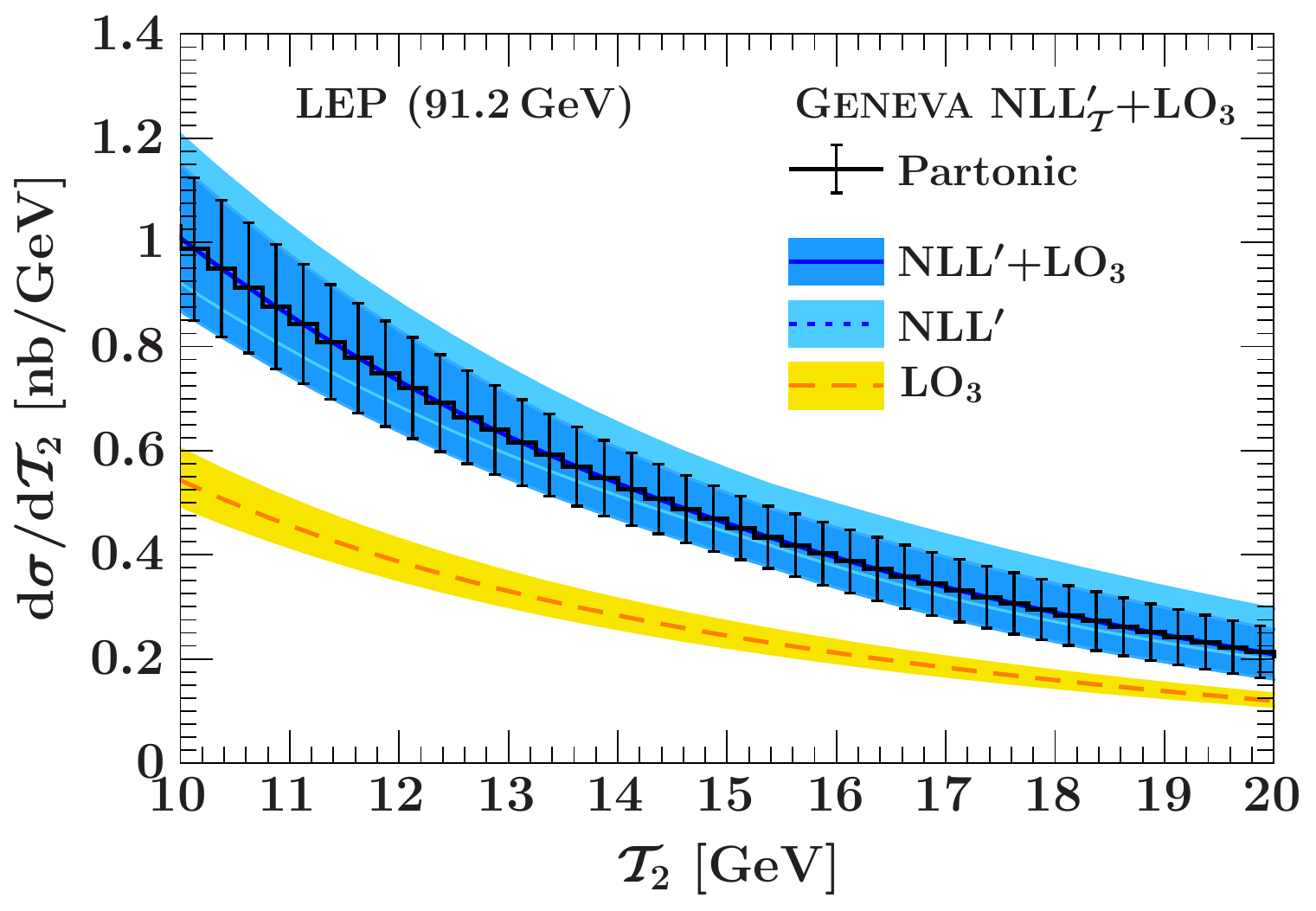}}%
\label{fig:TauNLLpTrans}}%
\hfill%
\subfigure[\hspace{1ex}Tail Region]{%
\parbox{0.333\columnwidth}{\includegraphics[scale=0.333]{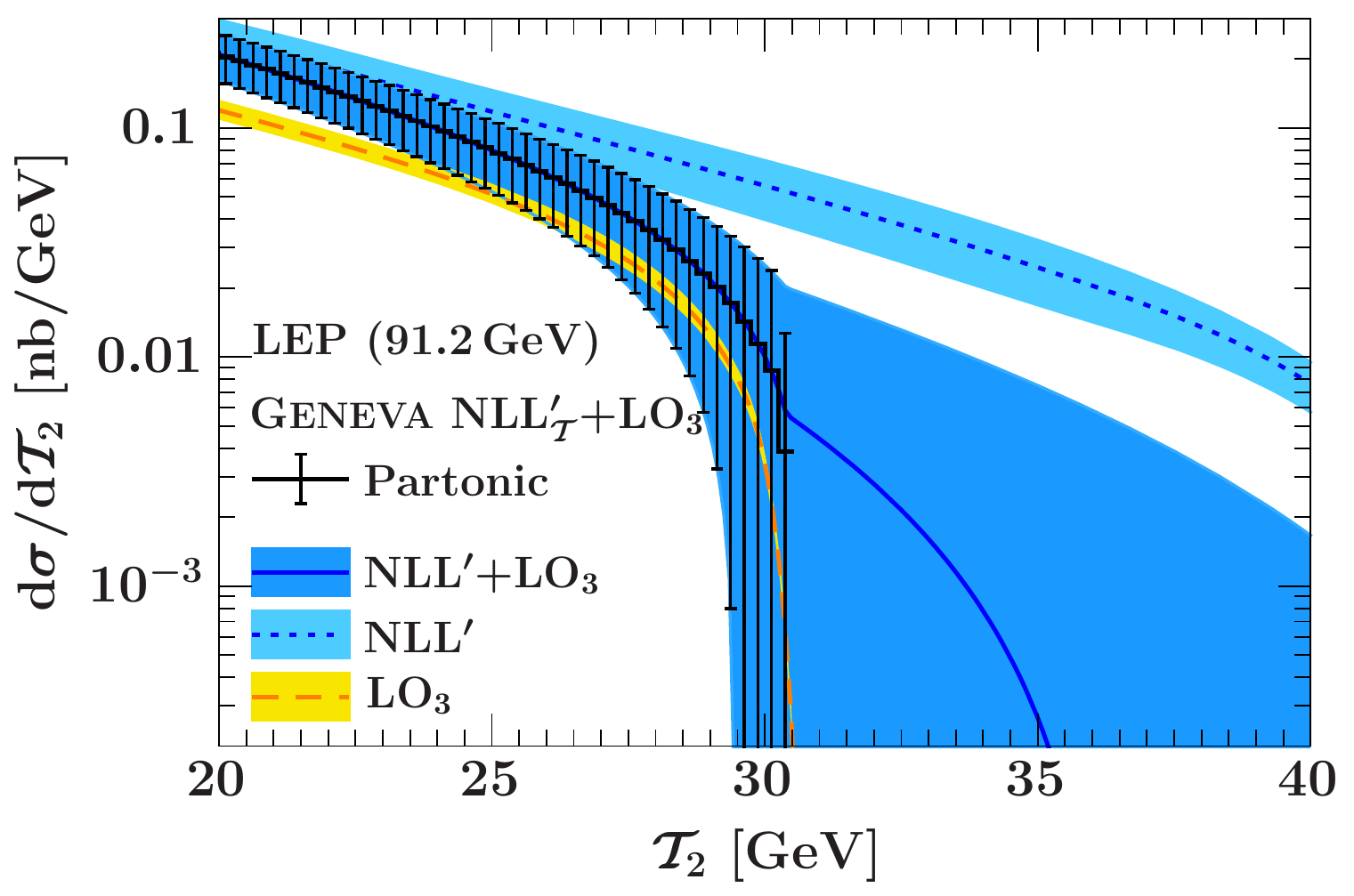}}%
\label{fig:TauNLLpTail}}%
\\%
\subfigure[\hspace{1ex} Peak Region]{%
\parbox{0.333\columnwidth}{\includegraphics[scale=0.333]{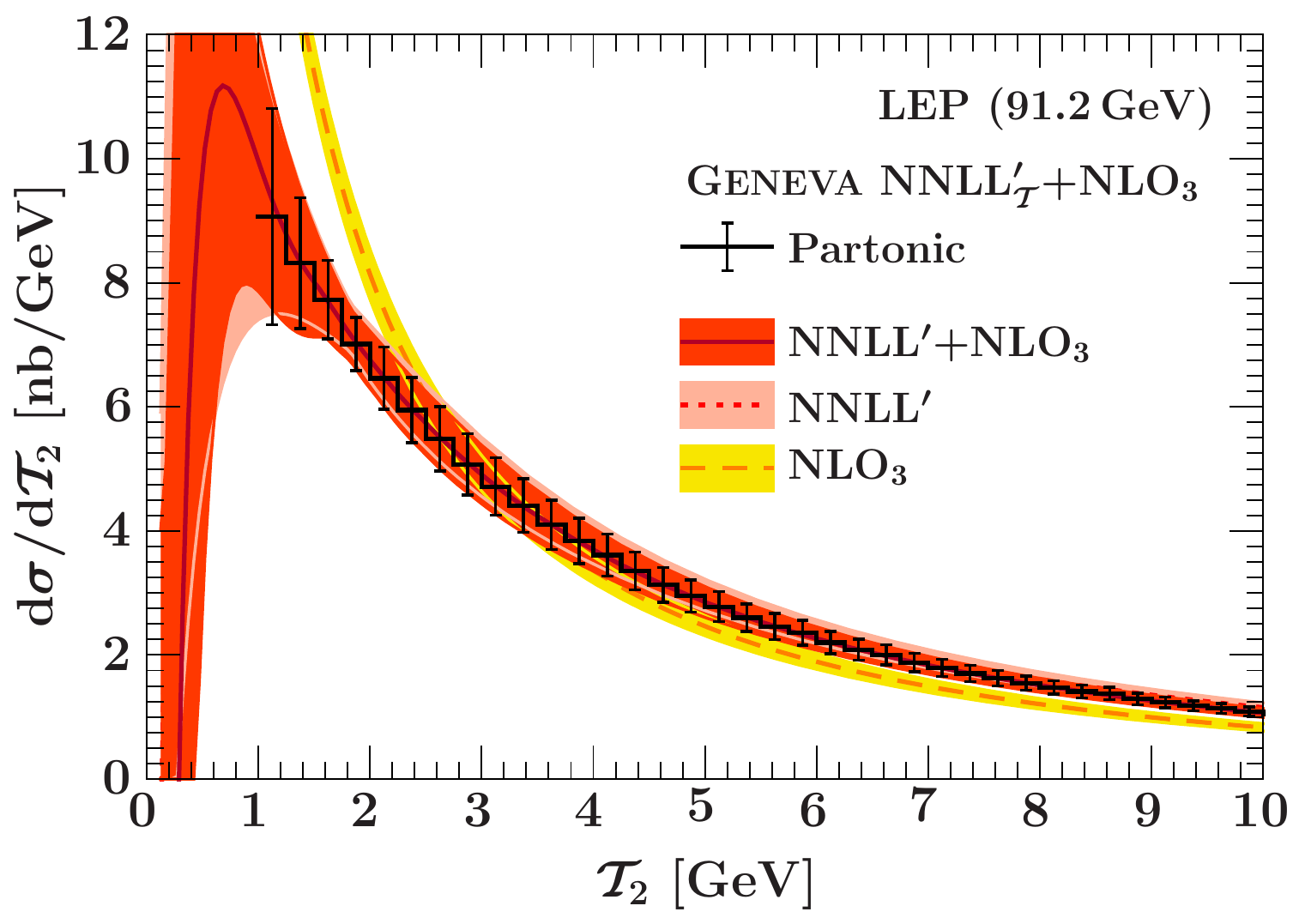}}%
\label{fig:TauNNLLpPeak}}%
\hfill%
\subfigure[\hspace{1ex}  Transition Region]{%
\parbox{0.333\columnwidth}{\includegraphics[scale=0.333]{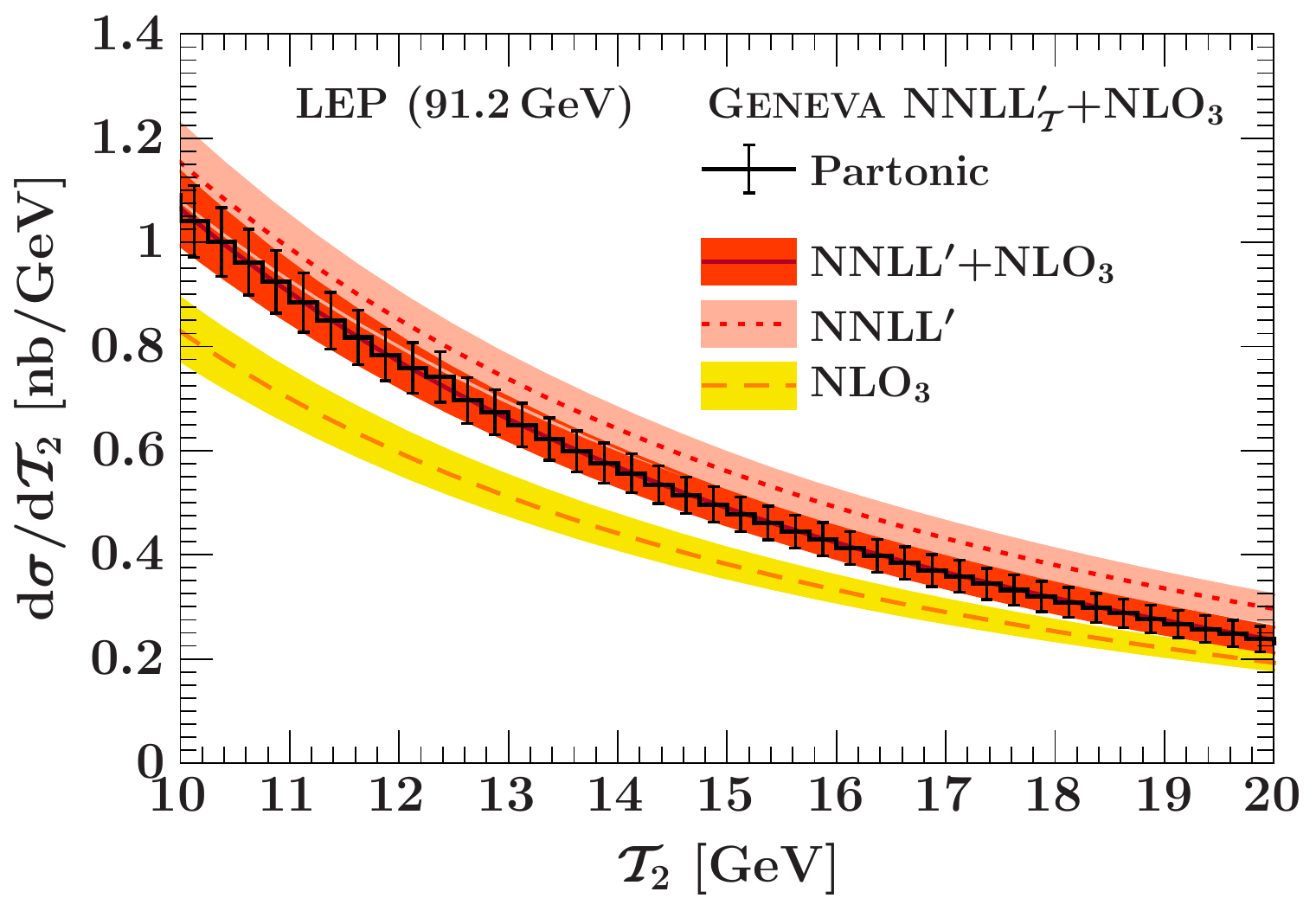}}%
\label{fig:TauNNLLpTrans}}%
\hfill%
\subfigure[\hspace{1ex}  Tail Region]{%
\parbox{0.333\columnwidth}{\includegraphics[scale=0.333]{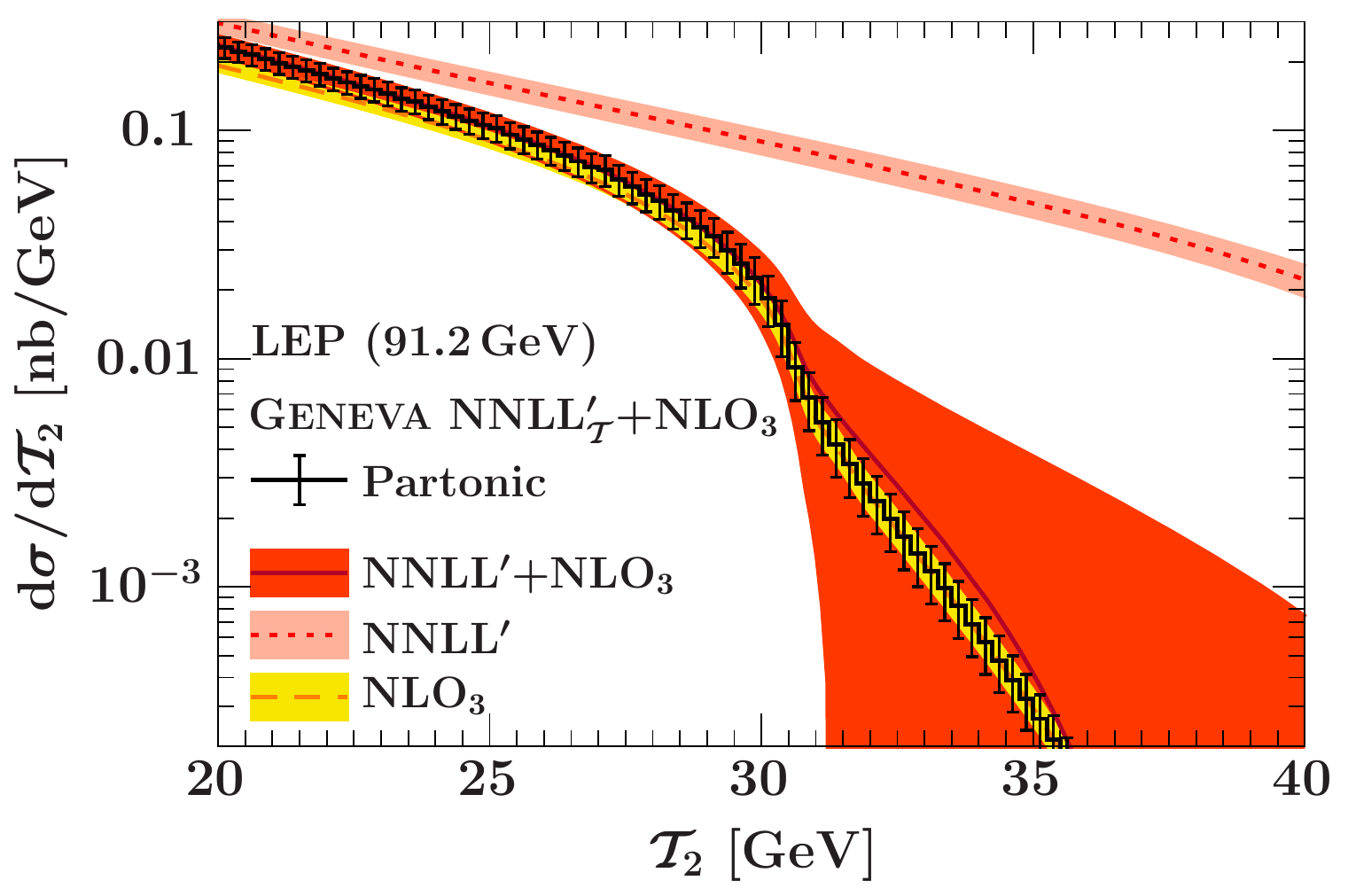}}%
\label{fig:TauNNLLpTail}}%
\vspace{-1ex}
\caption{ The \geneva partonic NLL$'$+LO$_3$  result is shown compared to the analytic resummation of $\Tau_2$ matched to fixed order at  NLL$'$+LO$_3$  in the (a) peak, (b) transition, and (c) tail regions. Also shown for comparison is the pure resummed result at NLL$'$  and the fixed-order LO$_3$ contribution. Figures (d), (e), and (f) show the \geneva partonic result at NNLL$'$+NLO$_3$ compared to the analytic resummation of $\Tau_2$ matched to fixed order at  NNLL$'$+NLO$_3$. The pure NNLL$'$ resummation and fixed-order NLO$_3$ result are also shown for comparison. }
\label{fig:TauVal}
\end{figure*}

In \fig{TauVal}, we compare the partonic $\Tau_2$ spectrum from \geneva with $\Tau_2^\cut=1 \GeV$ to the analytic resummed results from \fig{TauConv}. To illustrate the interpolation between resummed and fixed-order results, we also show the pure resummed results at NLL$'$ and NNLL$'$ and the pure fixed-order contribution at LO$_3$ and NLO$_3$. The latter are calculated using \event~\cite{Catani:1996jh, Catani:1996vz}, which serves as an independent crosscheck of our NLO$_3$ implementation.  Using the NLL$'$+LO$_3$ resummation of $\Tau_2$ and the LO$_3$ fixed-order contribution as inputs to our master formula for the spectrum in \eq{dsigma23}, the $\df\sigma^\FO/\df \Phi_3$ and $\df\sigma^\resum/\df \Phi_2 \df \Tau_2\big\vert_\FO$ contributions exactly cancel for the $\Tau_2$ spectrum. As a result, we see precise agreement between \geneva and the analytic NLL$'$+LO$_3$ result in \fig{TauNLLpPeak}-\ref{fig:TauNLLpTail} in the peak, transition, and tail regions. This result agrees well with the pure NLL$'$ resummed contribution in the peak, while in the tail, it is consistent within uncertainties with the LO$_3$ result, where the latter clearly underestimates the full perturbative uncertainties.

At next higher order, using as inputs to the master formula the NNLL$'$+LO$_3$ resummation of $\Tau_2$ and the NLO$_3$ fixed-order calculation, we see that the central value and event-by-event uncertainties in \geneva agree very well with the full analytic NNLL$'$+NLO$_3$ resummed prediction in the peak and transition regions, as shown in \figs{TauNNLLpPeak}{TauNNLLpTrans}. In the tail region, \fig{TauNNLLpTail}, \geneva has significantly smaller uncertainties of the same size as the pure fixed-order contribution. This is because there is a substantial cancellation between singular and nonsingular contributions in this region, which is incorporated differently in the analytic resummation and the master formula at NNLL$'$+NLO$_3$. For the former, the nonsingular $\alpha_s^2$ contributions are added. This preserves the absolute size of residual resummation uncertainties, which are very small relative to the singular contributions but large relative to the total result after cancellation. In the master formula in \eq{dsigma23}, the nonsingular contributions are incorporated multiplicatively through the ratio of $\df\sigma^\FO/\df \Phi_3$ and $\df\sigma^\resum/\df \Phi_2 \df \Tau_2\big\vert_\FO$. This preserves the relative size of residual resummation uncertainties, thus leading to much smaller absolute variations when compared to the final result. Comparing the \geneva prediction with the pure NNLL$'$ resummed and NLO$_3$ fixed-order results, we see that the master formula precisely interpolates as expected between the fixed-order and resummation regions, with the transition region properly describing the transition between the two, including uncertainties.

Combining the exclusive 2-jet cross section with the integral of the inclusive 3-jet cross section, the \geneva prediction at NNLL$'$+NLO$_3$ formally reproduces the total inclusive cross section at NLO. Numerically, we have $\sigma_{\rm tot}^{\rm NLO}= 44.1 \pm 0.2\,\mathrm{nb}$.  With $\Tau_2^\cut$ smeared between $0.5-1 \GeV$, the total inclusive cross section in \geneva is $\sigma_{\rm tot}^{\geneva}= 42.5 \pm 1.6\,\mathrm{nb}$, where the uncertainties are given by integrating over the different profile scale variations. The central value is $3.8\%$ low and agrees within the uncertainties of $\pm 3.8\%$. The uncertainty in \geneva that comes from integrating the spectrum over $\Tau_2>\Tau_2^\cut$, as in \eq{MCsinglediff}, is much larger than the fixed-order uncertainty. The reason is that, at any given point in the spectrum, but especially in the peak region, the relative uncertainties, reflecting both shape and normalization, are larger than in the total cross section. Hence, when integrating the spectrum to obtain the total cross section, the uncertainties in the spectrum must cancel each other, meaning there is a negative correlation in the uncertainties between different regions in the spectrum. When the resummation and matching to fixed order is performed for the spectrum, this correlation and cancellation is numerically not exact for the total cross section. This is a well-known limitation of analytic resummation~\cite{Abbate:2010xh}. In fact, the result from \geneva is completely consistent with the inclusive cross section obtained using the analytic resummed result in \eq{MCsinglediff} with $\sigma(\Tau_2^\cut)$ calculated at NNLL$'$+LO$_3$ and $\df \sigma/\df \Tau_2$ calculated at NNLL$'$+NLO$_3$. In this case, with $\Tau_2^\cut=1\GeV$, the central value  is $3.5\%$ low with uncertainties of $\pm 3.7\%$. One way to solve this problem would be to enforce a (highly nontrivial) constraint on the profile scale variation to reproduce the required correlation exactly, in which case the total cross section would come out exactly right. In practice, a simpler way to enforce this is to compute the result for the resummed cumulant as the difference between the total cross section and the integrated resummed spectrum. (This is similar in spirit to the method proposed in refs.~\cite{Lonnblad:2012ng, Platzer:2012bs, Lonnblad:2012ix}.) Since our focus in this paper is the differential spectrum, which serves as the primary input to the Monte Carlo, rather than the total cross section, we leave this for future improvement.

\subsubsection{Showered Results}

\begin{figure*}[t!]
\subfigure[\hspace{1ex}Peak Region]{%
\parbox{0.5\columnwidth}{\includegraphics[scale=0.5]{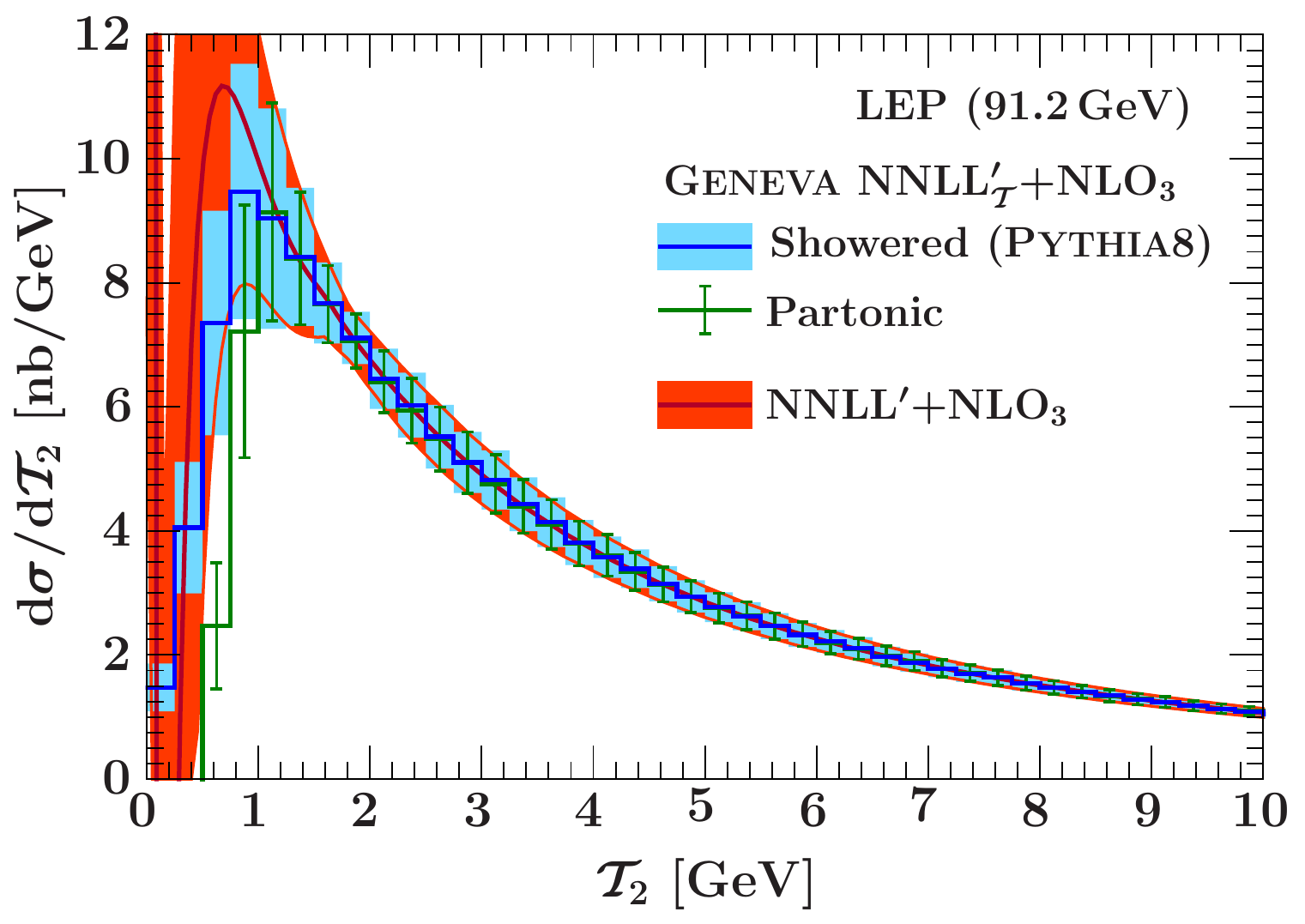}}%
\label{fig:TauPlot1a}}%
\hfill%
\subfigure[\hspace{1ex}Transition Region]{%
\parbox{0.5\columnwidth}{\includegraphics[scale=0.5]{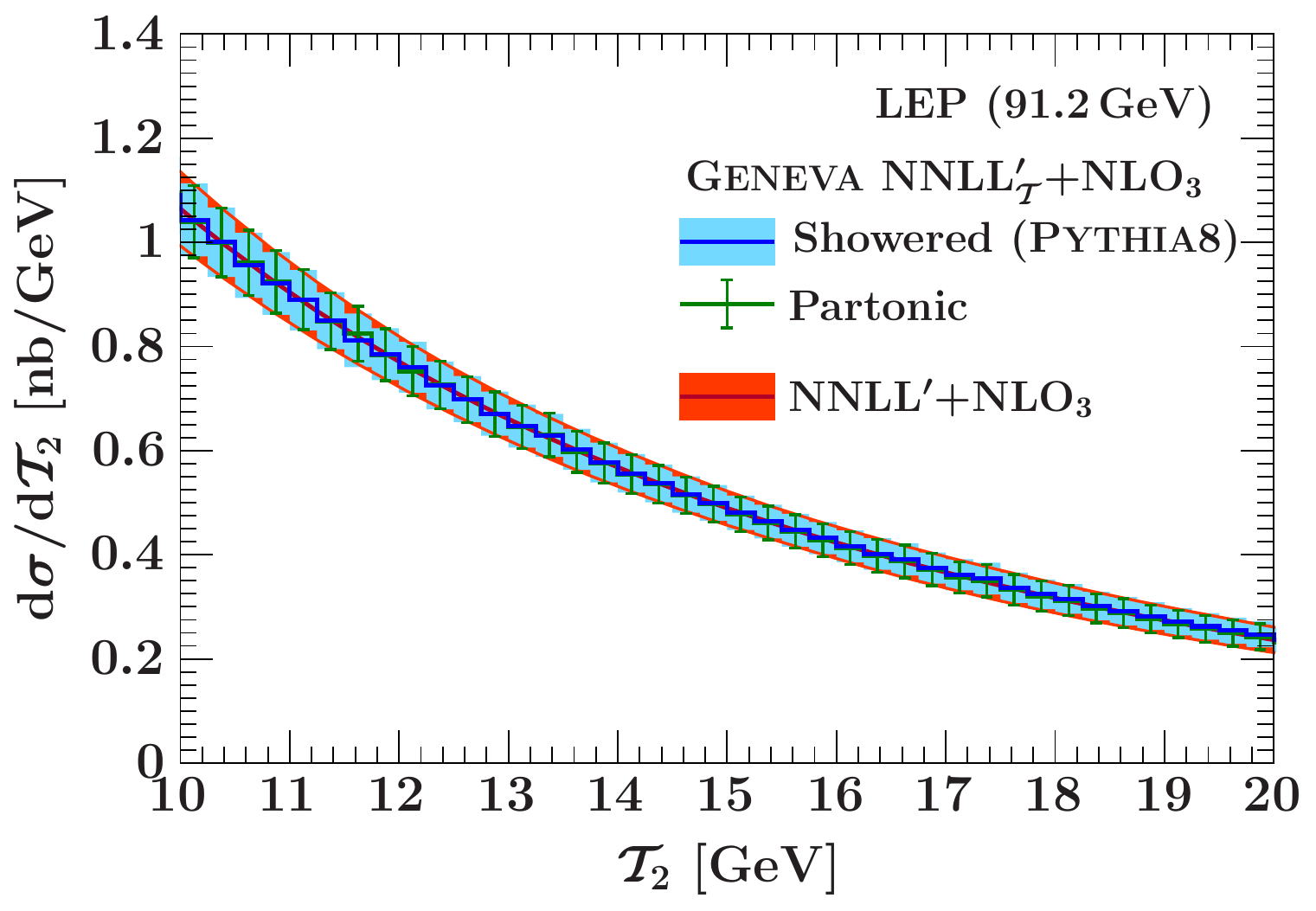}}%
\label{fig:TauPlot1b}}%
\\[-1ex]%
\begin{center}
\subfigure[\hspace{1ex} Tail Region]{%
\parbox{0.5\columnwidth}{\includegraphics[scale=0.5]{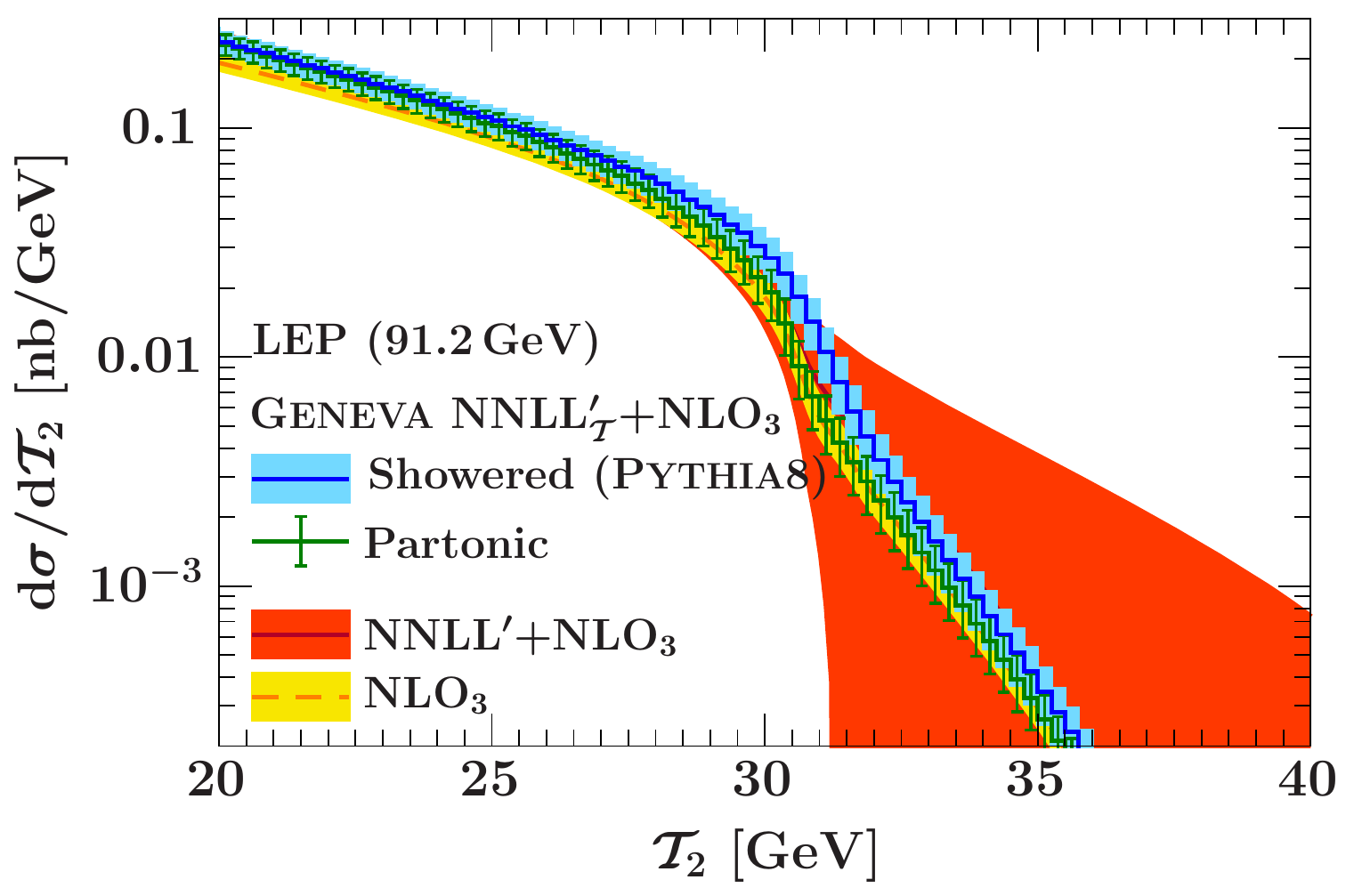}}%
\label{fig:TauPlot1c}}%
\end{center}
\vspace{-3ex}
\caption{The $\Tau_2$ distribution at NNLL$'$+NLO$_3$ from \geneva before and after showering with \pythia8 in the (a) peak, (b) transition, and (c) tail regions of the distribution. The analytic resummed result at NNLL$'$+NLO$_3$ and the fixed-order NLO$_3$ contribution are shown for comparison.}
\label{fig:TauPlot1}
\end{figure*}

\begin{figure*}[t!]
\begin{center}
\parbox{0.5\columnwidth}{\includegraphics[scale=0.5]{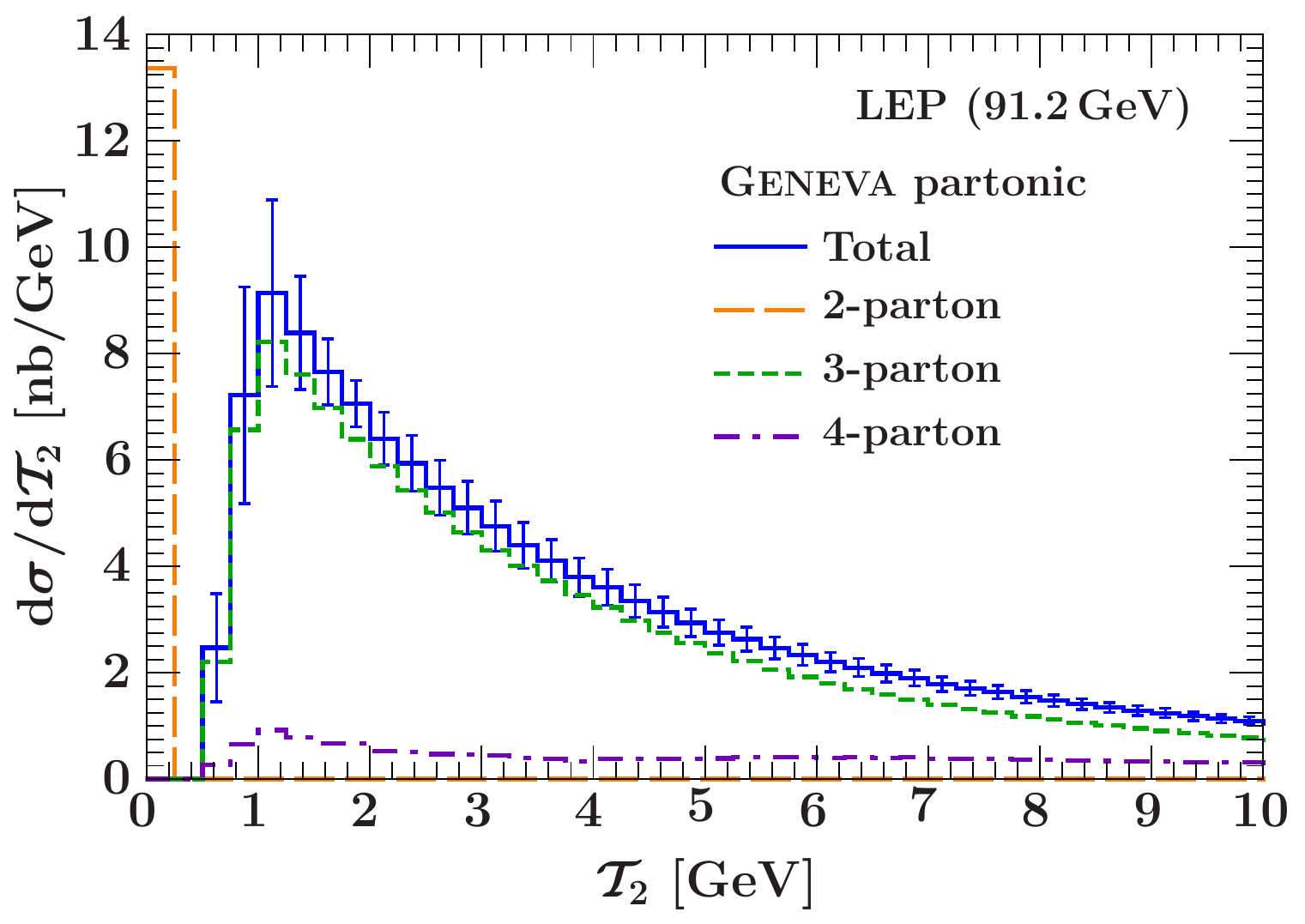}}%
\parbox{0.5\columnwidth}{\includegraphics[scale=0.5]{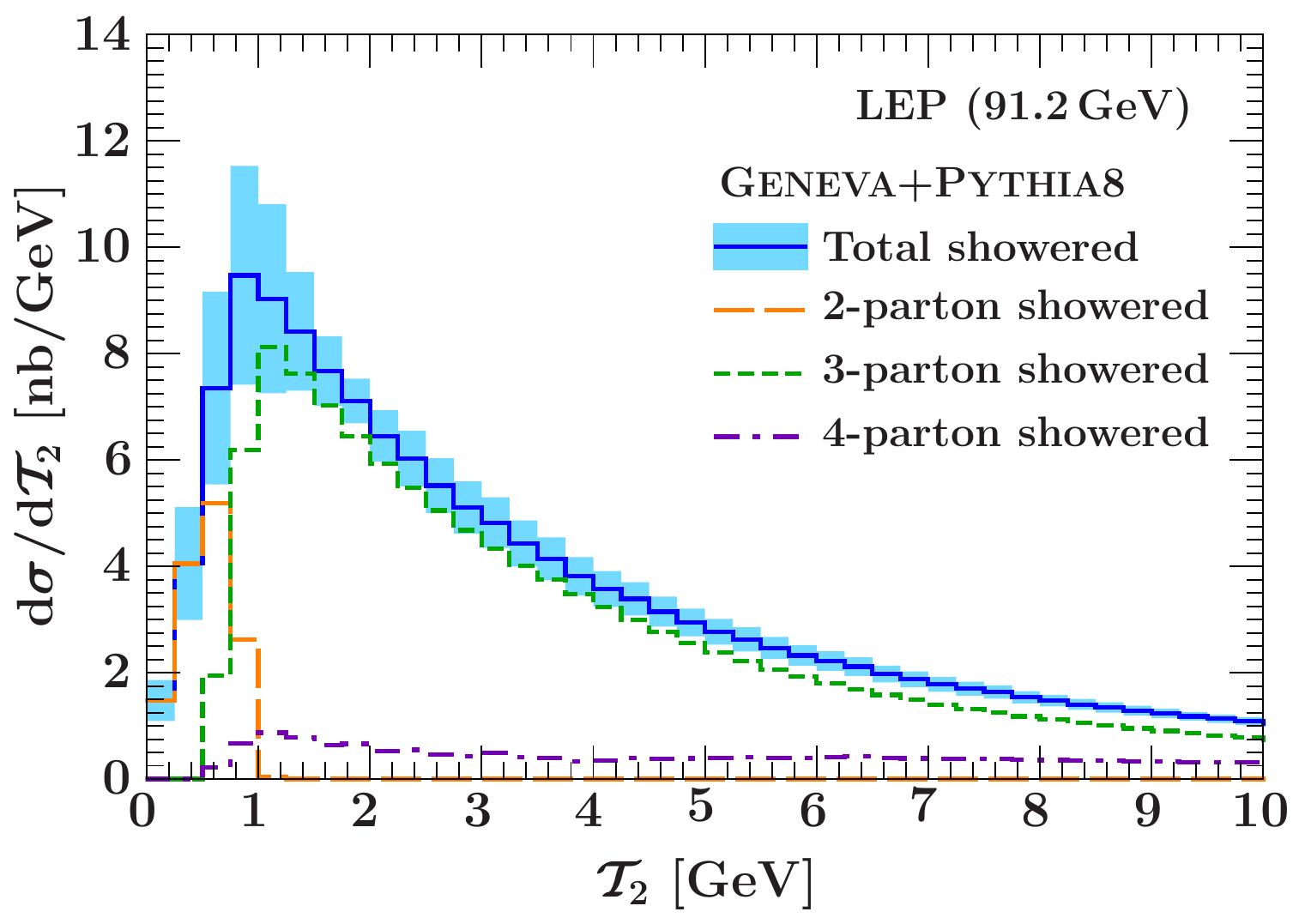}}
 \end{center}
\vspace{-3ex}
\caption{The peak region of the $\Tau_2$ distribution from \geneva partonic (left) and after showering with \pythia8 (right). The contribution from events originating from 2-, 3-, and 4-parton events is shown along with their sum (solid blue histogram), including the perturbative uncertainties shown by the error bars or band.}
\label{fig:Tauf234Plot}
\end{figure*}

Next, we validate our interface with the parton shower. In \fig{TauPlot1}, we compare the NNLL$'$+NLO$_3$ partonic and showered \geneva predictions with $\Tau_2^\cut$ smeared between $0.5-1 \GeV$. We also show the analytic resummed NNLL$'$+NLO$_3$ and pure fixed-order NLO$_3$ spectra for comparison. Before showering, the cumulant $\df \sigma/\df \Phi_2 (\Tau_2^\cut)$ is in the $\Tau_2=0 \GeV$ bin, and we see the effect of the smeared $\Tau_2^\cut$ on the spectrum in the \geneva partonic histogram in \fig{TauPlot1a}. The parton shower generates emissions inside the $2$-jet bin, which fills out and determines the shape of the \geneva showered result in the region below $\Tau_2^\cut$ and agrees remarkably well with the analytic resummed spectrum below the cut.
While the shape of the spectrum here is determined only by \pythia, the cross section below $\Tau_2^\cut$ is still accurate to NNLL$'+$LO$_3$. We can see this explicitly in \fig{Tauf234Plot} from the separate contribution of 2-, 3-, and 4-parton events before and after showering for the central value in the peak region. The shape of the 2-parton showered histogram is determined by \pythia, and the area under the histogram is the cumulant $\df \sigma/\df \Phi_2 (\Tau_2^\cut)$ calculated at NNLL$'$+LO$_3$. The relative contribution of 3-parton and 4-parton events is determined by $\Tau_3^\cut = 2\GeV$, for which the 4-parton contribution is well behaved, giving 15\% of the total cross section and no large cancellation with 3-parton events. These contributions all combine smoothly to generate the total \geneva showered result.

The action of the shower on 3-parton and 4-parton events, which make up the spectrum above $\Tau_2^\cut$, is restricted to not change $\Tau_2$ by more than a power suppressed amount $\lambda \,\Tau_2$, as discussed in \subsubsec{eePythia}. This controls the allowed shift from the \geneva partonic to showered histograms in \fig{TauPlot1}. We can see that there is excellent agreement, including uncertainties, between the two in the peak and transition regions. This validates that, with our choice of $\lambda$, the higher-order accuracy of the resummed $\Tau_2$ spectrum is not compromised by the shower. (Increasing $\lambda$, we do observe, at some point, a shift of showered results away from partonic.) The showering does shift the $\Tau_2$ spectrum in the far tail away from the partonic result, which matches the NLO$_3$ curve, as can be seen in \fig{TauPlot1c}. This is allowed, since our partonic prediction in this region becomes only leading order for 4 partons.

\subsubsection{Hadronized Results and Comparison to Data}

\begin{figure*}[t!]
\subfigure[\hspace{1ex} Peak Region]{%
\parbox{0.5\columnwidth}{\includegraphics[scale=0.5]{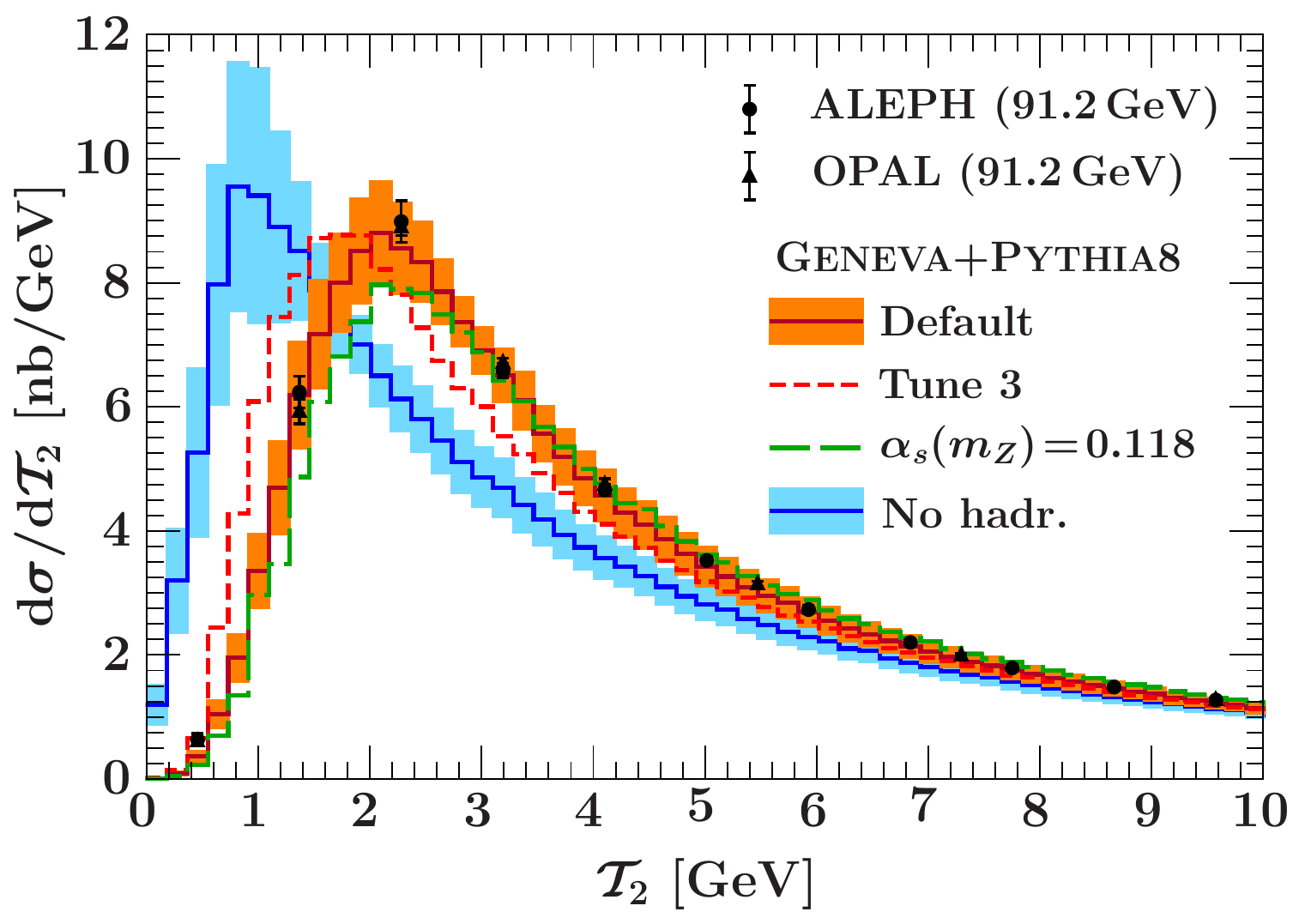}}%
\label{fig:TauPlot2a}}%
\hfill%
\subfigure[\hspace{1ex} Transition Region]{%
\parbox{0.5\columnwidth}{\includegraphics[scale=0.5]{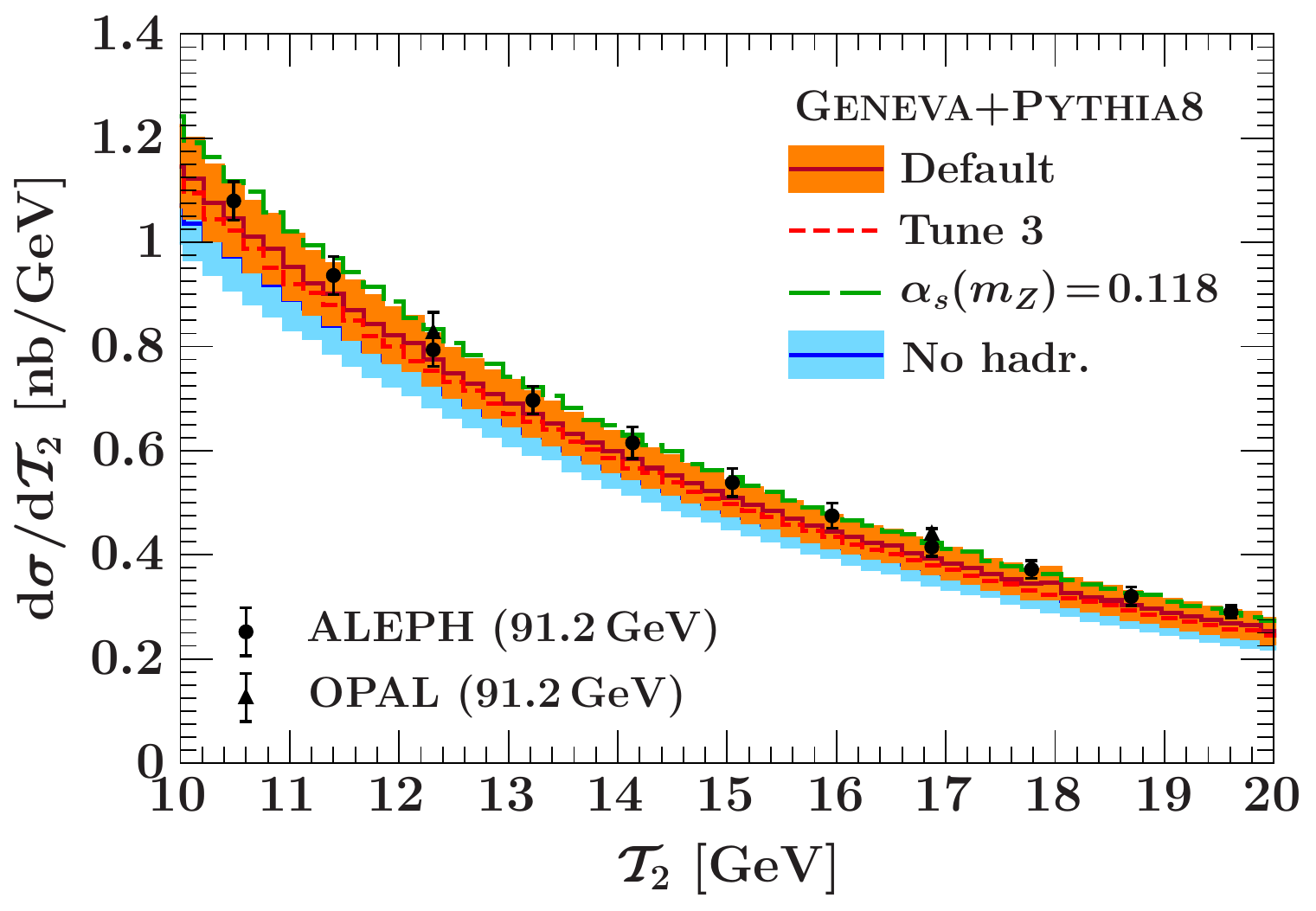}}%
\label{fig:TauPlot2b}}%
\\[-2ex]%
\subfigure[\hspace{1ex}  Tail Region]{%
\parbox{0.5\columnwidth}{\includegraphics[scale=0.5]{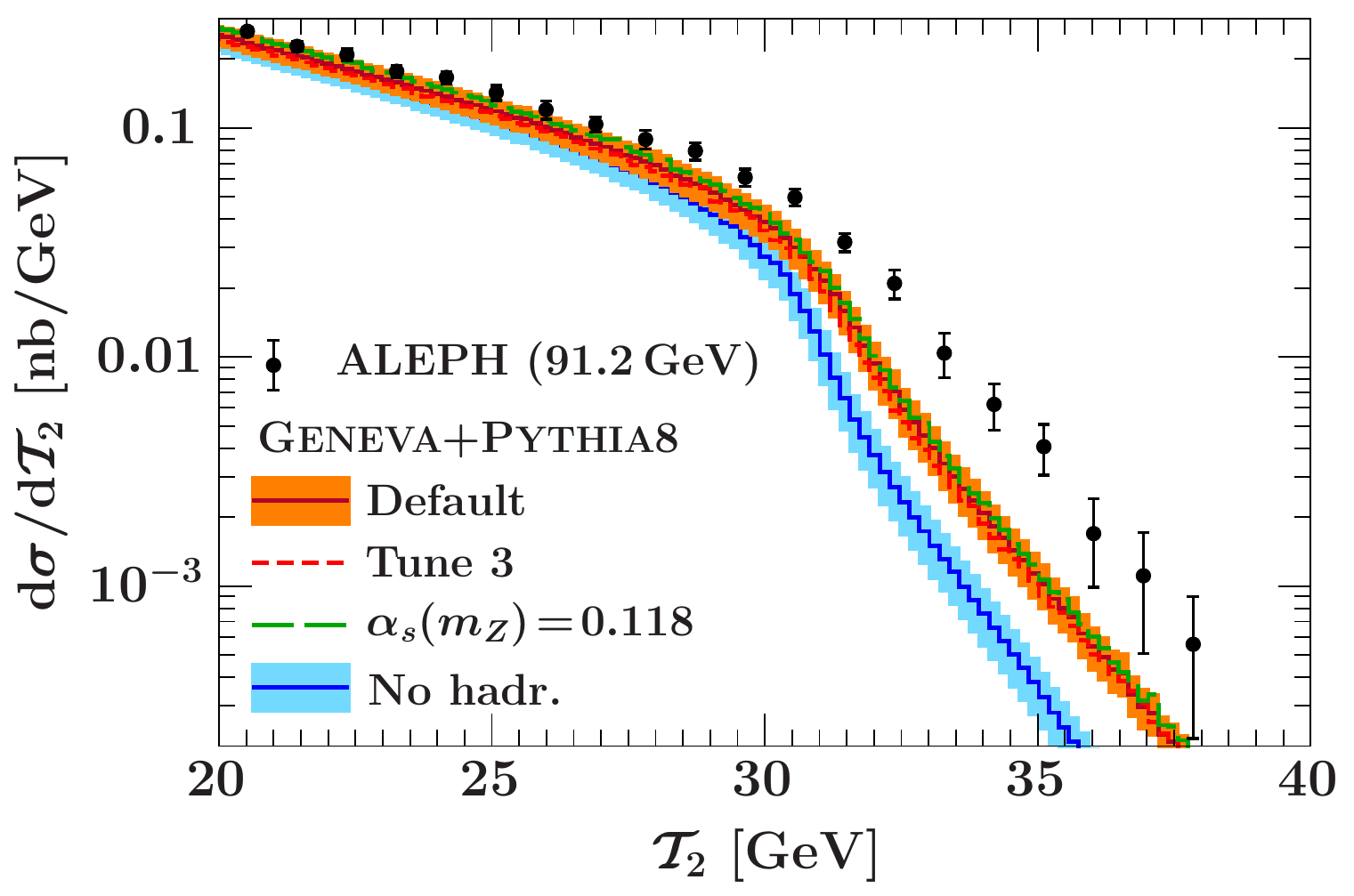}}%
\label{fig:TauPlot2c}}%
\subfigure[\hspace{1ex}  Ratio of \geneva to Data]{%
\parbox{0.5\columnwidth}{\includegraphics[scale=0.5]{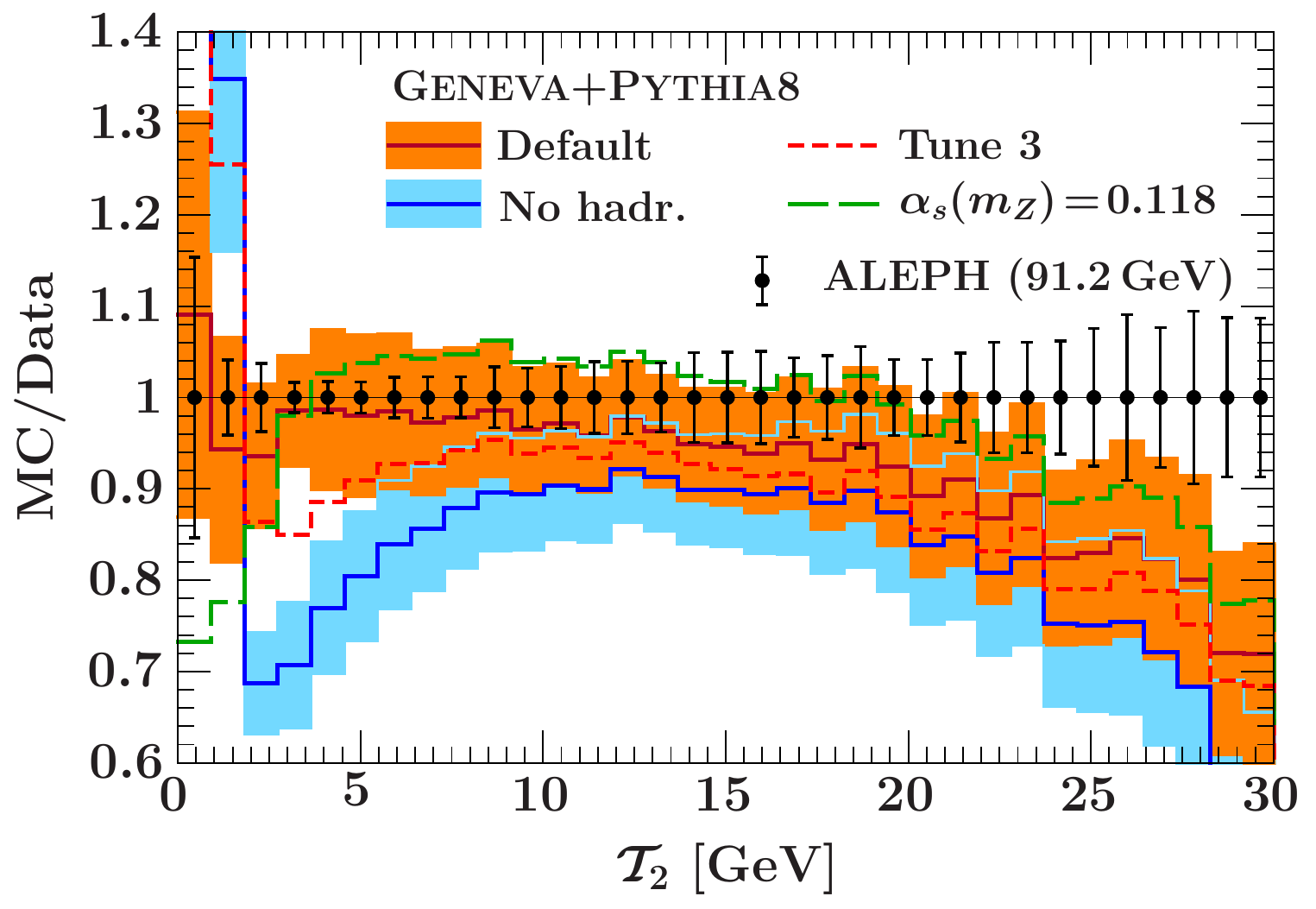}}%
\label{fig:TauPlot2d}}%
\vspace{-2ex}
\caption{The showered NNLL$'$+NLO$_3$ \geneva prediction with and without hadronization using the default values \pythia 8 $e^+e^-$ tune 1 and $\alpha_s (m_Z)=0.1135$ compared to data from \Aleph~\cite{Heister:2003aj} in the (a) peak, (b) transition, and (c) tail regions and to
\Opal~\cite{Abbiendi:2004qz} in the peak and transition regions. The ratio of \geneva predictions to the \Aleph data is shown in (d). Also shown is the \geneva prediction at the central scale with $\alpha_s(m_Z)=0.1184$ and $e^+e^-$ tune 3.}
\label{fig:TauPlot2}
\vspace{-2ex}
\end{figure*}

The full prediction for the jet resolution spectrum is obtained by turning on the hadronization in \pythia. This gives rise to a shift in the $\Tau_2$ spectrum, shown in \fig{TauPlot2}, where ``default'' refers to the default running parameters $\alpha_s(m_Z) = 0.1135$ and \pythia $e^+ e^-$ tune 1. As discussed in \subsubsec{eePythia}, we use the standard \pythia8 tunes without modifying any internal parameters. For comparison, we show the \geneva hadronized result for tune 3 with our default $\alpha_s$ value as well as for tune 1 with the world average value $\alpha_s(m_Z)=0.1184$. We also show a comparison to experimental data from \Aleph~\cite{Heister:2003aj} and \Opal~\cite{Abbiendi:2004qz}.  We only show \Aleph data in the tail since the \Opal data in this region is sparse. These measurements are fully corrected to the particle level, allowing us to directly compare to our hadronized predictions. Since the data are normalized to the total cross section, we rescale them to the total NNLO cross section and convert from thrust $T$ to $\Tau_2 = \Ecm(1 - T)$. This allows us to directly compare the data to the absolute cross section predictions in \geneva, unlike a comparison between normalized spectra, which would only test the shape. The \geneva prediction at the default values agrees impressively well with the data within uncertainties across the peak and transition regions and into the tail. The difference in the far tail is expected since here fixed-order contributions beyond LO$_4$ are important and are not yet included in our results.

The partonic \geneva prediction does not include nonperturbative effects in the soft function of $\ord{\lqcd/\Tau_2}$, nor power corrections of the form $\ord{\lqcd/\Ecm}$. Since we strongly constrain the action of the \pythia parton shower to not change the analytic resummed NNLL$'$+NLO$_3$ result, as discussed in \subsubsec{eePythia} and demonstrated in \fig{TauPlot1}, we expect the hadronization in \pythia to supply these missing nonperturbative effects.  In effect, \pythia provides a well-tested model of the nonperturbative soft function in \eq{S2nonpert}. We show the hadronized \geneva result with \pythia $e^+ e^-$ tune 3 at the central scale in \fig{TauPlot2} as a measure of the uncertainty from hadronization. Tune 3 turns out to give a smaller shift due to hadronization than tune 1, which makes a significant difference in the peak below $\lesssim 3 \GeV$, where nonperturbative corrections are $\ord{1}$ and depend on the details of the hadronization model. In the transition and tail regions, we see a smaller difference, with tune 3 being systematically lower than tune 1. This is consistent with the fact that the transition and tail regions are sensitive only to the first nonperturbative power correction in the soft function of $\ord{\lqcd/\Tau_2}$. 

There is an important interplay between the effect of hadronization and the value of $\alpha_s(m_Z)$, as discussed in ref.~\cite{Abbate:2010xh}, where a simultaneous fit to $\alpha_s(m_Z)$ and the first nonperturbative correction to the soft function of $\ord{\lqcd/\Tau_2}$ was carried out. Generically, larger nonperturbative corrections shift the partonic spectrum to larger values of $\Tau_2$, while a smaller value of $\alpha_s(m_Z)$ shifts the 2-jettiness spectrum downward. This gives rise to compensating effects. Since tune 3 gives a smaller shift due to hadronization than tune 1, the combination of tune 3 and $\alpha_s (m_Z)=0.1135$ gives an estimate of the lower bound on the combined uncertainty of these two effects in the transition and tail regions, while the combination of tune 1 and $\alpha_s(m_Z)=0.1184$ gives an estimate of the upper bound. This is illustrated very well in the ratio of \geneva to \Aleph data in \fig{TauPlot2d}. Both are, however, still within the perturbative uncertainties from \geneva across most of the transition and tail regions.

We have also checked that the nonperturbative shift from \pythia tune 1 is of similar size as expected from the fit results in ref.~\cite{Abbate:2010xh}. This is consistent with the fact that it gives a good description of the data when used together with their fitted value of $\alpha_s (m_Z)$.
Hence, we use tune 1 with $\alpha_s (m_Z)=0.1135$ as the default since it agrees best with the data in the peak and provides a consistent description of the data across larger values of $\Tau_2$.

\pagebreak[3]
\subsection{Predictions for Other Event Shapes}
\label{subsec:eeEvShapes}

In this section, we present \geneva's predictions for a variety of dijet event shape variables.  Examining observables other than the jet resolution variable we use as input serves to validate our master formula at the fully differential $\Phi_{2,3}$ level [see \eq{MCfullydiff23}] rather than its projection onto the $\Tau_2$ spectrum [see \eq{MCPhiNdiff}].  Event shapes are particularly useful to consider because there exist both higher-order resummed results and precision LEP data with which we can compare.

By construction, \geneva correctly predicts other observables at NLO$_3$, while maintaining the correct inclusive cross section.  However, as discussed in \subsubsec{ordercounting}, it is an important open question to what extent the NNLL$'_{\Tau}$+NLO$_3$ resummation of the $\Tau_2$ spectrum increases the accuracy of resummed predictions for the other observables (beyond the partial NLL order naively expected by interfacing with the parton shower). While other observables will not be predicted at the same resummed order as $\Tau_2$, the accuracy of the predictions for event shapes is expected to increase as a function of their correlation with 2-jettiness. The comparison of \geneva to the higher-order analytic resummation of event shapes plays a crucial role in numerically testing the accuracy achieved in our approach and validating the event-by-event perturbative uncertainties.

We present results for the $C$-parameter~\cite{Parisi:1978eg,Donoghue:1979vi,Ellis:1980wv}, heavy jet mass ($\rho$)~\cite{Clavelli:1979md,Catani:1991bd}, and jet broadening ($B$)~\cite{Rakow:1981qn,Catani:1992jc} event shapes. These are defined as follows:
\begin{align}\label{eq:defCrhoB}
C &= \frac{3}{2}\frac{1}{ \left(\sum_k \abs{\vec{p}_k } \right)^2}\sum_{i,j} \abs{\vec{p}_i } \abs{\vec{p}_j }\sin^2 \theta_{ij} 
\,, \nn \\[-1ex]
\rho &= \frac{1}{\Ecm^2} \max \left( M_1^2,M_2^2  \right) \,,
\quad \text{where}\quad
M_i^2 = \biggl(\sum_{k \in {\rm hemi}_i} p_k\biggr)^2
\quad \text{for} \quad i=1,2
\,, \nn\\[-1ex]
B &= \frac{1}{2\sum_k \abs{\vec{p}_k } } \sum_i \abs{ \vec{p}_i \times \hat{n}_T } \,,
\end{align}
where $\hat{n}_T$ is the thrust axis and is used in heavy jet mass to divide the event into two hemispheres, hemi$_{1,2}$, with respect to which the masses $M_{1,2}$  are measured. $C$, $\rho$, and $B$ provide a useful range of event shapes to compare to since their resummation structure is increasingly different from that of $\Tau_2$. The resummation of $C$-parameter is precisely the same as $\Tau_2$ to NLL and has the same convolution structure as \eq{tau2fact} beyond. Heavy jet mass has a different convolution structure from $\Tau_2$. Both $\rho$ and $\Tau_2$ are projections of the same doubly differential spectrum $\df \sigma/\df M_1^2 \df M_2^2$, where $\Tau_2$ is related to the sum and heavy jet mass to the maximum of the hemisphere masses. Of the event shapes we consider, jet broadening is most different from $\Tau_2$; it measures momentum transverse to the thrust axis and, in the dijet limit, is sensitive to the recoil of the thrust axis due to soft emissions~\cite{Dokshitzer:1998kz}, unlike $\Tau_2$. This complicates the higher-order resummation of jet broadening, which was only recently extended to NNLL$_B$~\cite{Becher:2012qc} and gives a logarithmic structure that is very different from $\Tau_2$. As a result, jet broadening provides a highly nontrivial test of the accuracy and theory uncertainties of the \geneva prediction.

\pagebreak[3]
For each of these observables, we compare to analytic resummed predictions as well as the NLO$_3$ fixed-order contribution from \event. We present new results for the analytic resummation of $C$-parameter at NNLL$'_C$+NLO$_3$, extending the previous NLL$_C$ resummation of~\cite{Catani:1998sf}.\footnote{We thank Vicent Mateu and Iain Stewart for pointing out to us the relationship between thrust and $C$-parameter in SCET.} Note the subscript on the order of resummation indicates the observable for which analytic resummation was carried out. Since resummed results for jet broadening do not exist at NNLL$'_B$, we compare to the highest available resummation NNLL$_B$+LO$_3$, where we use the results of~\cite{Becher:2012qc}, which we extend to include fixed-order matching that is necessary to describe the tail and transition regions. Finally, for heavy jet mass, N$^3$LL$_\rho$ resummed results exist~\cite{Chien:2010kc}; however, we show the NNLL$'_\rho$+NLO$_3$ resummation since this is consistent with the highest $\Tau_2$ resummation we use.

It is important to note that all running parameters were set based on the $\Tau_2$ spectrum alone, and no further optimization was carried out for other observables. This ensures that our results for other observables are true predictions of the \geneva framework.

\subsubsection{$C$-parameter}
\label{subsubsec:eeCparam}

\begin{figure*}[t!]
\subfigure[\hspace{0.5 ex}\geneva NLL$'_\Tau$+LO$_3$ Peak Region]{%
\parbox{0.5\columnwidth}{\includegraphics[scale=0.5]{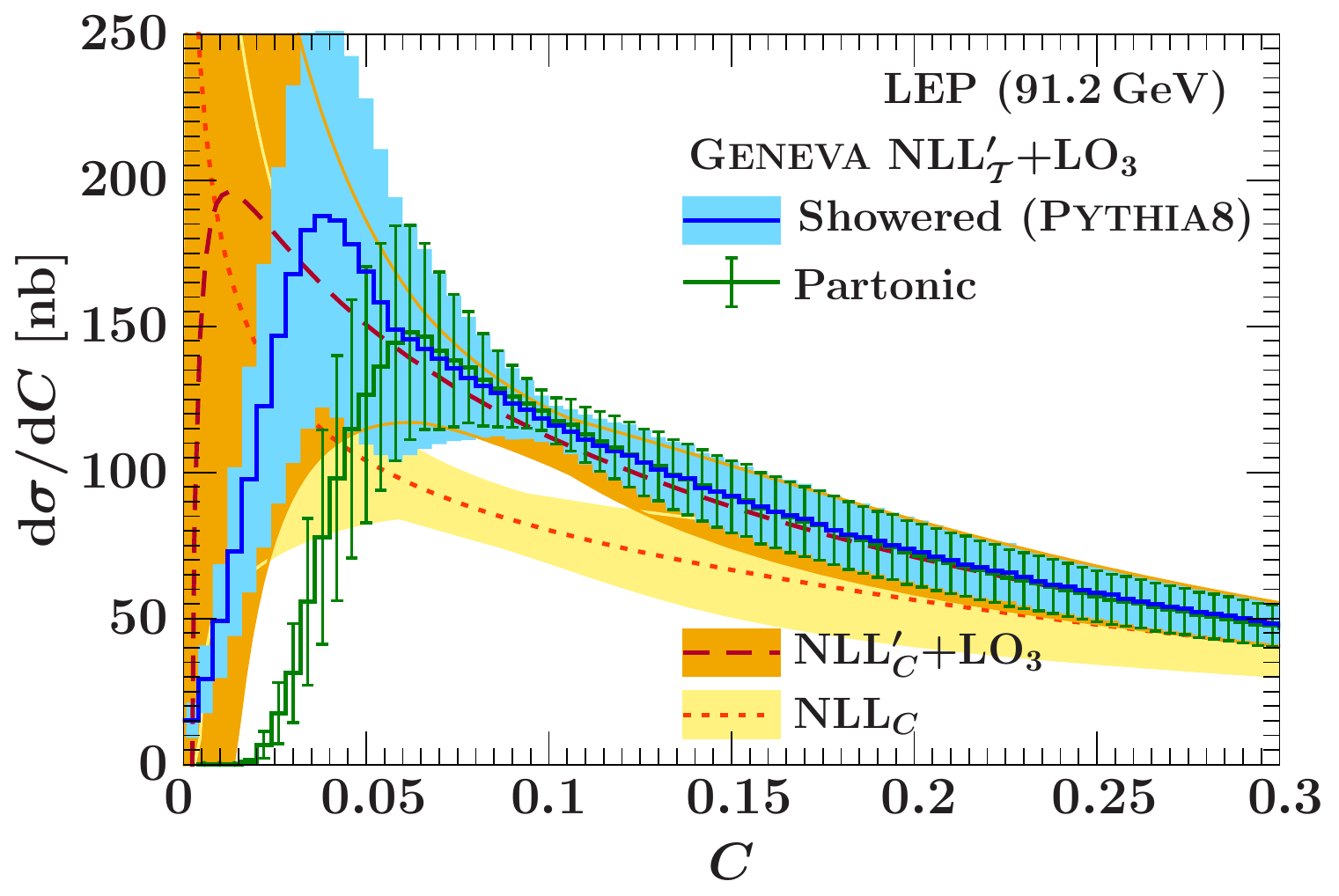}}%
\label{fig:Cplot1a}}%
\hfill%
\subfigure[\hspace{0.5ex} \geneva NLL$'_\Tau$+LO$_3$ Transition Region]{%
\parbox{0.5\columnwidth}{\includegraphics[scale=0.5]{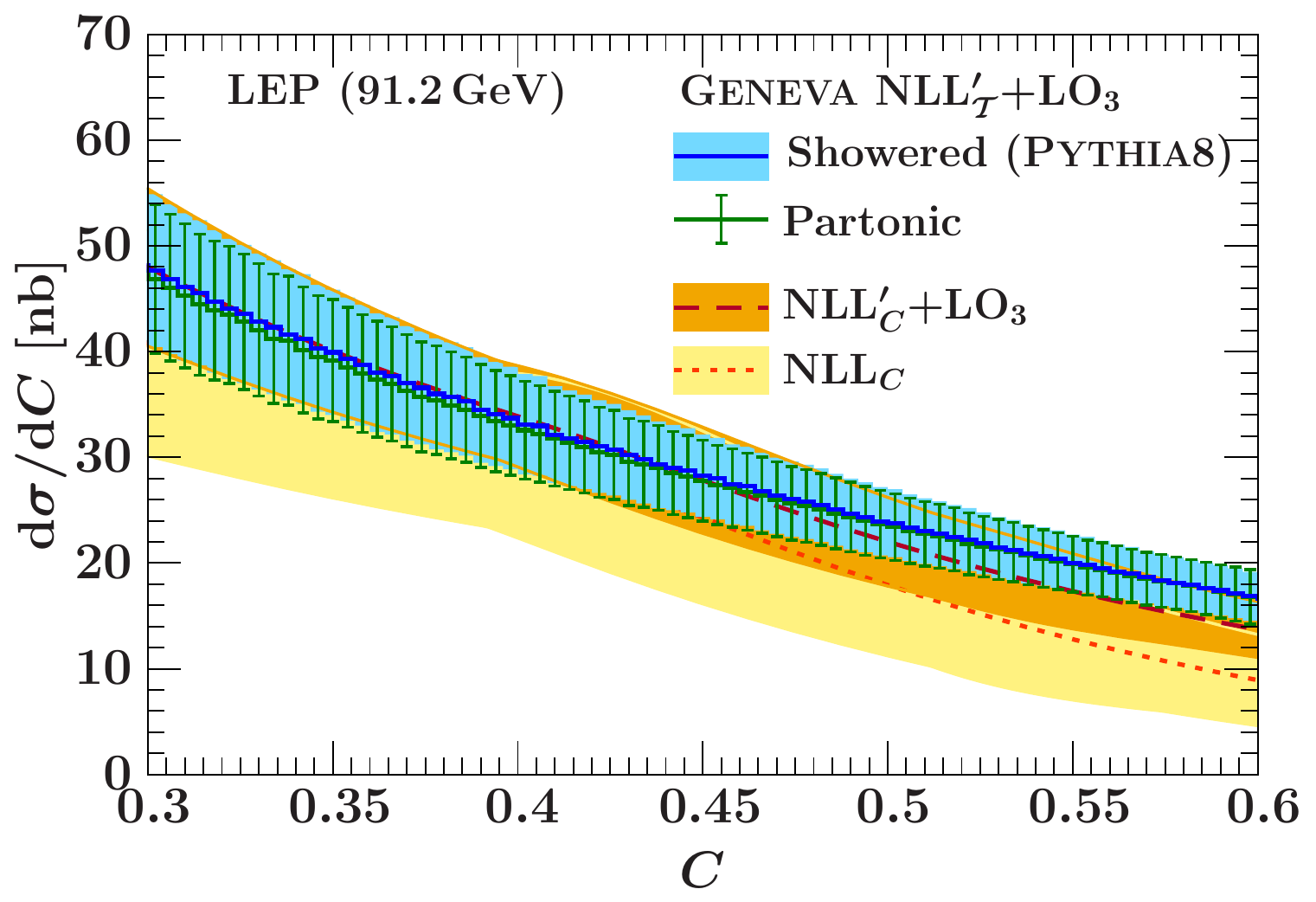}}%
\label{fig:Cplot1b}}%
\\%
\subfigure[\hspace{0.5ex} \geneva NNLL$'_\Tau$+NLO$_3$ Peak Region]{%
\parbox{0.5\columnwidth}{\includegraphics[scale=0.5]{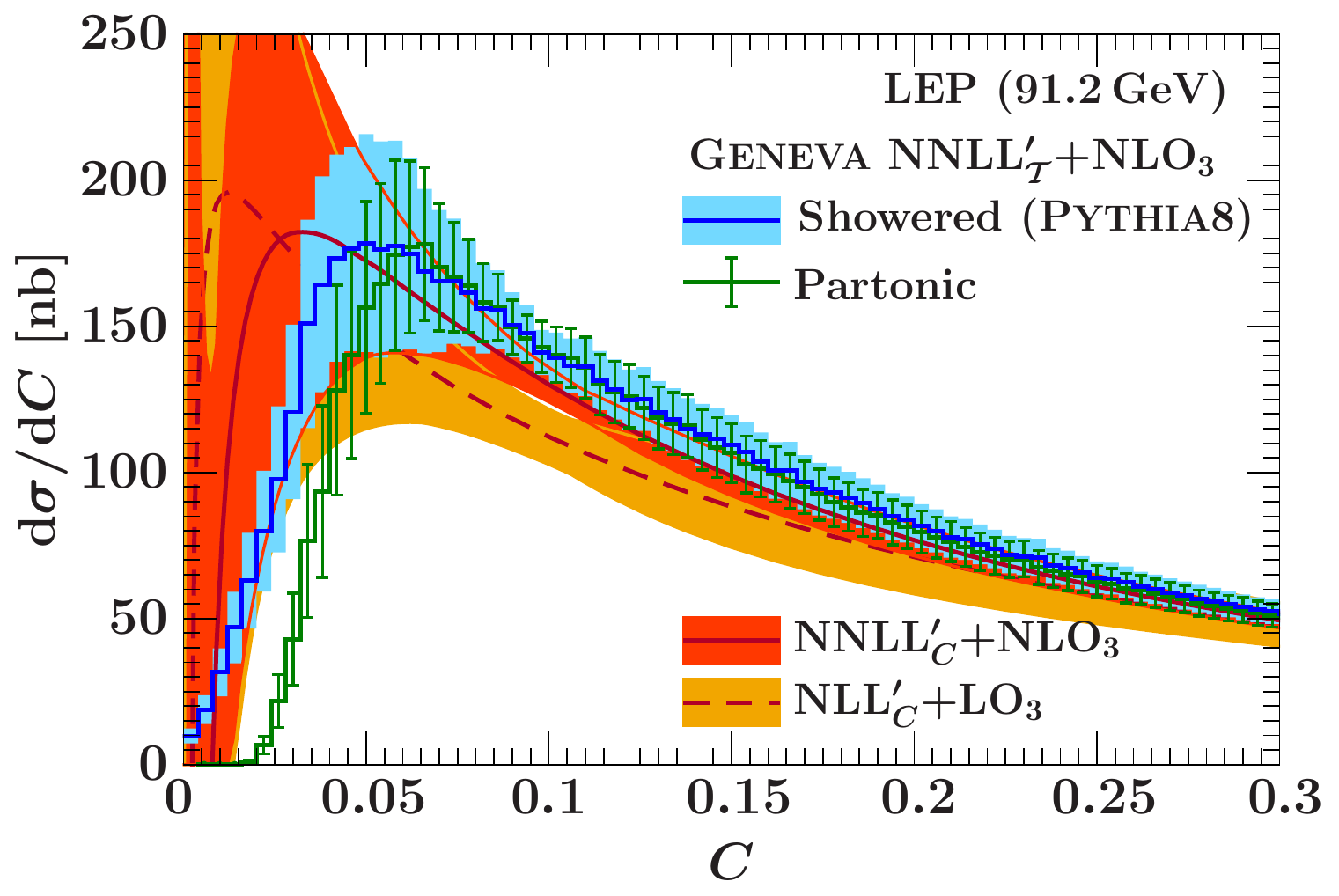}}%
\label{fig:Cplot1c}}%
\hfill%
\subfigure[\hspace{0.5ex} \geneva NNLL$'_\Tau$+NLO$_3$ Transition Region]{%
\parbox{0.5\columnwidth}{\includegraphics[scale=0.5]{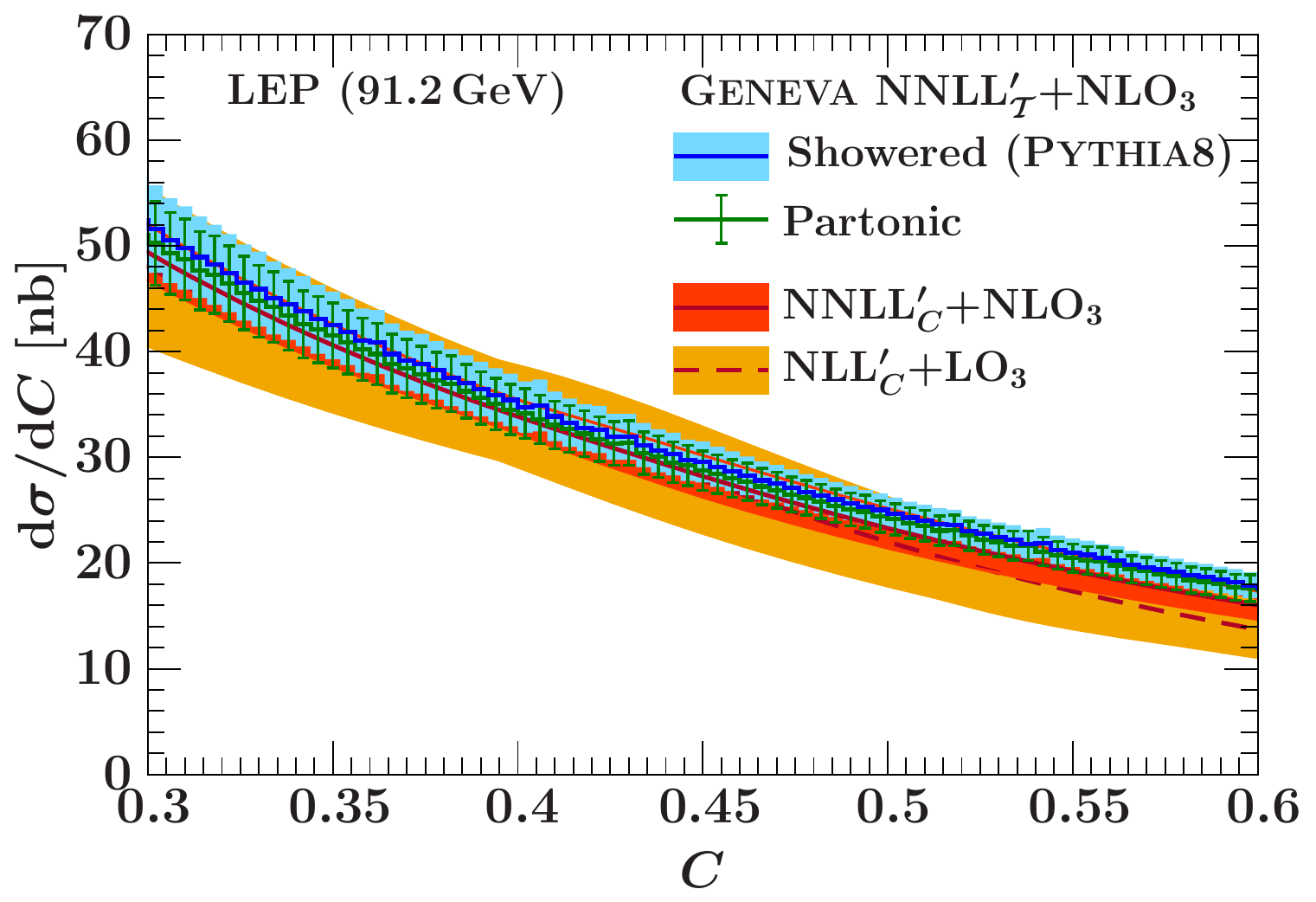}}%
\label{fig:Cplot1d}}%
\\%
\begin{center}
\subfigure[\hspace{0.5ex} \geneva NNLL$'_\Tau$+NLO$_3$ Tail Region]{%
\parbox{0.5\columnwidth}{\includegraphics[scale=0.5]{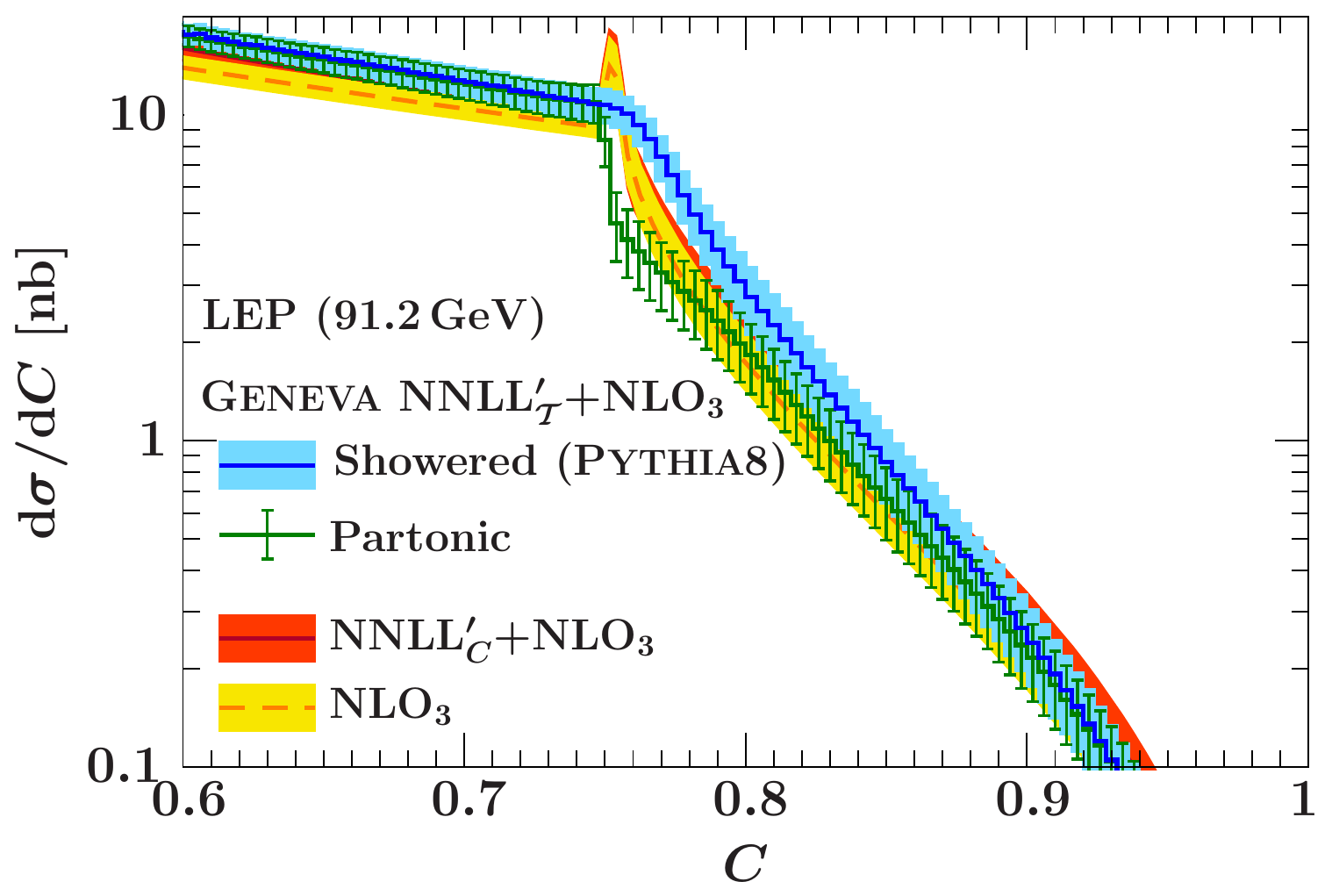}}%
\label{fig:Cplot1e}}%
\end{center}
\vspace{-0.5ex}
\caption{The $C$-parameter partonic and showered \geneva predictions are shown compared to the analytic resummation of $C$-parameter at different orders. The \geneva result at NLL$'_\Tau$+LO$_3$ is compared to NLL$_C$ and NLL$'_C$+LO$_3$ in (a) and (b). In (c) and (d), the \geneva prediction at one order higher, NNLL$'_\Tau$+NLO$_3$, is compared to NLL$'_C$+LO$_3$ and NNLL$'_C$+NLO$_3$, while in the tail (e), we also show the fixed-order NLO$_3$ prediction from \event.}
\label{fig:Cplot1}
\end{figure*}

In \fig{Cplot1}, we show the \geneva prediction for $C$-parameter both at the partonic level and showered, using NLL$'_{\Tau}$+LO$_3$ resummation as input to our master formula in \figs{Cplot1a}{Cplot1b} and at next higher order NNLL$'_{\Tau}$+NLO$_3$ in \figs{Cplot1c}{Cplot1d}. We compare this to the analytic resummed $C$-parameter prediction at the same order as the $\Tau_2$ resummation we input, as well as one order lower. The comparison of the \geneva prediction at different orders in the peak and transition regions is useful because it highlights the features of resummation that are consistently captured by our implementation. In the tail region, \fig{Cplot1e}, where the comparison to the NLO$_3$ fixed-order result is most relevant, we only show our highest order NNLL$'_\Tau$+NLO$_3$ \geneva result.

We see the effect of the cut on 2-jettiness of $\Tau_2^\cut = 0.5-1\GeV$ up to $C=0.066$ in the partonic prediction from \geneva in \figs{Cplot1a}{Cplot1c} since $C\leq 6 \Tau_2/Q$~\cite{Catani:1998sf}. Interfacing with the shower generates emissions inside the jets and fills out the region below $C=0.066$. The action of the parton shower is restricted based on the constraints on $\Tau_2$ discussed in \subsubsec{eePythia}. This effectively constrains the $C$-parameter distribution as well, giving very little change from the partonic to showered predictions at both resummation orders except in the multijet region of the far tail. Here, the constraints on the shower are looser, reflecting the fact that our prediction is correct at LO$_4$. The size of the shift from the partonic to the showered result in the peak and transition regions is a measure of the correlation between the $C$ and $\Tau_2$ event shapes, where, although the two differ beyond NLL, their logarithmic structure is the same.
It is worth noting however, that despite the similarity of the resummation structure between $C$ and $\Tau_2$, the shape of the $C$-parameter spectrum is very different, with the singular terms dominating the nonsingular for a much larger region of the spectrum.
\clearpage

\begin{figure*}[t!]
\subfigure[\hspace{1ex} Peak Region]{%
\parbox{0.5\columnwidth}{\includegraphics[scale=0.5]{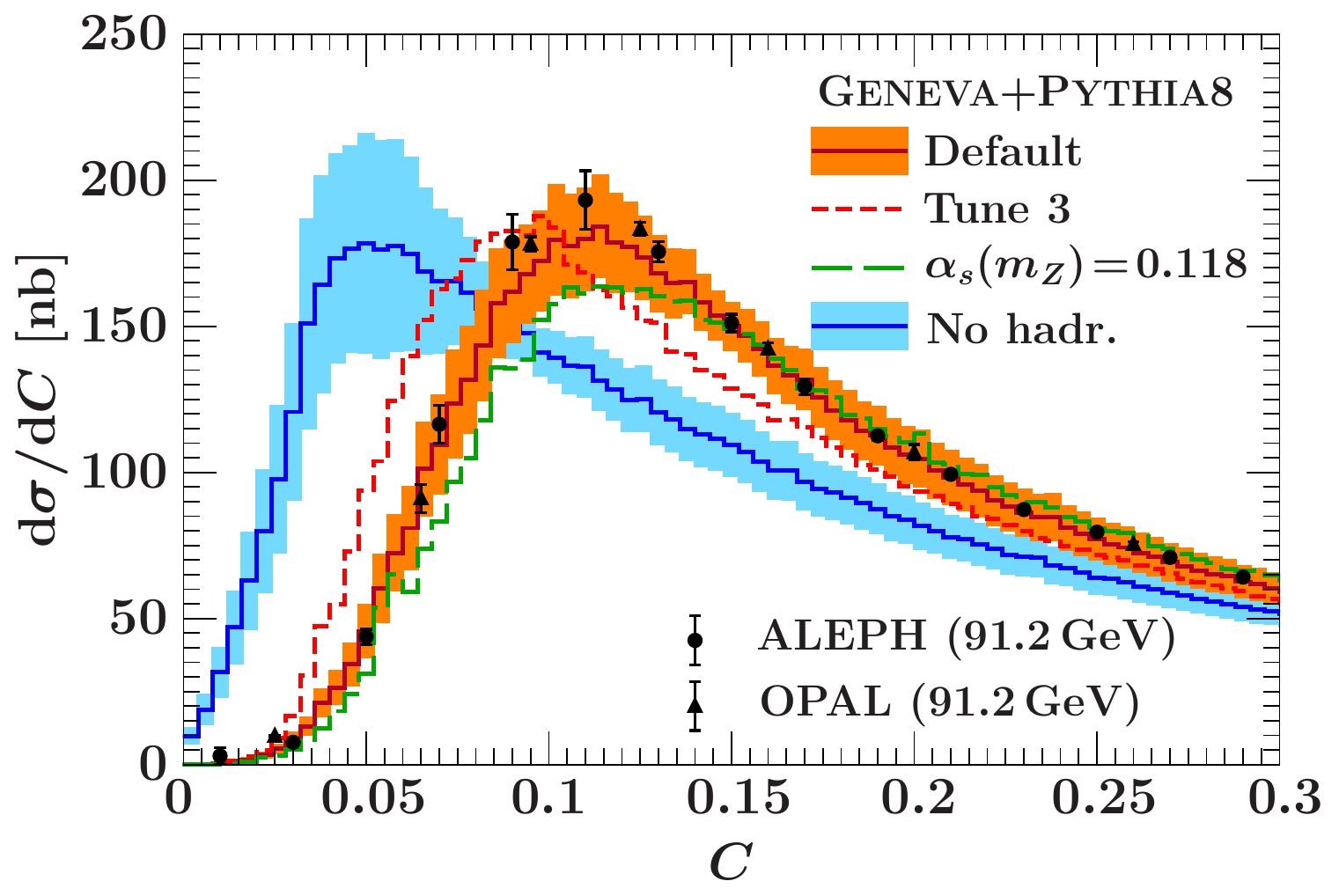}}%
\label{fig:CPlot2a}}%
\hfill%
\subfigure[\hspace{1ex} Transition Region]{%
\parbox{0.5\columnwidth}{\includegraphics[scale=0.5]{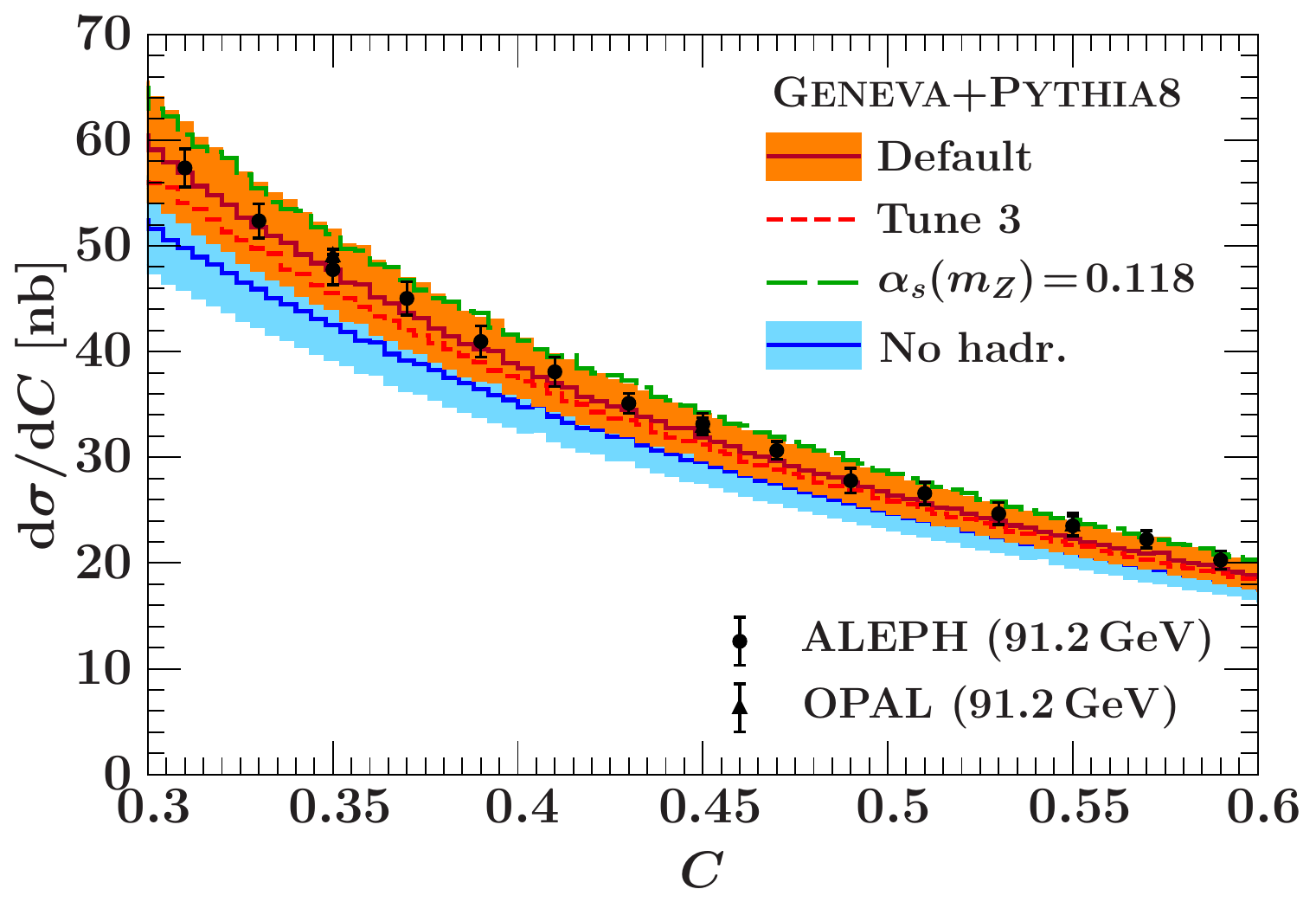}}%
\label{fig:CPlot2b}}%
\\%
\subfigure[\hspace{1ex} Tail Region]{%
\parbox{0.5\columnwidth}{\includegraphics[scale=0.5]{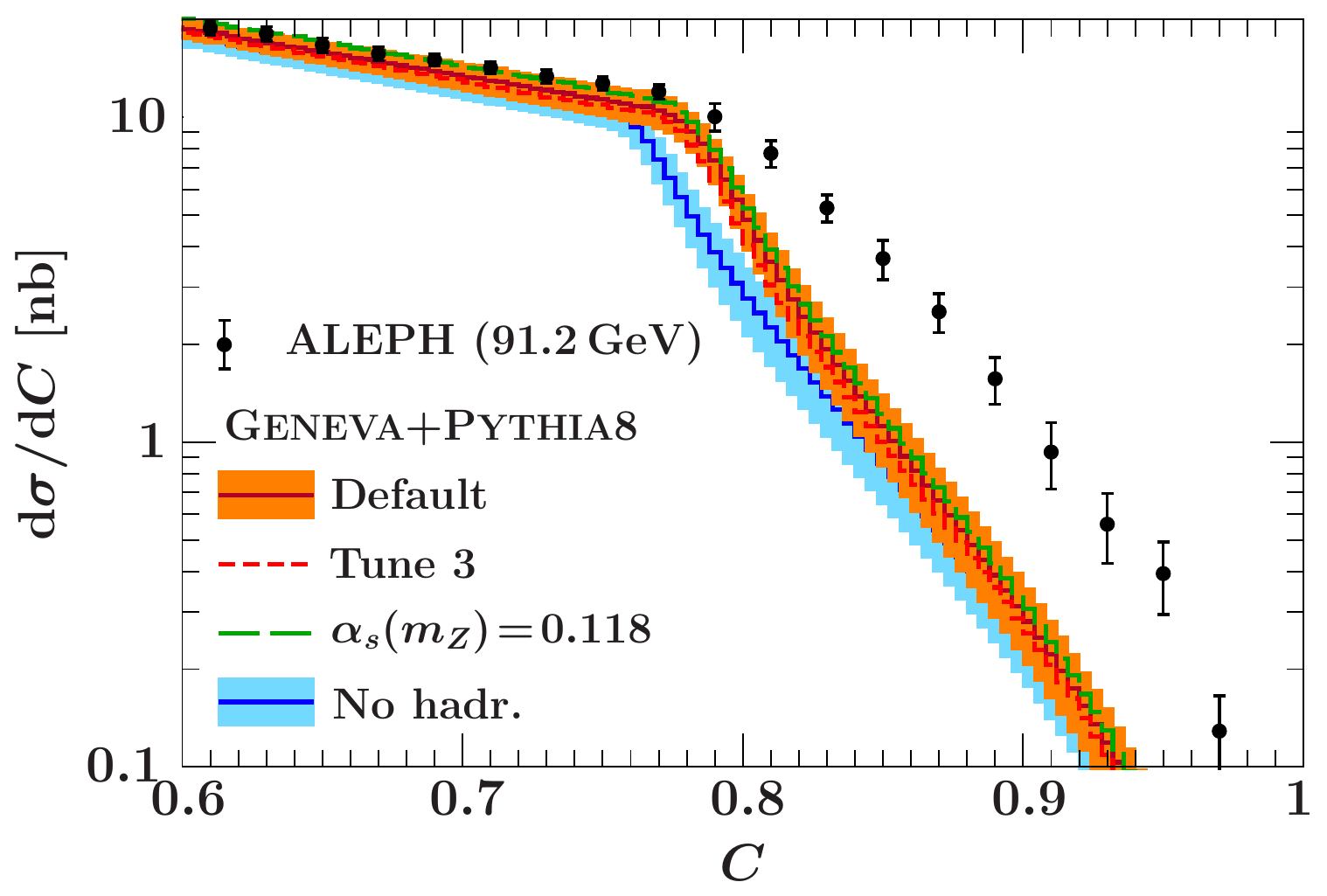}}%
\label{fig:CPlot2c}}%
\subfigure[\hspace{1ex} Ratio of \geneva to Data]{%
\parbox{0.5\columnwidth}{\includegraphics[scale=0.5]{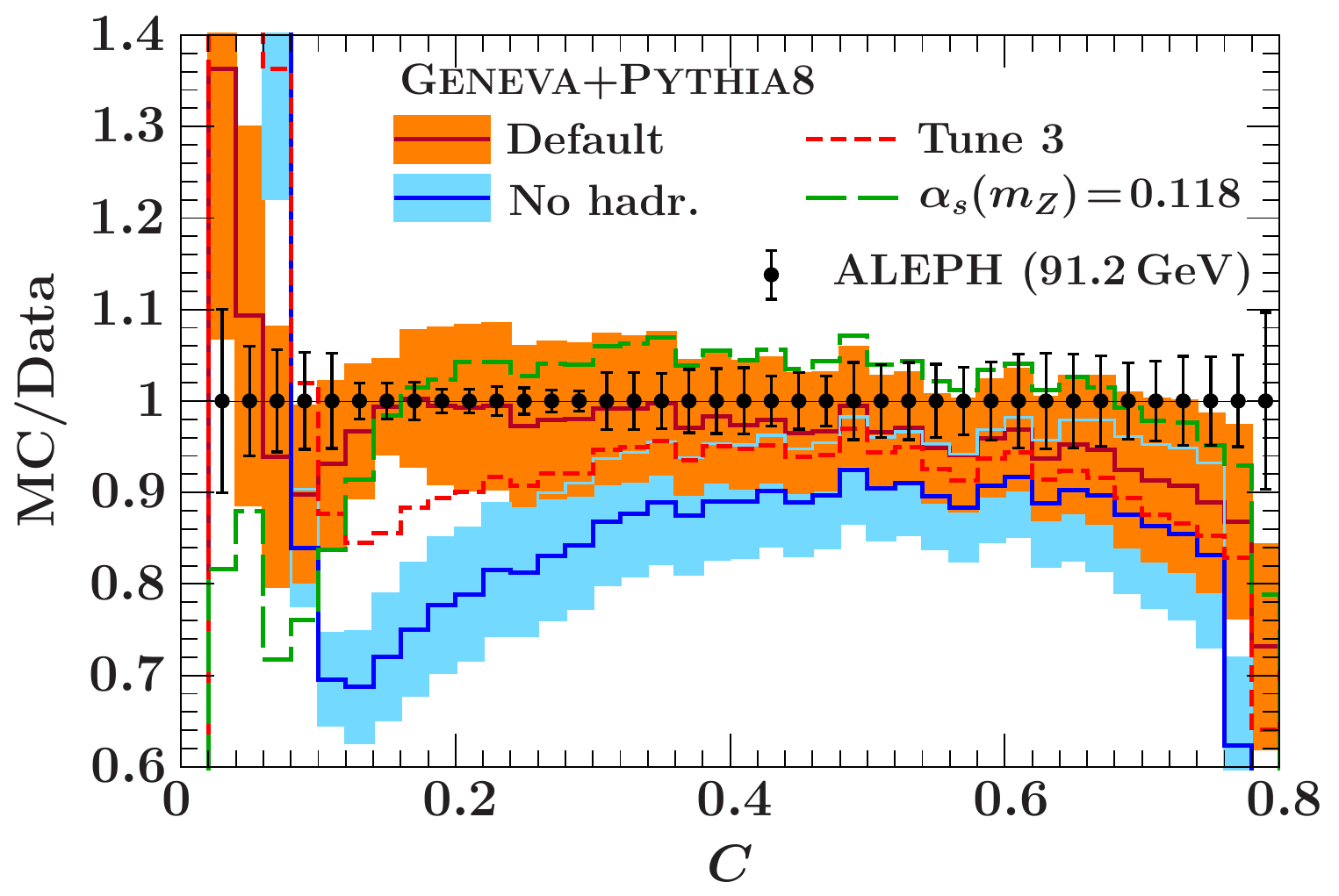}}%
\label{fig:CPlot2d}}%
\vspace{-0.5ex}
\caption{The $C$-parameter distribution comparing \geneva with and without hadronization using \pythia 8 $e^+e^-$ tune 1 and $\alpha_s(m_Z)=0.1135$ is shown compared to \Aleph data in the (a) peak, (b) transition, and (c) tail regions and to \Opal data in the peak and transition regions. The ratio of the \geneva predictions to \Aleph data is shown in (d). Also shown are the \geneva predictions at the central scale with $\alpha_s(m_Z)=0.1184$ and $e^+e^-$ tune 3.}
\label{fig:Cplot2}
\end{figure*}

One might naively expect that the accuracy of the resummation achieved in \geneva for any observable other than $\Tau_2$ would only be the partial NLL of the parton shower. However, it is clear that the \geneva prediction at NLL$'_\Tau$+LO$_3$ in the peak and into the transition region, \figs{Cplot1a}{Cplot1b}, is much more consistent with NLL$'_C$+LO$_3$ than NLL$_C$ resummation, both in its central value and also in the size of the perturbative uncertainties it predicts. This appears to hold even in the peak region below $C\sim0.05$, where the parton shower determines the shape of the spectrum. Going to one higher order in \figs{Cplot1c}{Cplot1d}, we see that the same pattern holds: the \geneva prediction is consistent with the higher-order NNLL$'_C$+NLO$_3$ resummation rather than NLL$'_C$+LO$_3$, including uncertainties. This is particularly clear in the peak region where the central values of the two analytic resummation orders are significantly different and \geneva tracks the NNLL$'_C$+NLO$_3$ prediction. The convergence of the \geneva result for $C$-parameter from NLL$'_\Tau$+LO$_3$ to NNLL$'_\Tau$+NLO$_3$ demonstrates the consistency of the \geneva implementation including the event-by-event uncertainties for this observable. Although the accuracy of the \geneva prediction for $C$-parameter is not formally of the same order as the $\Tau_2$ resummation we used as input to the master formula, the fact that it matches the analytic $C$-parameter resummation remarkably well both at NLL$'_C$+LO$_3$ and at NNLL$'_C$+NLO$_3$ shows that numerically the accuracy achieved is very close.

The \geneva uncertainties in the transition region start to shrink relative to the analytic resummation as we interpolate to the fixed-order NLO$_3$ result. In the tail region, the partonic \geneva prediction matches smoothly to the fixed-order NLO$_3$ result past the Sudakov shoulder at $C=0.75$~\cite{Catani:1997xc}, demonstrating the validity of the multiplicative implementation of $\df \sigma/\df \Phi_3(\Tau_2)$ in \eq{dsigma23} in this limit.

The \geneva prediction including hadronization with the default running values of \pythia $e^+e^-$ tune 1 and $\alpha_s(m_Z)= 0.1135$ is shown in \fig{Cplot2} compared to \Aleph and \Opal data rescaled to the NNLO inclusive cross section. \geneva agrees with the data remarkably well across the entire distribution up to the multijet region in the tail.  We show the effect of $\alpha_s(m_Z)=0.1135$ with tune 3, which gives a smaller correction from hadronization than tune 1, as seen from the size of the shift from the \geneva unhadronized result to the hadronized in \fig{Cplot2}. We also show the \geneva prediction at the central scale using the world average $\alpha_s(m_Z) = 0.1184$ and tune 1. These two combinations provide an estimate of the upper and lower bounds on the combined uncertainty of the nonperturbative effect and $\alpha_s(m_Z)$ value in the transition and tail regions, as discussed in \subsec{eeTau2}. The ratio of the Monte Carlo to data in~\fig{CPlot2d} shows that they are both largely within the perturbative uncertainties from \geneva in these regions. Of the values we consider, the default tune 1 with $\alpha_s(m_Z)=0.1135$ gives the best agreement with the data across the $C$ spectrum and is consistent with our findings for the $\Tau_2$ distribution.

\subsubsection{Heavy Jet Mass}
\label{subsubsec:eeHJM}
\begin{figure*}[t!]
\subfigure[\hspace{0.5ex} \geneva NLL$'_\Tau$+LO$_3$ Peak Region]{%
\parbox{0.5\columnwidth}{\includegraphics[scale=0.5]{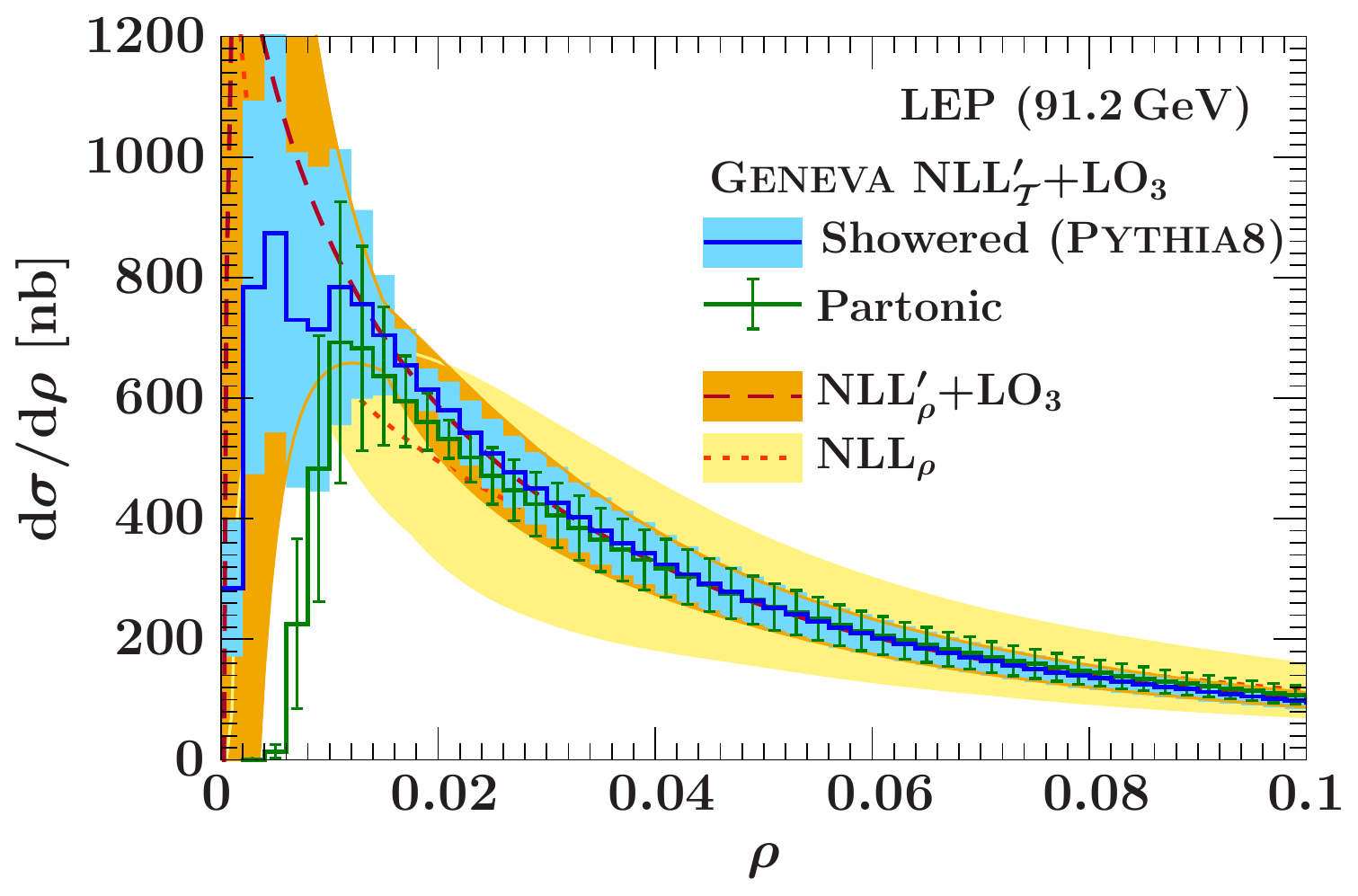}}%
\label{fig:rhoplot1a}}%
\hfill%
\subfigure[\hspace{0.5ex} \geneva NLL$'_\Tau$+LO$_3$ Transition Region]{%
\parbox{0.5\columnwidth}{\includegraphics[scale=0.5]{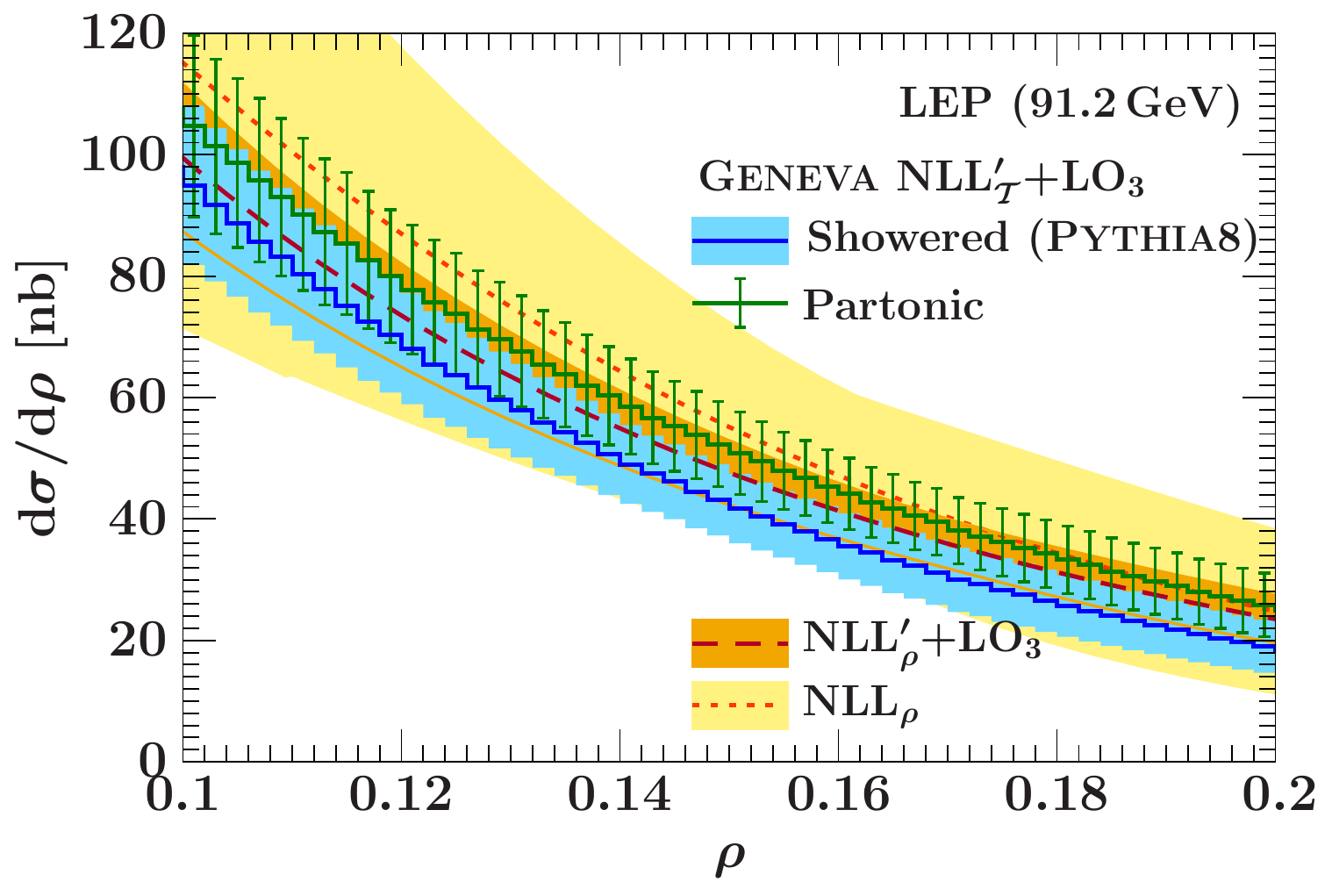}}%
\label{fig:rhoplot1b}}%
\\%
\subfigure[\hspace{0.5ex} \geneva NNLL$'_\Tau$+NLO$_3$ Peak Region]{%
\parbox{0.5\columnwidth}{\includegraphics[scale=0.5]{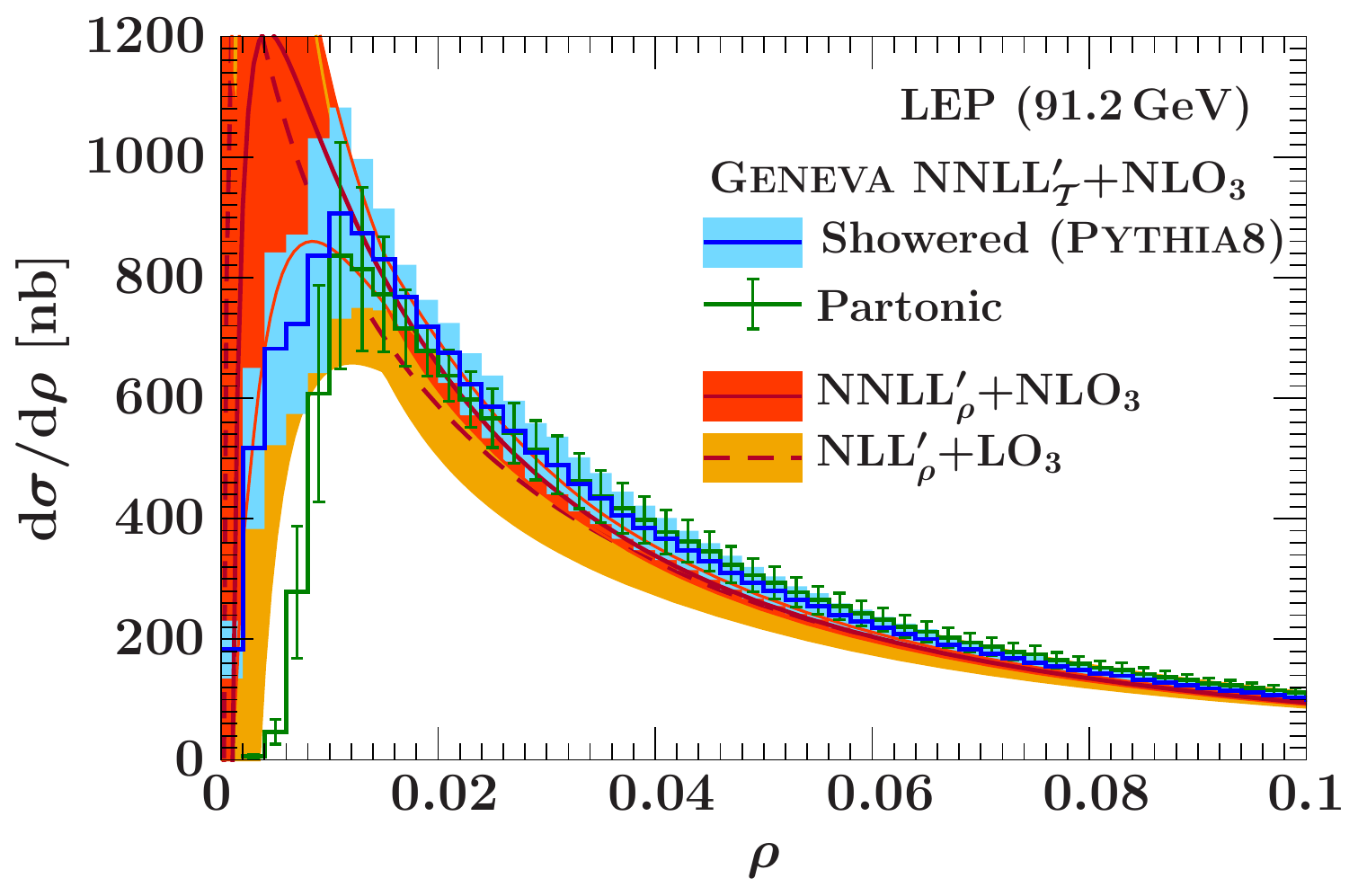}}%
\label{fig:rhoplot1c}}%
\hfill%
\subfigure[\hspace{0.5ex} \geneva NNLL$'_\Tau$+NLO$_3$ Transition Region]{%
\parbox{0.5\columnwidth}{\includegraphics[scale=0.5]{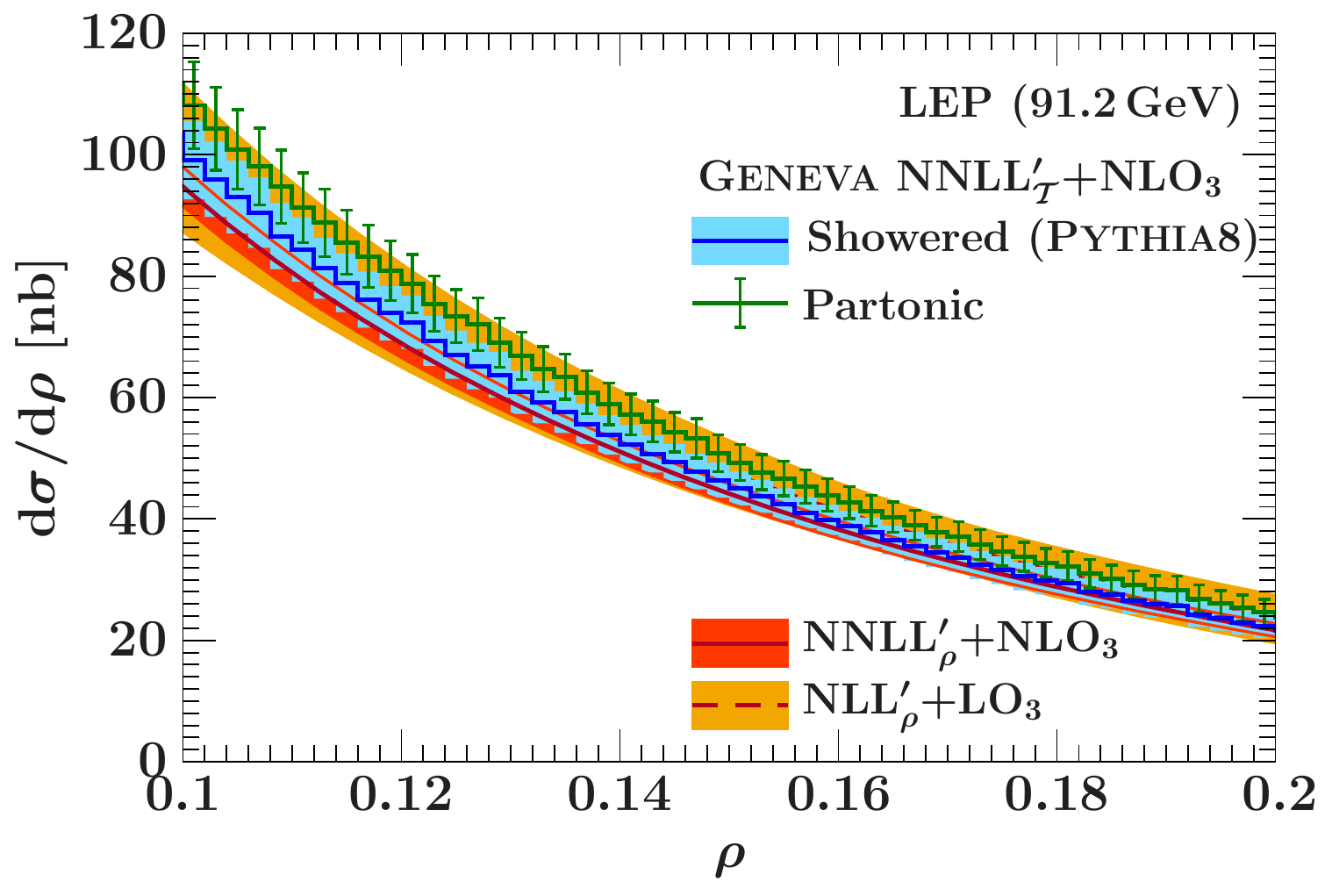}}%
\label{fig:rhoplot1d}}%
\\%
\centering
\subfigure[\hspace{0.5ex} \geneva NNLL$'_\Tau$+NLO$_3$ Tail Region]{%
\parbox{0.5\columnwidth}{\includegraphics[scale=0.5]{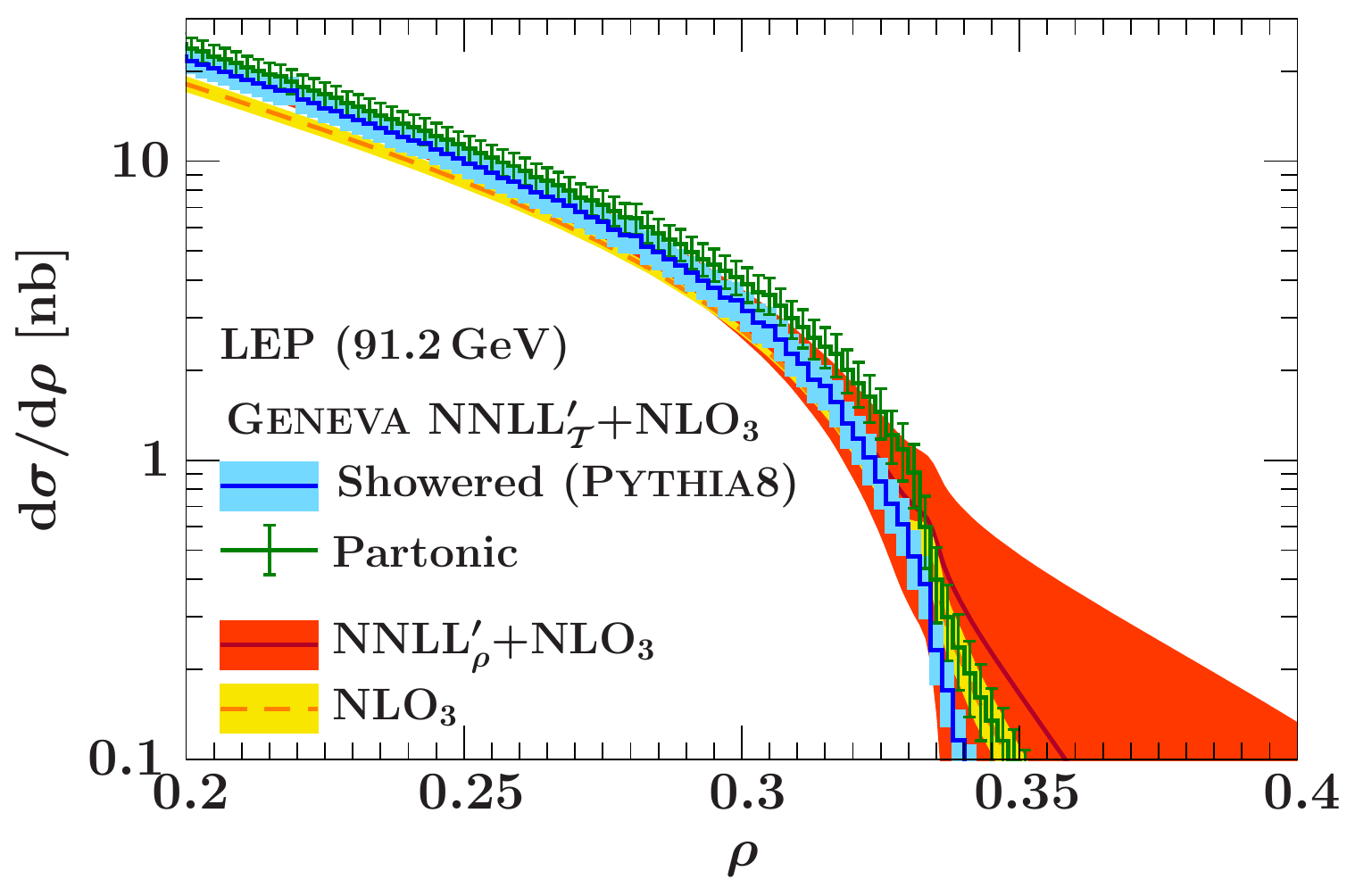}}%
\label{fig:rhoplot1e}}%
\vspace{-0.5ex}
\caption{The heavy jet mass partonic and showered \geneva predictions are shown compared to the analytic resummation of $\rho$ at different orders. The \geneva result at NLL$'_\Tau$+LO$_3$ is compared to NLL$_\rho$ and NLL$'_\rho$+LO$_3$ in (a) and (b). In (c) and (d), the \geneva prediction at one order higher, NNLL$'_\Tau$+NLO$_3$, is compared to NLL$'_\rho$+LO$_3$ and NNLL$'_\rho$+NLO$_3$, while in the tail (e), we also show the fixed-order NLO$_3$ prediction from \event.}
\label{fig:rhoplot1}
\end{figure*}

\begin{figure*}[t!]
\subfigure[\hspace{1ex} Peak Region]{%
\parbox{0.5\columnwidth}{\includegraphics[scale=0.5]{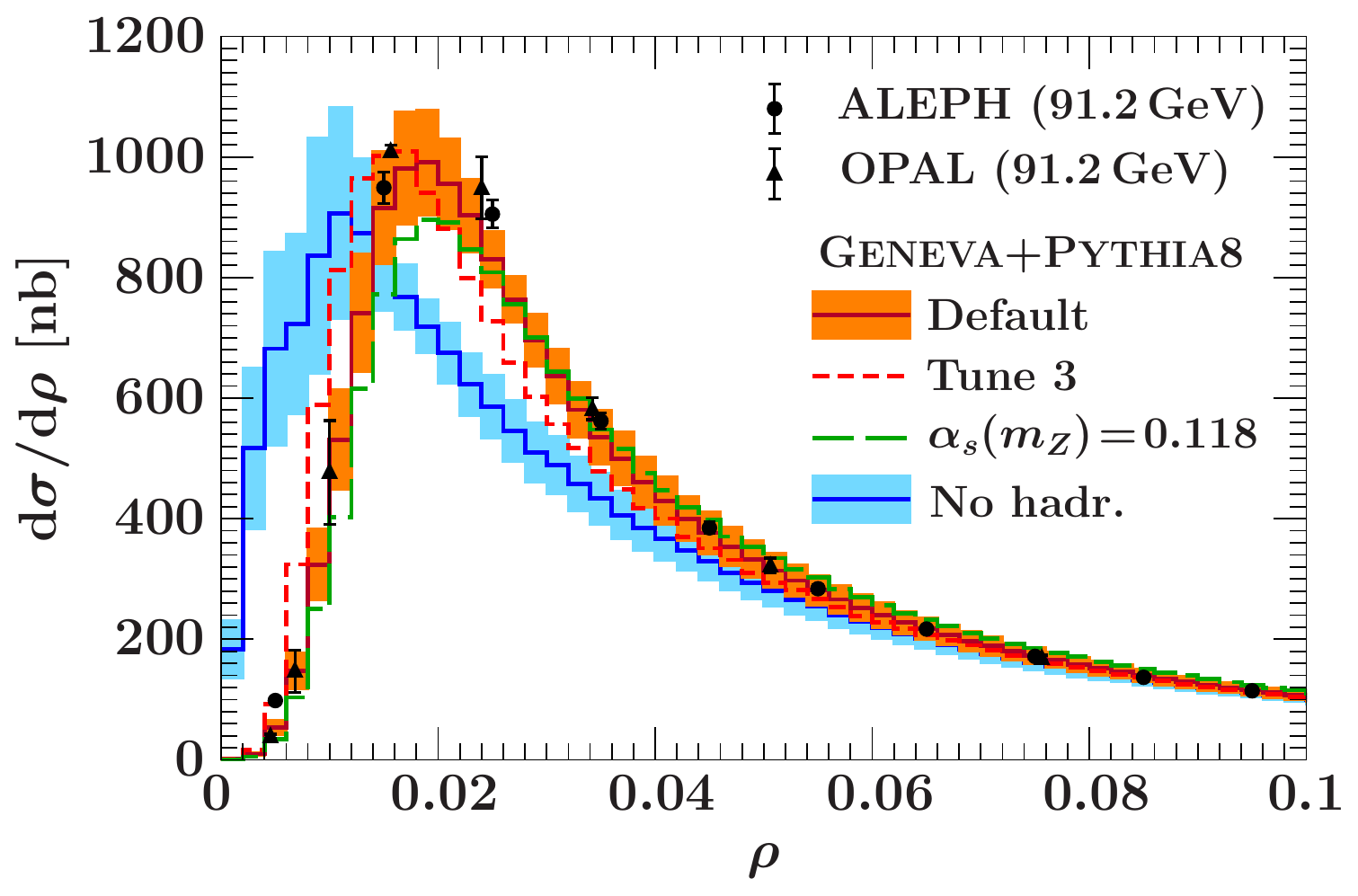}}%
\label{fig:rhoPlot2a}}%
\hfill%
\subfigure[\hspace{1ex} Transition Region]{%
\parbox{0.5\columnwidth}{\includegraphics[scale=0.5]{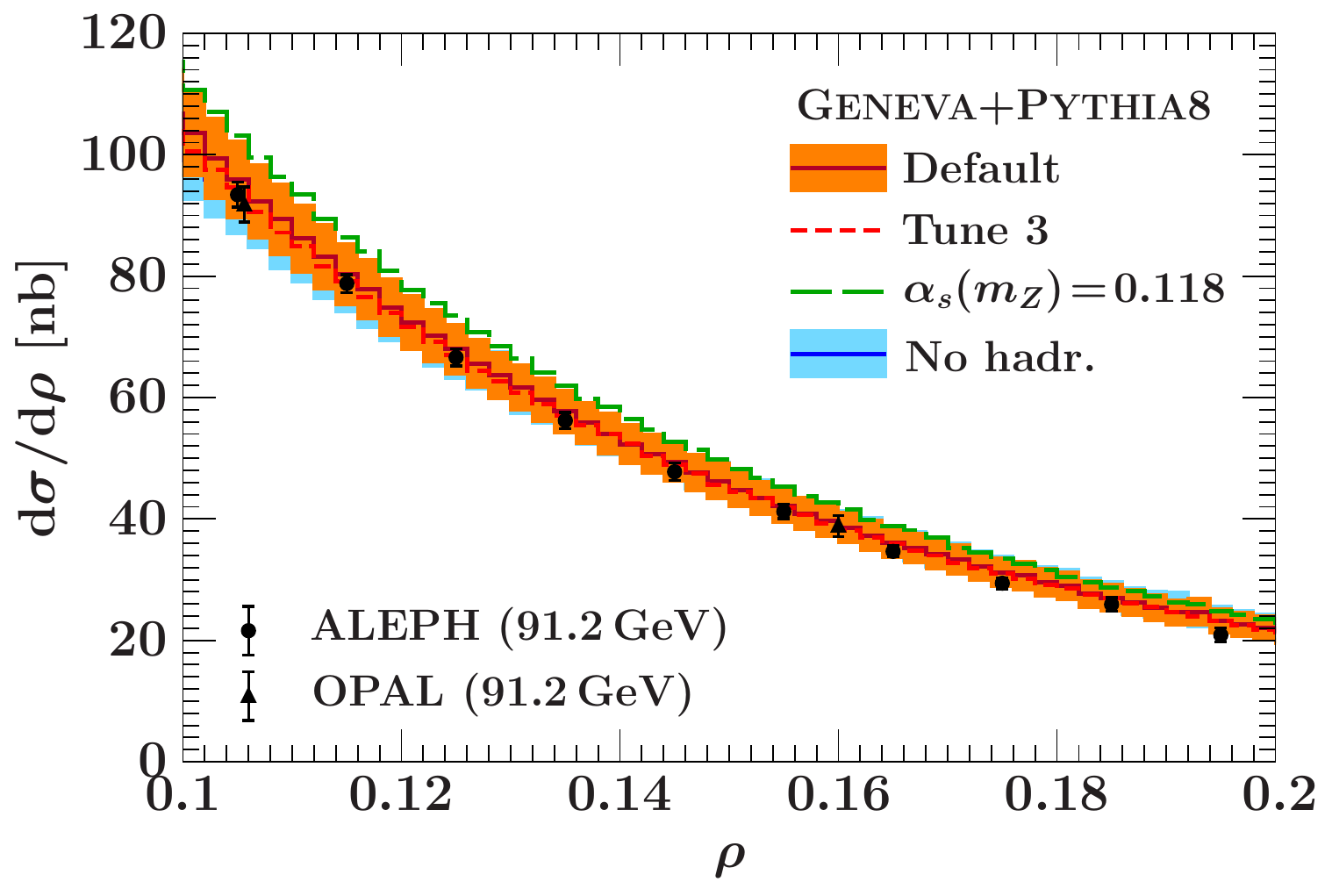}}%
\label{fig:rhoPlot2b}}%
\\%
\subfigure[\hspace{1ex} Tail Region]{%
\parbox{0.5\columnwidth}{\includegraphics[scale=0.5]{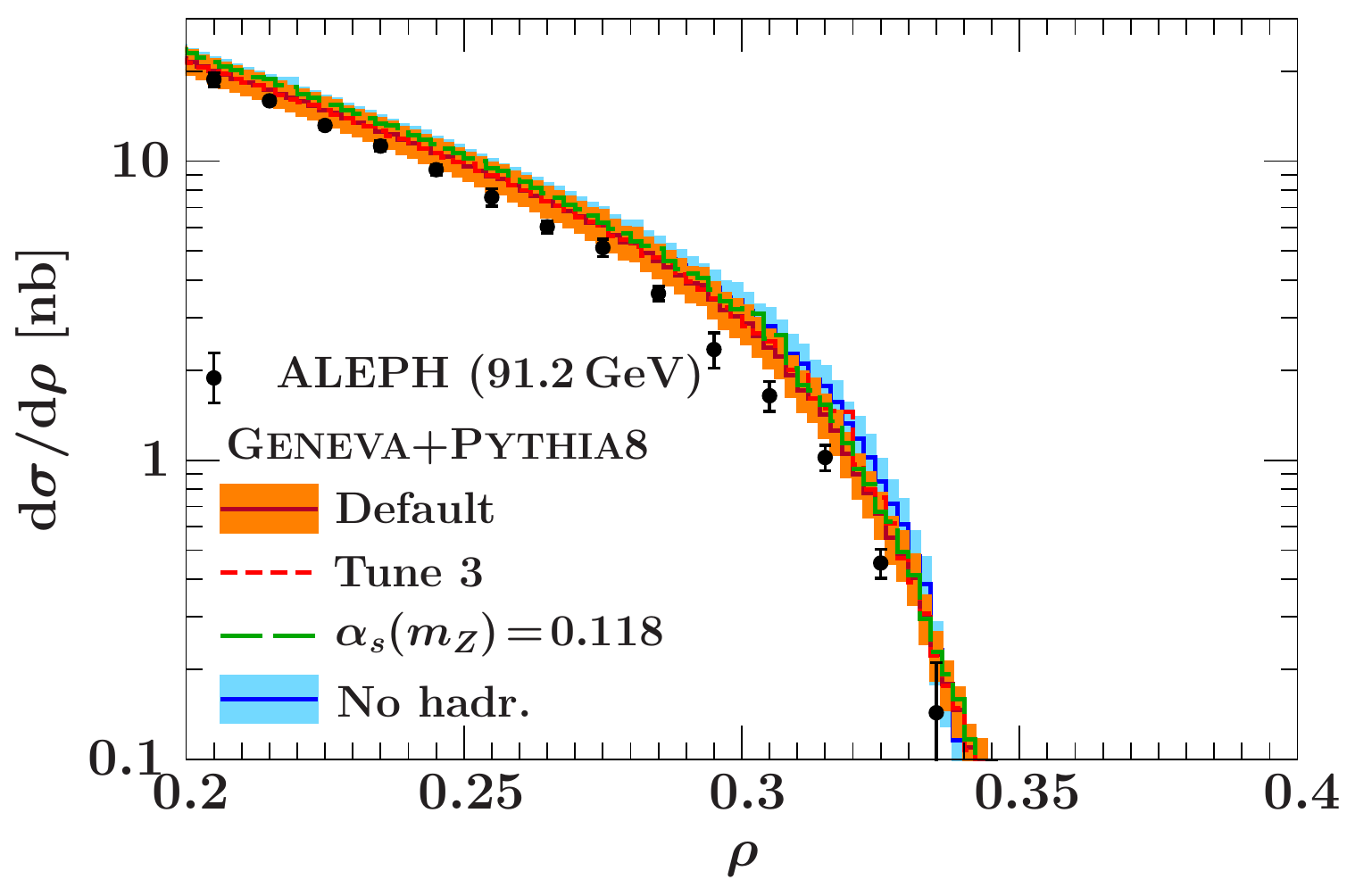}}%
\label{fig:rhoPlot2c}}%
\subfigure[\hspace{1ex} Ratio of \geneva to Data]{%
\parbox{0.5\columnwidth}{\includegraphics[scale=0.5]{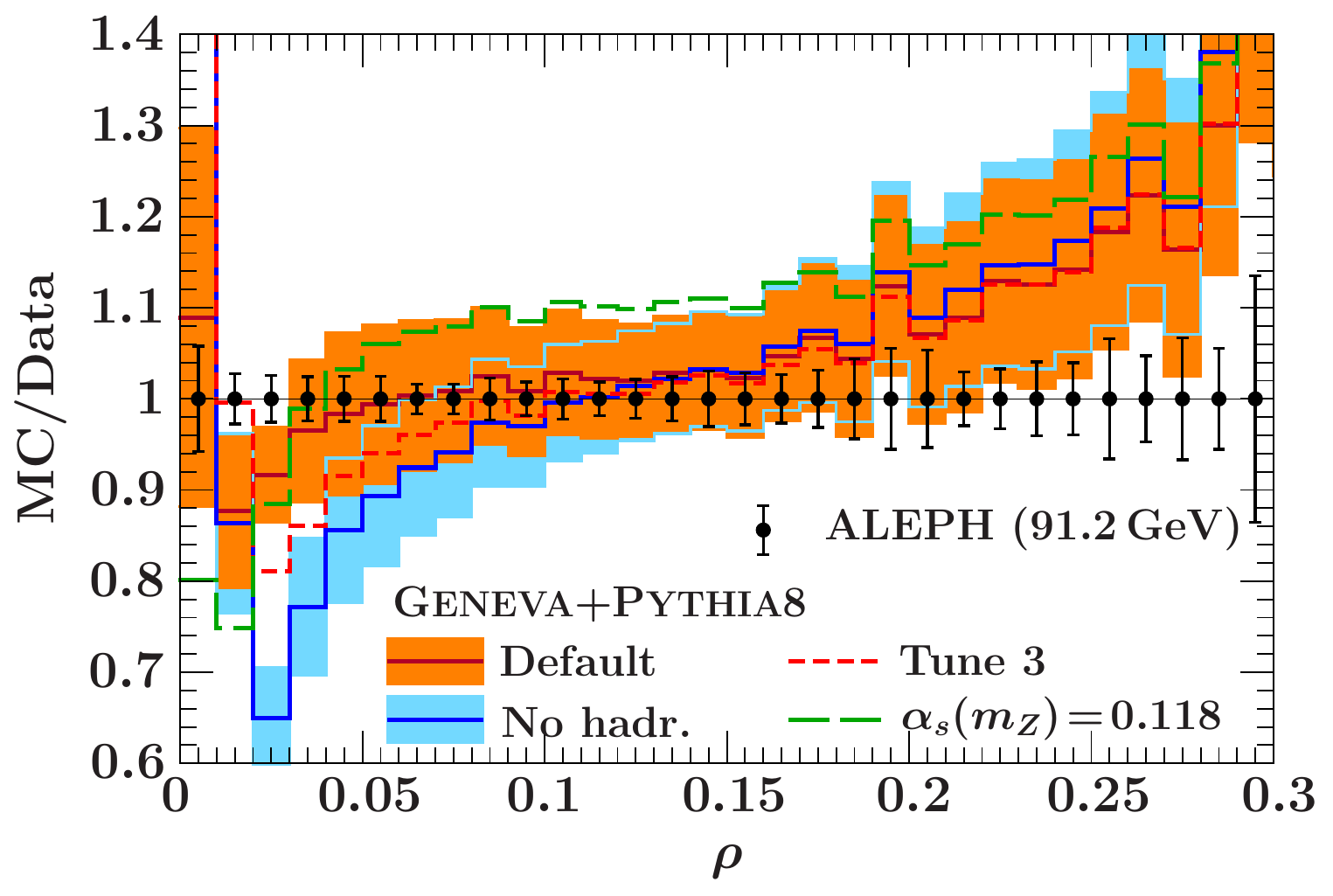}}%
\label{fig:rhoPlot2d}}%
\vspace{-0.5ex}
\caption{The heavy jet mass distribution comparing \geneva with and without hadronization using \pythia 8 $e^+e^-$ tune 1 and $\alpha_s(m_Z)=0.1135$ is shown compared to \Aleph data in the (a) peak, (b) transition, and (c) tail regions and to \Opal data in the peak and transition regions. The ratio of the \geneva predictions to \Aleph data is shown in (d). Also shown are the \geneva predictions at the central scale with $\alpha_s(m_Z)=0.1184$ and $e^+e^-$ tune 3.}
\label{fig:rhoplot2}
\end{figure*}

The \geneva prediction for heavy jet mass is shown in \fig{rhoplot1}, where we compare the partonic and showered results using NLL$'_\Tau$+LO$_3$ resummation in the master formula in \figs{rhoplot1a}{rhoplot1b} to the analytic resummation of $\rho$ at NLL$_\rho$ and NLL$'_\rho$+LO$_3$. In \figs{rhoplot1c}{rhoplot1d}, we show the same results at one order higher, comparing NNLL$'_\Tau$+NLO$_3$ \geneva results to NLL$'_\rho$+LO$_3$ and NNLL$'_\rho$+NLO$_3$ analytic $\rho$ resummation. In the tail, we show only the highest-order \geneva and resummed results along with the pure fixed-order NLO$_3$ contribution, since this is sufficient to demonstrate the behavior in this region.

In the peak region, \figs{rhoplot1a}{rhoplot1c}, we see the effect of $\Tau_2^\cut$ on the partonic $\rho$ spectrum up to $\rho=1\GeV/\Ecm = 0.011$, which is smoothly filled out by interfacing with the parton shower. The \geneva showered prediction in \fig{rhoplot1a} shows impressive agreement with the NLL$'_\rho$+LO$_3$ resummed result in the peak region, including uncertainties. The improvement in accuracy of the \geneva prediction for heavy jet mass over the partial NLL resummation provided by the parton shower is clear. Going to one higher order in \fig{rhoplot1c}, the \geneva prediction is more consistent with the NNLL$'_\rho$+NLO$_3$, with which it agrees within uncertainties, rather than the NLL$'_\rho$+LO$_3$ result.

The perturbative uncertainties of the \geneva showered prediction are larger than those at NNLL$'_\rho$+NLO$_3$ and smaller than at NLL$'_\rho$+LO$_3$. This is consistent with the fact that, while the \geneva prediction for $\rho$ is not formally of the same order as the $\Tau_2$ resummation that is input into the master formula, there is a gain in accuracy going from \geneva at NLL$'_\Tau$+LO$_3$ to NNLL$'_\Tau$+NLO$_3$ that is transferred to the $\rho$ prediction. 

In the transition region, both at NLL$'_\Tau$+LO$_3$ and NNLL$'_\Tau$+NLO$_3$, adding the parton shower gives rise to a larger shift from the partonic spectrum than for the $C$-parameter, because heavy jet mass is less correlated with $\Tau_2$ than $C$. This shift is necessary to obtain agreement with the NNLL$'_\rho$+NLO$_3$ resummation within uncertainties in \fig{rhoplot1d}. The partonic \geneva prediction in this region is more consistent with NLL$'_\rho$+LO$_3$ analytic resummation, which is higher than the NNLL$'_\rho$+NLO$_3$ result. By restricting the change in $\Tau_2$ due to the shower, we are constraining the sum of the hemisphere masses, $M_{1,2}^2$ in \eq{defCrhoB}. For a given value of $\Tau_2$, $\rho$ is largest when either hemisphere mass is zero, and so $\rho=\Tau_2$. Adding the parton shower tends to make this mass nonzero (while constraining the sum) and therefore gives an overall shift of the spectrum to lower values of $\rho$. In the tail region, the partonic spectrum interpolates to the fixed-order NLO$_3$ result, as expected, with the shower giving rise to a larger shift in the multijet region where our constraints are looser.

In \fig{rhoplot2}, we show the showered \geneva prediction with and without hadronization, with our default parameters. As before, we compare to \Aleph and \Opal data, which shows impressive agreement with the data within uncertainties across all three regions of the $\rho$ spectrum, with the expected deviation in the multijet region of the far tail. As discussed previously, tune 3 with $\alpha_s(m_Z)=0.1135$ and tune 1 with $\alpha_s(m_Z)=0.1184$ provide bounds on the estimate of the combined uncertainty from these two inputs. 
It is interesting to note that heavy jet mass is relatively insensitive to hadronization in the transition and tail regions. This is demonstrated by the shift from the shower-only to the fully hadronized result in \figs{rhoPlot2b}{rhoPlot2c}, as well as the small change in the default central value from using tune 1 to tune 3 above $\rho\sim0.1$ seen in \fig{rhoPlot2d}. This breaks the coupling in some respect between the nonperturbative effects and the value of $\alpha_s$ in this region and suggests that $\alpha_s(m_Z)=0.1135$ provides better agreement with the data.

\subsubsection{Jet Broadening}
\label{subsubsec:eeJB}

\begin{figure*}[t!]
\subfigure[\hspace{1ex} \geneva NLL$'_\Tau$+LO$_3$ Peak Region]{%
\parbox{0.5\columnwidth}{\includegraphics[scale=0.5]{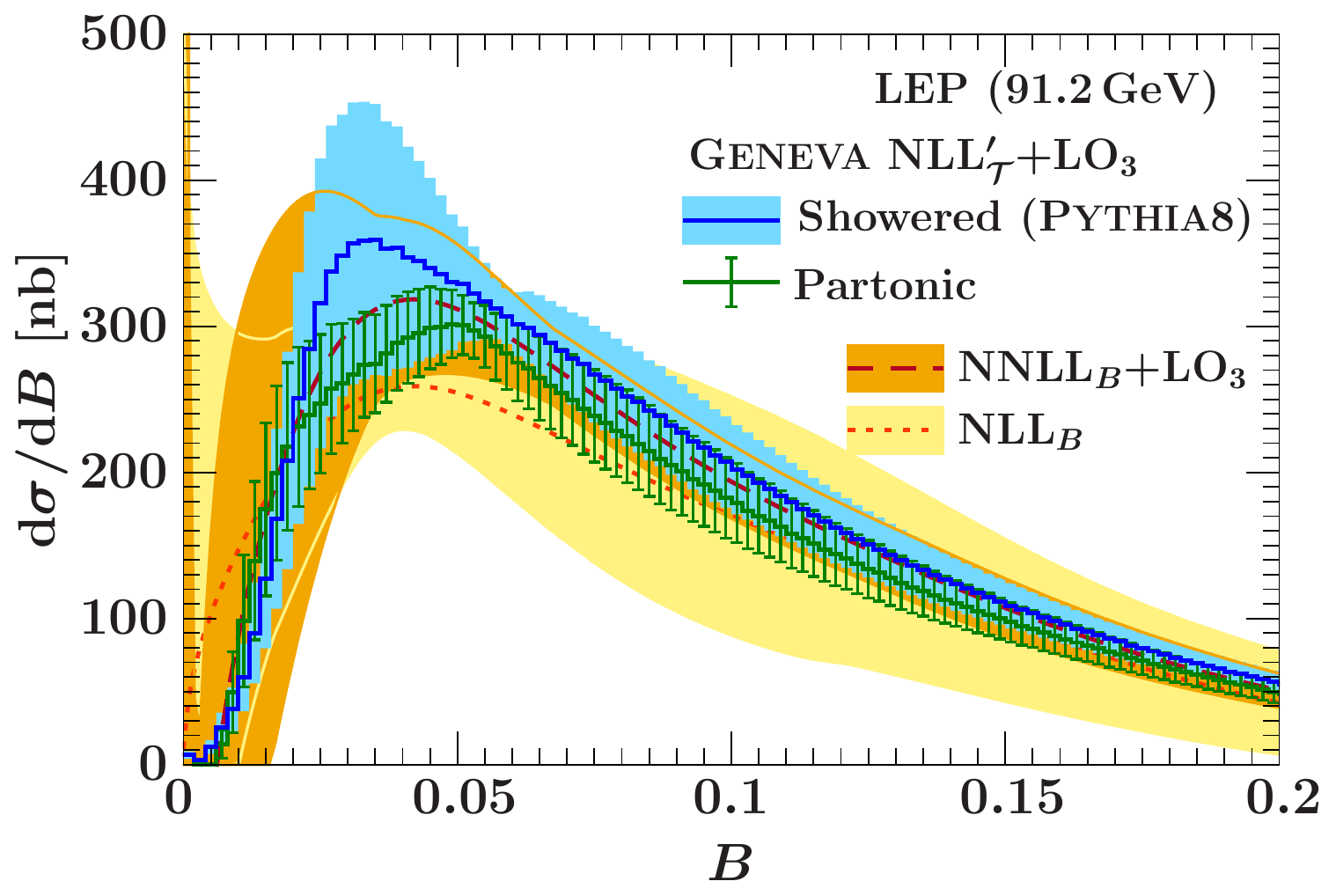}}%
\label{fig:JetBplot1a}}%
\hfill%
\subfigure[\hspace{1ex} \geneva NLL$'_\Tau$+LO$_3$ Transition Region]{%
\parbox{0.5\columnwidth}{\includegraphics[scale=0.5]{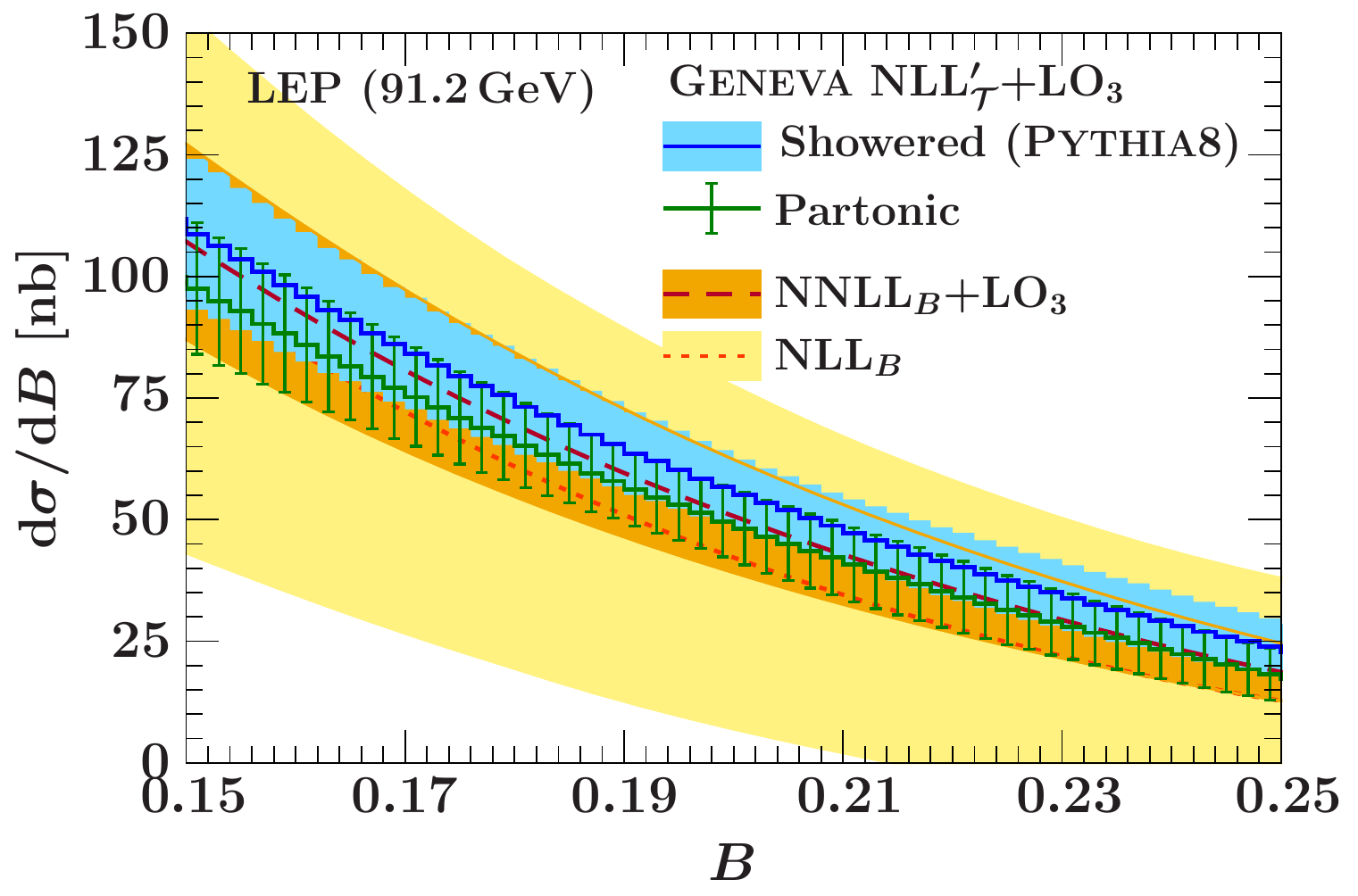}}%
\label{fig:JetBplot1b}}%
\\%
\subfigure[\hspace{1ex} \geneva NNLL$'_\Tau$+NLO$_3$ Peak Region]{%
\parbox{0.5\columnwidth}{\includegraphics[scale=0.5]{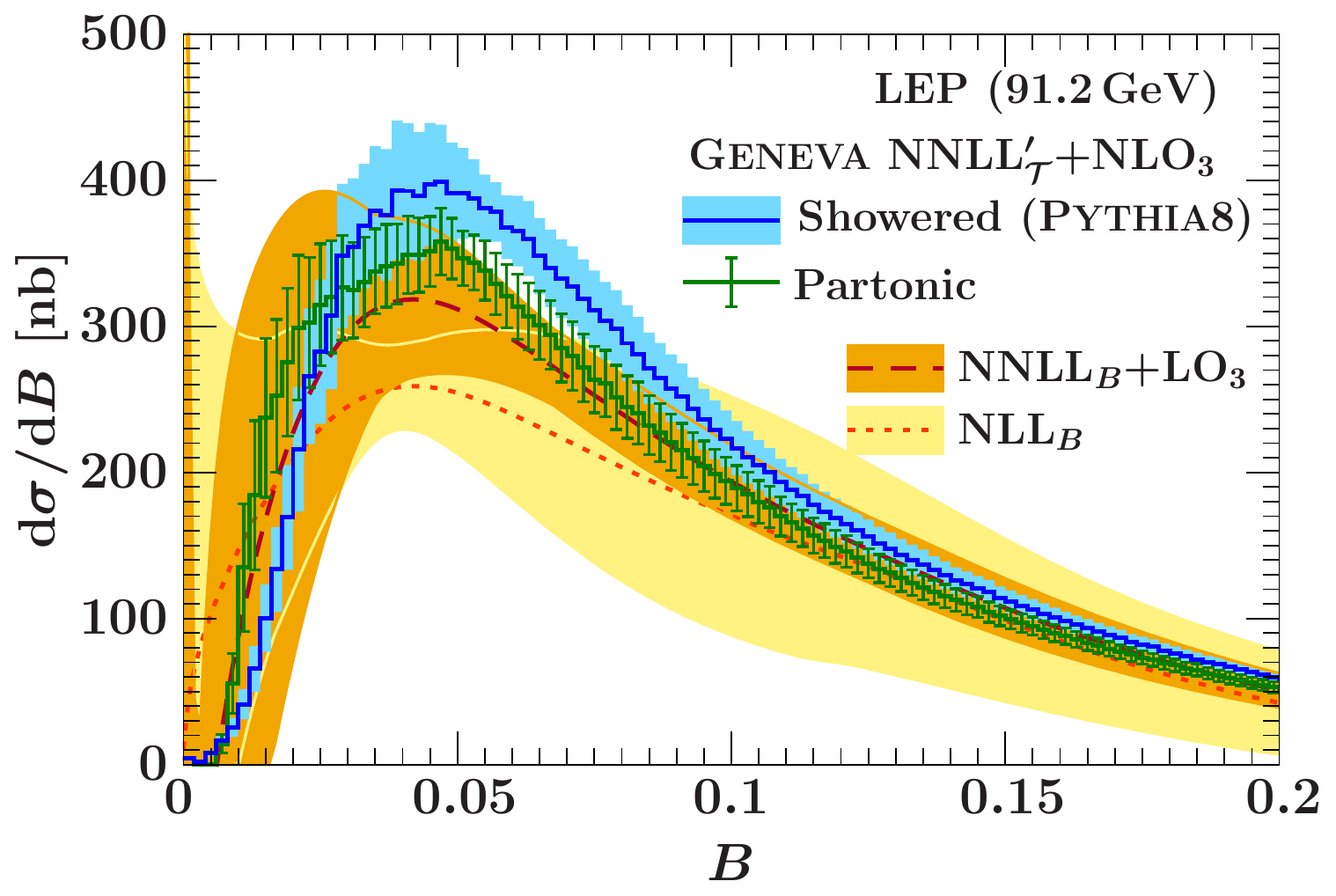}}%
\label{fig:JetBplot1c}}%
\hfill%
\subfigure[\hspace{1ex}  \geneva NNLL$'_\Tau$+NLO$_3$ Transition Region]{%
\parbox{0.5\columnwidth}{\includegraphics[scale=0.5]{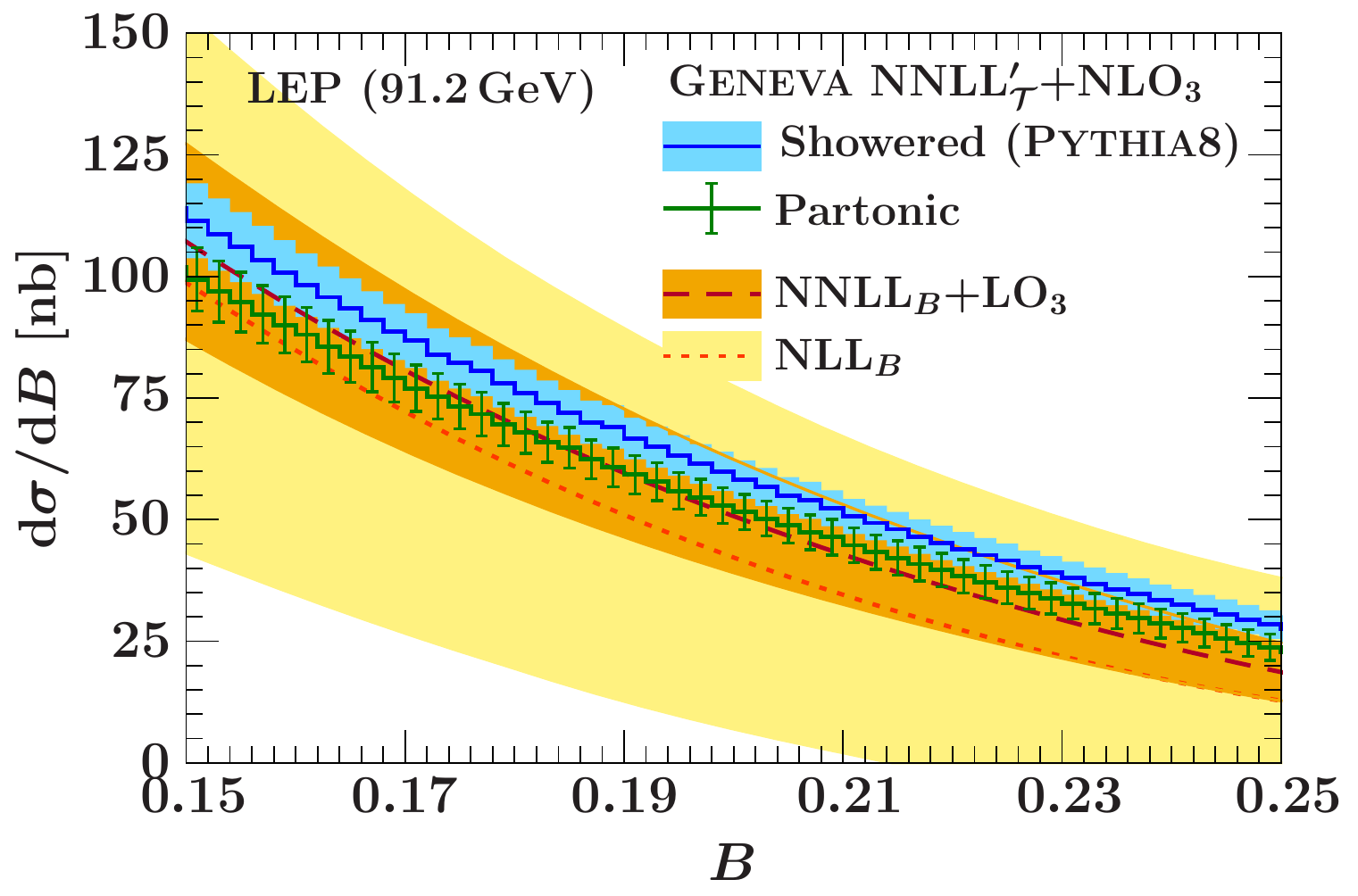}}%
\label{fig:JetBplot1d}}%
\\%
\centering
\subfigure[\hspace{1ex} \geneva NNLL$'_\Tau$+NLO$_3$ Tail Region]{%
\parbox{0.5\columnwidth}{\includegraphics[scale=0.5]{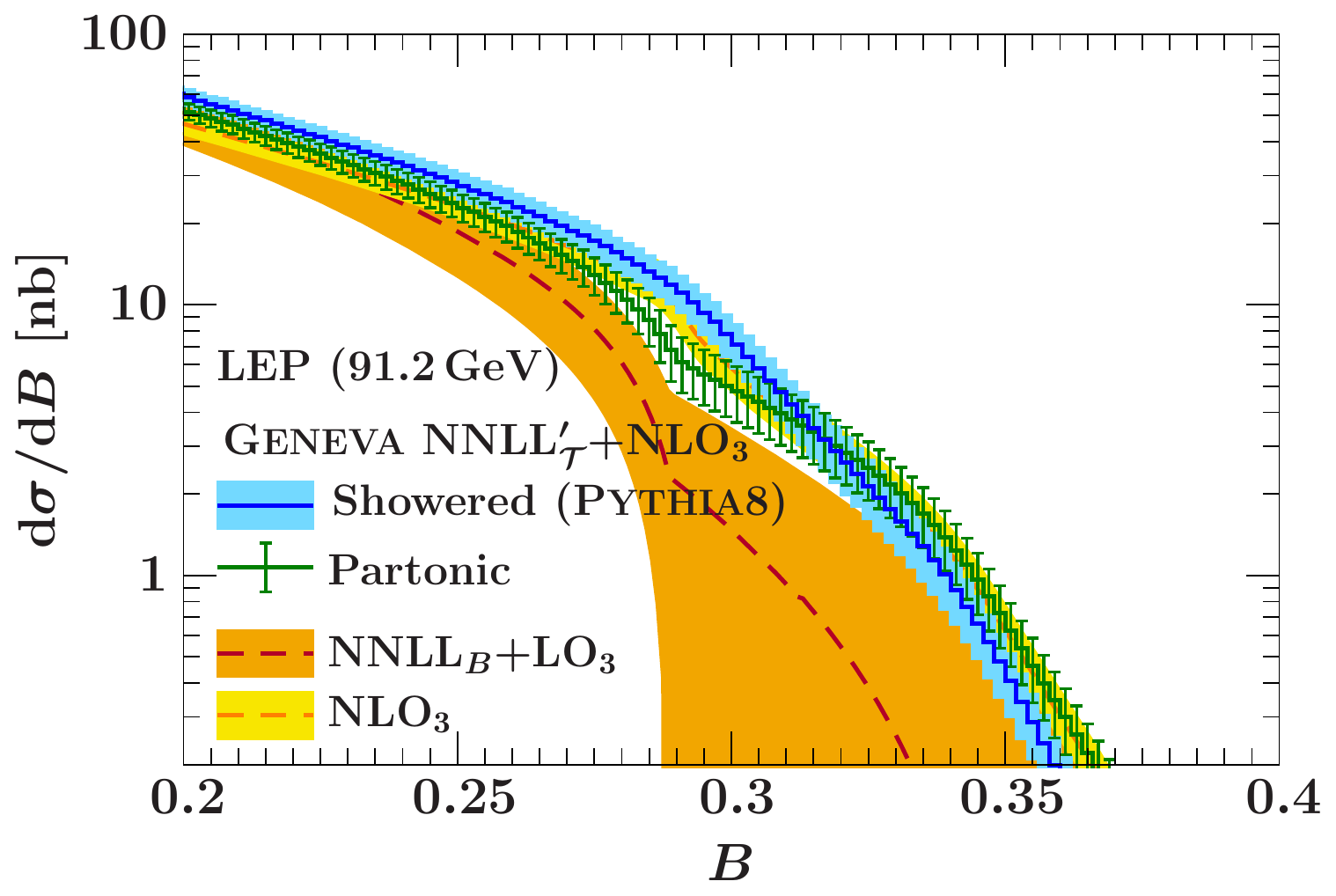}}%
\label{fig:JetBplot1e}}%
\vspace{-0.5ex}
\caption{The jet broadening partonic and showered \geneva predictions are shown compared to the analytic resummation of $B$ at NLL$_B$ and NNLL$_B$+LO$_3$. The \geneva result at NLL$'_\Tau$+LO$_3$ is shown in (a) and (b), and at one order higher, NNLL$'_\Tau$+NLO$_3$, in (c) and (d). In the tail (e), we also show the fixed-order NLO$_3$ prediction from \event.}
\label{fig:JetBplot1}
\end{figure*}

Finally, we turn to the results of \geneva for jet broadening, which is the most orthogonal event shape to our jet resolution variable that we consider. In \fig{JetBplot1}, we show the \geneva partonic and showered results using NLL$'_\Tau$+LO$_3$ resummation in \figs{JetBplot1a}{JetBplot1b} and NNLL$'_\Tau$+NLO$_3$ resummation in \figs{JetBplot1c}{JetBplot1d}. We compare these to the analytic NLL$_B$ and the best available NNLL$_B$+LO$_3$ resummed prediction. 
Note that we would like to compare the NLL$'_\Tau$+LO$_3$ resummation in \geneva to the resummation of $B$ at the same order. However, since going from NLL$'_B$ to NNLL$_B$ (which incorporates $\alpha_s^2 \ln B$ terms into the resummation) is a comparatively small effect in this case, we will find it useful to compare the NLL$'_\Tau$+LO$_3$ prediction with the analytic NNLL$_B$+LO$_3$ result. In the tail, we compare to the fixed-order NLO$_3$ result from \event, which is the more relevant comparison in this region.

The effect of the cut on $\Tau_2$ extends up to $B\simeq0.055$ in the peak region and is smoothly removed by attaching the parton shower. Both at NLL$'_\Tau$+LO$_3$ and NNLL$'_\Tau$+NLO$_3$, there is a significant shift induced by the parton shower across the jet broadening spectrum toward larger values of $B$. The size of this shift is a measure of the lack of correlation between $B$ and $\Tau_2$, the variable used to constrain the parton shower, and is therefore progressively larger for $C$, $\rho$, and $B$, as we have seen.
\clearpage
\begin{figure*}[t!]
\subfigure[\hspace{1ex} Peak Region]{%
\parbox{0.5\columnwidth}{\includegraphics[scale=0.5]{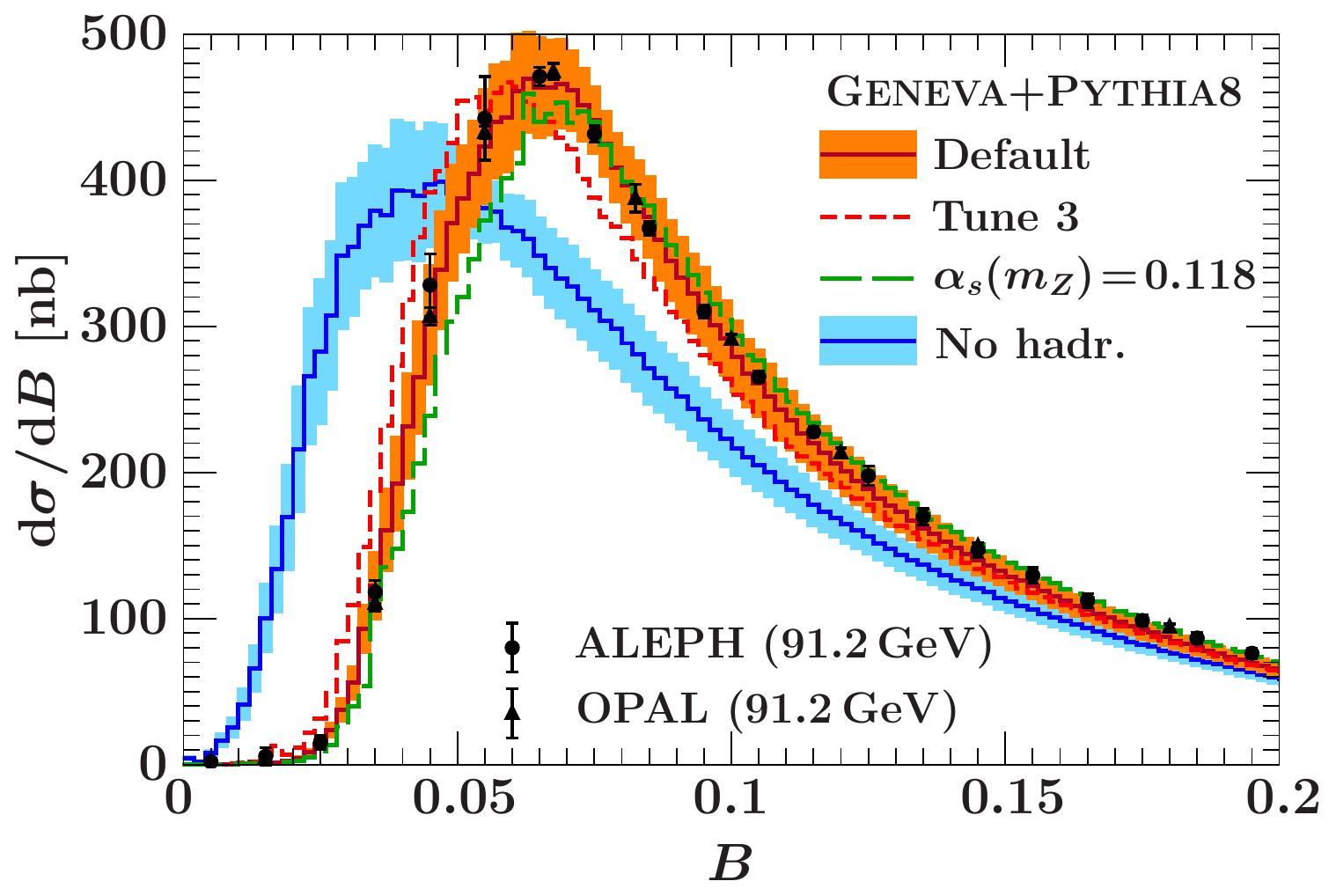}}%
\label{fig:JetBPlot2a}}%
\hfill%
\subfigure[\hspace{1ex} Transition Region]{%
\parbox{0.5\columnwidth}{\includegraphics[scale=0.5]{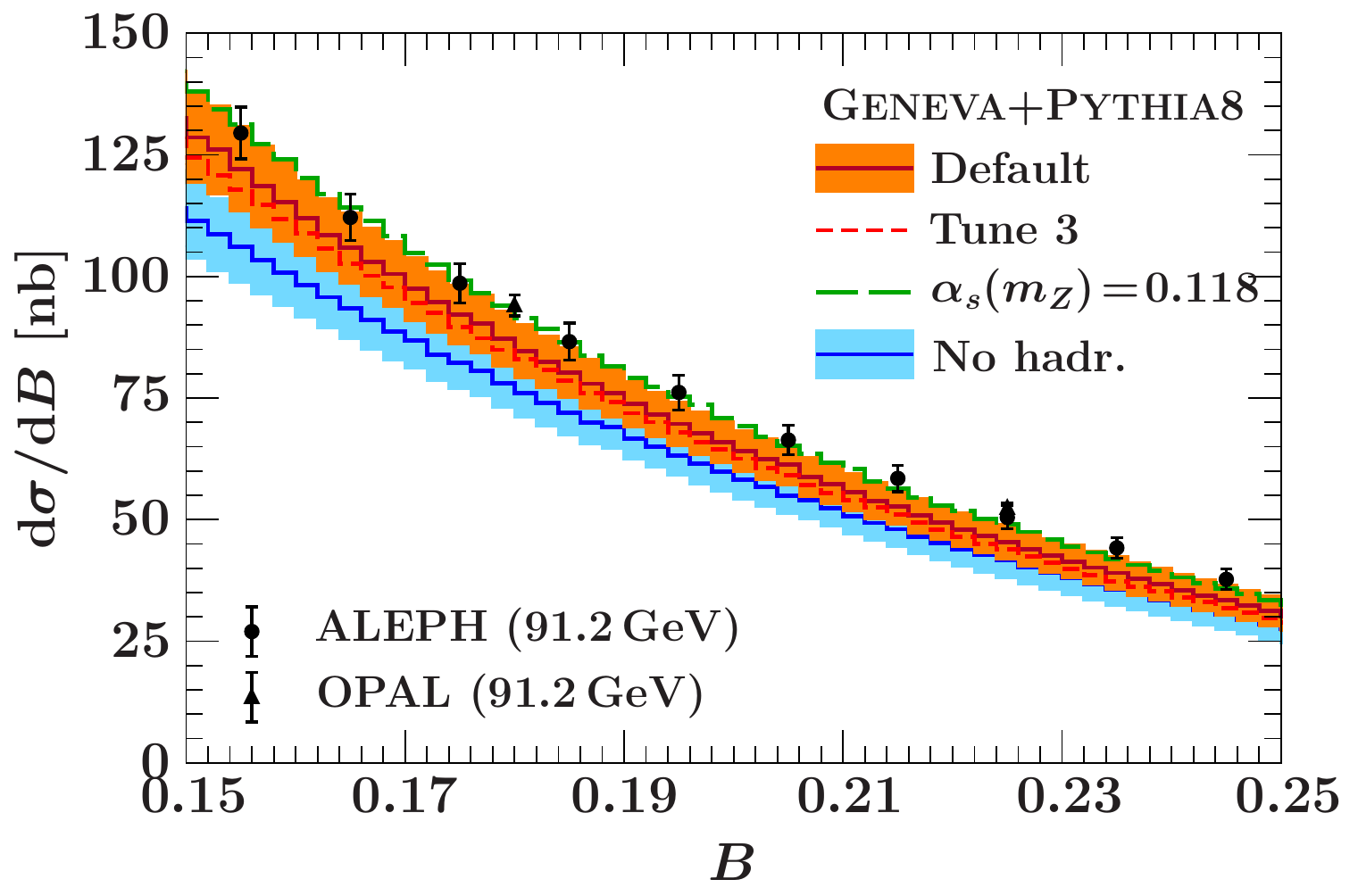}}%
\label{fig:JetBPlot2b}}%
\\%
\subfigure[\hspace{1ex} Tail Region]{%
\parbox{0.5\columnwidth}{\includegraphics[scale=0.5]{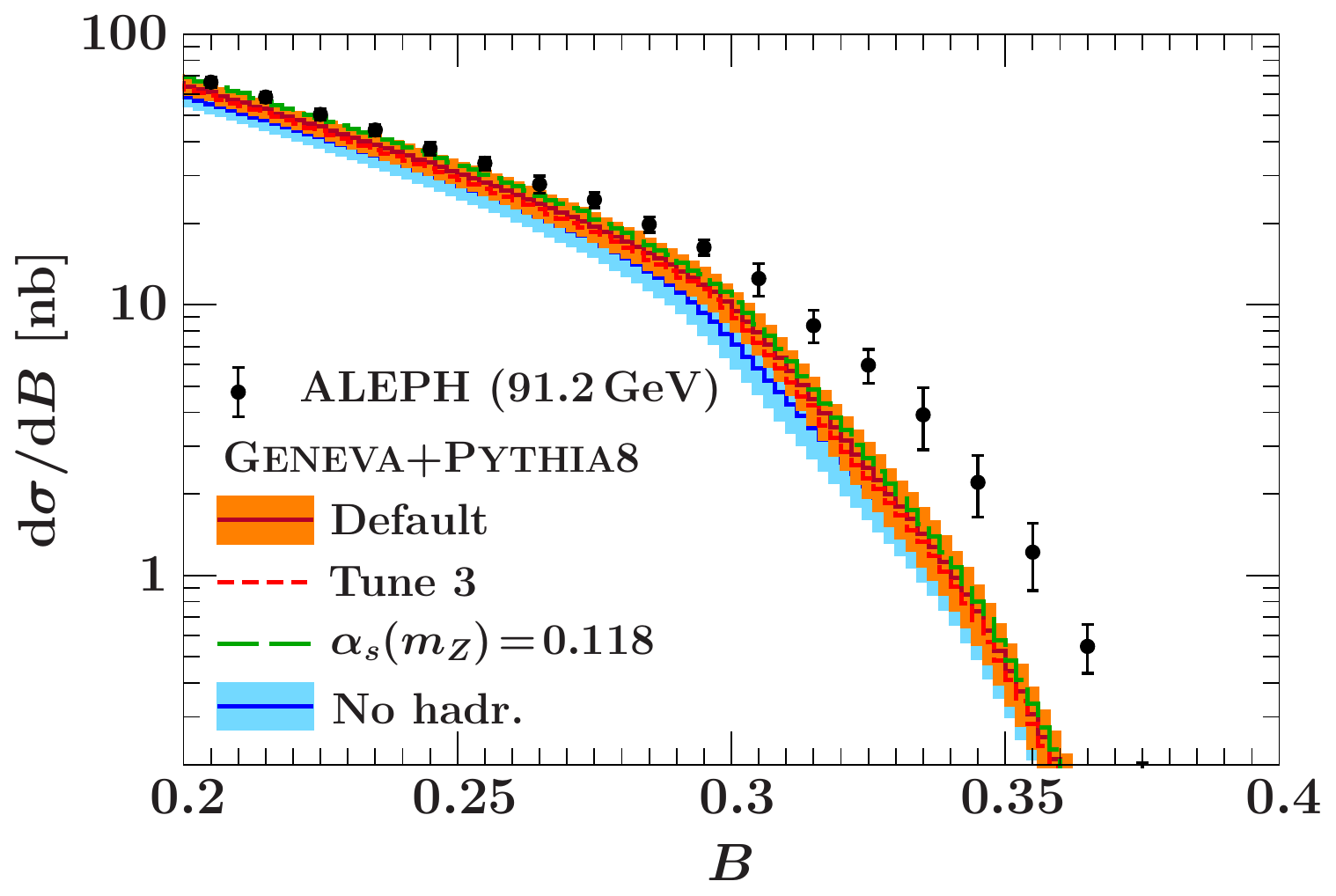}}%
\label{fig:JetBPlot2c}}%
\subfigure[\hspace{1ex} Ratio of \geneva to Data]{%
\parbox{0.5\columnwidth}{\includegraphics[scale=0.5]{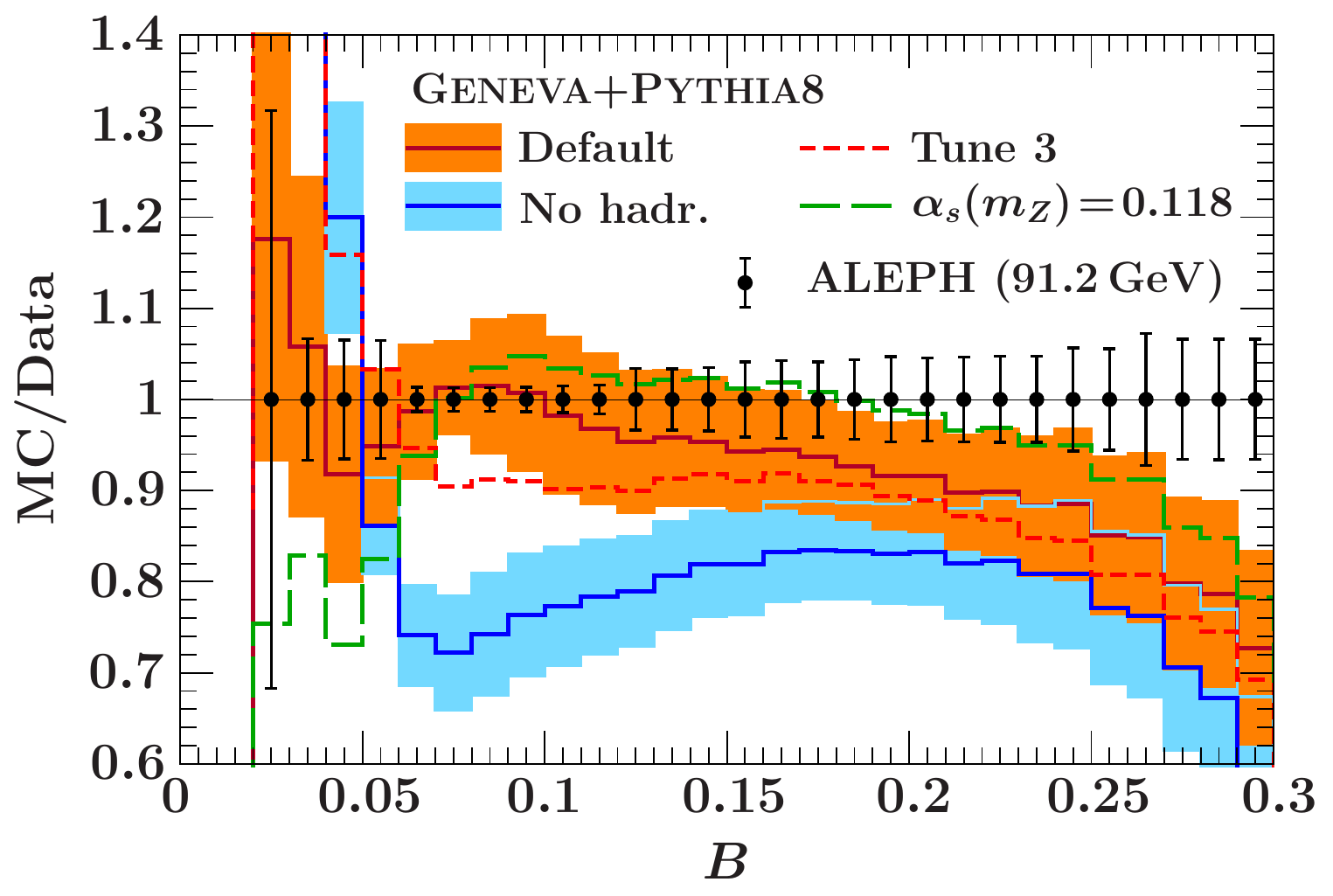}}%
\label{fig:JetBPlot2d}}%
\vspace{-0.5ex}
\caption{The jet broadening distribution comparing \geneva with and without hadronization using \pythia 8 $e^+e^-$ tune 1 and $\alpha_s(m_Z)=0.1135$ is shown compared to \Aleph data in the (a) peak, (b) transition, and (c) tail regions and \Opal data in the peak and transition regions. The ratio of the \geneva predictions to \Aleph data is shown in (d). Also shown are the \geneva predictions at the central scale with $\alpha_s(m_Z)=0.1184$ and $e^+e^-$ tune 3.}
\label{fig:JetBplot2}
\end{figure*}

 As discussed in \subsubsec{ordercounting}, in the IR limit where both $\Tau_2, B\to 0$, one might expect to see some improved accuracy of the \geneva prediction over the partial NLL of the parton shower, since the higher-order resummation of $\Tau_2$ provides a better description in this region. This is consistent with \fig{JetBplot1a}, where the showered \geneva prediction agrees well with the NNLL$_B$+LO$_3$ resummed result, including uncertainties in the peak region. In the transition region in \fig{JetBplot1b}, the central value of the \geneva showered prediction agrees with the NNLL$_B$+LO$_3$ resummed result within uncertainties; however, the uncertainties from \geneva are smaller than the corresponding analytic ones, which suggests that in this region they may be underestimated.

In the far transition region and into the tail, the uncertainties in \geneva generically are smaller than the corresponding analytic resummation and of order the NLO$_3$ scale variation, as seen for example in the $\Tau_2$, $C$, and $\rho$ spectra. 
This difference arises because \geneva multiplicatively interpolates to the fixed-order result, while the analytic resummation does an additive matching, as discussed in \subsubsec{partonicresultsee}. For an observable such as jet broadening, the lack of correlation with $\Tau_2$ means that larger values of the $\Tau_2$ spectrum contribute at smaller values of $B$. This can lead to an underestimate of the uncertainties from \geneva at intermediate values of $B$ where the resummation is still important.

Going to higher order in \geneva in \figs{JetBplot1c}{JetBplot1d}, the uncertainties of the showered prediction decrease and overlap with the NNLL$_B$+LO$_3$ uncertainties over much of the peak and transition regions. It would be interesting to compare the \geneva prediction to the next higher-order analytic resummed jet broadening prediction to numerically test the accuracy achieved; however, this is not yet available. Determining the formal accuracy of the \geneva prediction for a given observable and systematically including the uncertainty associated to the lack of correlation with $\Tau_2$ are next steps that we leave for future work.
As mentioned in \subsubsec{eePythia}, one possibility would be to include the size of the shift from the partonic to the showered results as a conservative estimate of the uncertainty due to the remaining showering.

The partonic \geneva jet broadening prediction interpolates smoothly to the fixed-order NLO$_3$ result in the tail, with uncertainties that match the fixed-order result in this region. As for the other observables, this validates the behavior of the fully differential master formula in \eq{dsigma23} in this limit.

In \fig{JetBplot2}, we show the hadronized \geneva prediction for jet broadening compared to data from \Aleph and \Opal, which shows good agreement within uncertainties across the peak and transition regions and is low as expected in the far tail. The uncertainty from the \pythia tune and value of $\alpha_s$ are indicated by the central values of the tune 3 and $\alpha_s(m_Z)=0.1184$  histograms, which agree within the perturbative uncertainties of the default \geneva prediction across most of the transition and tail regions. As for the other observables, we see better agreement in the peak (below $B\sim 0.1$) with \pythia tune 1 and $\alpha_s(m_Z)=0.1135$.

\section{Application to Hadronic Collisions}
\label{sec:pp}

In this section, we apply the framework developed in \sec{master} to hadronic collisions and present first results from the implementation in the \geneva Monte Carlo.  To accommodate hadrons in the initial state, special consideration is required for each component of our master formula. Our goal is to demonstrate that the same methods can be applied in a hadronic environment to obtain a consistent description at the next higher perturbative accuracy. We use Drell-Yan production $pp\to Z/\gamma^* \to \ell^+\ell^-$ as a concrete example to study the framework, deferring a detailed comparison with Tevatron and LHC data to later work.

In hadronic collisions, $N$-jettiness can be used as a jet resolution variable, with the observable taking the initial states into account.  The theoretical framework exists to resum $N$-jettiness at hadron colliders, and this resummation has been applied to several processes~\cite{Stewart:2010pd,Berger:2010xi,Liu:2012zg}.  Similarly, the techniques required to perform the next-to-leading-order calculations at hadron colliders are known.  A phenomenological study additionally requires \geneva to be interfaced with a parton shower and hadronization model that includes multiple parton interactions.

In the Drell-Yan example, the 0-jet resolution variable is beam thrust, $\Tau_0$, which is the analog to 2-jettiness, $\Tau_2$, used in the $e^+e^-$ application.\footnote{Since we are mainly interested in QCD corrections, we have chosen the subscript on $\Tau$ to indicate the multiplicity of jets in the final state.} The resummation of beam thrust is performed to NNLL, and the jet multiplicities at fixed order are calculated at NLO$_0$ and LO$_1$.  The prediction of \geneva is compared to the analytic resummation of $\Tau_0$ at NNLL+LO$_1$ as a demonstration that the program correctly describes the matching between 0- and 1-jet multiplicities.  Finally, we discuss how the accuracy of these ingredients can be improved and the challenges present in applying \geneva to hadron collisions.

Beam thrust is defined as a sum of contributions from particles in the final state~\cite{Stewart:2010tn},
\begin{align} \label{eq:tauB}
\Tau_0 &= \sum_k \min \bigl\{ n_a\cdot p_k , n_b \cdot p_k \bigr\} \,,
\end{align}
where the observable is evaluated in the center-of-mass frame of the hard partonic collision.  The $n_{a,b}$ are light cone vectors along the beam ($\hat{z}$) axis, with $n_a = (1,\hat{z})$ and $n_b = (1, -\hat{z})$.  Beam thrust can be evaluated in any frame by performing a longitudinal boost on \eq{tauB}.

With the addition of more final-state jets, the $N$-jettiness definition can be generalized from the 0-jet case,
\begin{equation}
\label{eq:tauNhadronic}
\Tau_N = \sum_k \min \bigl\{ n_a \cdot p_k ,\, n_b \cdot p_k ,\, n_1 \cdot p_k ,\, \ldots ,\, n_N \cdot p_k \bigr\} \,.
\end{equation}
As for beam thrust, this observable is evaluated in the partonic center-of-mass frame. The $n_i = (1, \hat{n}_i)$ for $i = 1, \ldots, N$ are light cone vectors along the jet directions.  Note that the above definition of $\Tau_N$ is a simple extension of the observable for $e^+e^-$ collisions, which has no contribution from the beam directions but is otherwise identical.

\subsection{Master Formula and Ingredients for Hadronic Collisions}
\label{subsec:masterpp}

As in the $e^+e^-$ case, the master formula for the cross section in \geneva is given by \eq{MCfullydiff}, \eq{sigmaNTaucut}, and \eq{mastermult}.  To match the 0- and 1-jet multiplicities for a general process, the master formula is
\begin{align} \label{eq:pp0jmaster}
\frac{\df\sigma_{\incl}}{\df\Phi_0}
&= \frac{\df \sigma}{\df\Phi_0}(\Tau_0^\cut)
+ \int\!\frac{\df\Phi_1}{\df\Phi_0}\, \frac{\df\sigma}{\df\Phi_1}(\Tau_0)\, \theta(\Tau_0 > \Tau_0^\cut)
\,,\end{align}
where the 0-jet cumulant, $\df\sigma/\df\Phi_0 (\Tau_0^\cut)$, and the 1-jet spectrum, $\df\sigma/\df\Phi_1 (\Tau_0)$, are
\begin{align} \label{eq:dsigma01}
\frac{\df\sigma}{\df\Phi_0}(\Tau_0^\cut)
&= \frac{\df\sigma^\resum}{\df\Phi_0}(\Tau_0^\cut)
+ \biggl[\frac{\df\sigma^\FO}{\df\Phi_0}(\Tau_0^\cut)
- \frac{\df\sigma^\resum}{\df\Phi_0}(\Tau_0^\cut)\bigg\vert_\FO \biggr]
\,, \nn \\
\frac{\df\sigma}{\df\Phi_1}(\Tau_0)
&= \frac{\df\sigma_\incl}{\df\Phi_1}
\biggl[\frac{\df\sigma^\resum}{\df\Phi_0 \df\Tau_0}\bigg/\frac{\df\sigma^\resum}{\df\Phi_0\,\df\Tau_0}\bigg\vert_\FO \biggr] 
\,.\end{align}
$\Phi_0$ is the phase space for the hard scattering that produces the 0-jet final state, and $\Phi_1$ includes the additional phase space for the final-state jet.

In \geneva, the contributions to these cross sections are calculated separately for each parton subprocess.  Because the fixed-order matching in the 0-jet cumulant is performed additively, the net weight in the 0-jet cumulant after summing over events is the same as the flavor-summed cumulant.  In the 1-jet spectrum, the fixed-order matching is performed multiplicatively, meaning the sum over all events for a given $\Tau_0$ has a different cross section than if we used flavor-summed components for the different terms in the matching formula.  The two approaches agree up to higher-order corrections, but the former approach is natural in the Monte Carlo.  In the following subsections, we will discuss how the resummed and fixed-order contributions to the master formula are obtained.

\subsubsection{Resummation}
\label{subsubsec:ppResumm}

Like $\Tau_2$ for $e^+e^-$ collisions, beam thrust can be factorized in SCET and the resummation can be carried out systematically to higher orders. The factorized beam thrust spectrum for Drell-Yan is given by~\cite{Stewart:2009yx}
\begin{equation} \label{eq:Tau0fact}
\frac{\df\sigma}{\df\Phi_0 d\Tau_0} = \frac{\df\sigma_B}{\df\Phi_0} \sum_{ij} H_{ij} (Q, \mu) \int \df t_a \df t_b \, B_i (t_a, x_a, \mu) \, B_j (t_b, x_b, \mu) \, S_{ij} \Bigl(\Tau_0 - \frac{t_a + t_b}{Q}, \mu \Bigr)
\,,\end{equation}
where $Q$ is the dilepton invariant mass, and $\Phi_0$ is the phase space for the $q\bar{q} \to \ell^+ \ell^-$ hard scattering.  The momentum fractions $x_{a,b}$ are defined in terms of the total rapidity $Y$ of the final state from the hard scattering,
\begin{equation}
x_a = \frac{Q}{\Ecm} e^{Y} \,, \qquad x_b = \frac{Q}{\Ecm} e^{-Y} \,.
\end{equation}
Comparing \eq{Tau0fact} to the $e^+e^-$ analog, \eq{tau2fact}, it is clear that the factorization theorems are structurally similar.  The chief difference is that, while the jet functions in \eq{tau2fact} parametrize the collinear evolution of final-state jets, the beam functions in \eq{Tau0fact} parametrize the collinear evolution of the incoming partons as well as the nonperturbative process of extracting high-energy partons from the proton.  The beam functions can be further factorized into a convolution between the parton distribution functions $f_j$ and perturbatively calculable Wilson coefficients $\mathcal{I}_{ij}$~\cite{Stewart:2009yx, Stewart:2010qs},
\begin{equation} \label{eq:beamfunc}
B_i (t, x, \mu) = \sum_j \int_x^1\! \frac{\df\xi}{\xi}\, \mathcal{I}_{ij} \Bigl(t, \frac{x}{\xi}, \mu\Bigr) f_j (\xi, \mu)
\,.\end{equation}
Note that due to initial-state radiation, the $x_{a,b}$ are distinct from the Bjorken variable $\xi$ appearing in the convolution above that gives the momentum fraction of the energetic partons that are liberated from the proton.  The sum over partons $i,j$ in \eq{Tau0fact} is a sum over flavor singlet quark-antiquark combinations, such as $u\bar{u}$ or $\bar{b}b$.  For each flavor, the beam functions are different, and the hard function $H_{ij}$ differs for up- and down-type quarks due to the different electroweak couplings to the gauge boson.  This flavor sum is an important consideration when implementing the master formula in \geneva, since the Monte Carlo generates events for each flavor combination, and the flavor sum is performed in the sum over events.

For processes with final-state jets, the extension of the beam thrust factorization theorem to $N$-jettiness is known and has the schematic form~\cite{Stewart:2010tn, Jouttenus:2011wh}
\begin{equation} \label{eq:ppTauNfact}
\frac{\df\sigma}{\df\Phi_N d\Tau_N} = \frac{\df\sigma_B}{\df\Phi_N} \, \textrm{Tr} \sum_{\kappa} H_{\kappa} \biggl[ \Bigl( B_a^\kappa \, B_b^\kappa \, J_1^\kappa \, \cdots \, J_N^\kappa \Bigr) \otimes S^{\kappa}_{N+2} \biggr] (\Tau_N)
\,.\end{equation}
The trace is over the nontrivial color structures that can exist in the hard and soft functions.  Additionally, there is a sum over parton channels for the hard scattering that are labeled by the index $\kappa$.  The additional jets are associated with additional collinear sectors in SCET, and the factorization theorem reflects this by including additional jet functions.  The soft function also changes to account for the soft radiation between the final-state jets and the initial-state radiation from the colliding partons. 

The factorization theorems in \eqs{Tau0fact}{ppTauNfact} can be used to perform the resummation for both the spectrum and cumulant.  Although these factorization theorems directly describe the spectrum in $\Tau_0$ or $\Tau_N$, they can be integrated over the observable to obtain the cumulant.  The perturbative part of each function in the factorization theorem is calculable, and for many processes the functions are known to high order.  Each function is associated with a scale that is connected to the physical degrees of freedom that the function describes.  As in the $e^+e^-$ case, renormalization group evolution resums the large logarithms of ratios of these scales (see \tab{expcounting}).  

\subsubsection{Fixed Order}
\label{subsubsec:ppFixedOrder}

Following~\eq{pp0jmaster},  we need to define the 0-jet cumulant $\df
\sigma / \df \Phi_0 (\Tau_0^{\rm cut})$ and the  1-jet spectrum $\df\sigma / \df\Phi_1(\Tau_0)$. At the order we are interested in, the 1-jet spectrum will be given by the tree-level cross section $B_1(\Phi_1)$ for the process  $pp\to Z/\gamma^*j \to \ell^+ \ell^- j$. 
The 0-jet cumulant is given by 

\begin{eqnarray} \label{eq:dsigma0NLOpp}
\frac{\df \sigma}{\df \Phi_0} (\Tau_0^{\rm cut})& =& B_0(\Phi_0) + V_0(\Phi_3) + \int \frac{\df \Phi_1}{\df \Phi_0} B_1(\Phi_1) \theta(\Tau_0 < \Tau_0^\cut) + \nn\\
 &&\int \frac{\df \Phi_{1,a}}{\df \Phi_0} G_a(\Phi_{1,a}) \theta(\Tau_0 < \Tau_0^\cut)   + \int \frac{\df \Phi_{1,b}}{\df \Phi_0} G_b(\Phi_{1,b}) \theta(\Tau_0 < \Tau_0^\cut)   
\,,\end{eqnarray}
where the corresponding parton distribution functions have been included into the definitions of the Born, virtual, and real emission cross sections,
\begin{align}
 B_N (\Phi_N) &=  f_a(x_a,\mu_F) f_b(x_b,\mu_F)  \mathcal{B}_N(\Phi_N) \,, \nn \\
 V_N (\Phi_N) &=  f_a(x_a,\mu_F) f_b(x_b,\mu_F)  \mathcal{V}_N(\Phi_N) \,.
\end{align}
In addition, in order to account for the incomplete cancellations of initial-state collinear singularities, we have included one collinear counterterm $G_{a,b}$ for each initial-state parton,
\begin{align}
 G_a (\Phi_N)  &=  f_a(x_a,\mu_F) f_b(x_b,\mu_F)  \mathcal{G}_a(\Phi_N) \,, \nn \\
 G_b (\Phi_N)  &=  f_a(x_a,\mu_F) f_b(x_b,\mu_F)  \mathcal{G}_b(\Phi_N) \,.
\end{align}
Assuming the UV divergences of $V_0$ have already been taken care of by a proper renormalization procedure, the remaining  divergences present in $B_{1}$, $V_0$, and $G_{a,b}$ are of IR origin.  We handle these divergences with the FKS subtraction procedure.  After having partitioned the phase space into nonoverlapping regions $m$, which contain at most one collinear and one soft singularity, by means of a set of $\theta^\Tau_m$-functions, the final formula, including the subtraction counterterms, is
\begin{align}
\label{eq:subtraction}
\frac{\df \sigma}{\df \Phi_0} (\Tau_0^{\rm cut})&= B_0(\Phi_0) + V_0(\Phi_0) + I(\Phi_0) + \sum_m  \int \df \Phi^m_{\rm rad} \bigg[ B_1(\Phi_1^{m}) \theta^\Tau_m(\Phi_0,\Phi^m_{\rm rad}) \theta(\Tau_0 < \Tau_0^\cut)
\nn\\
&
- \frac{\df \Phi_{\rm rad}^{m,{\rm s}}}{\df \Phi^m_{\rm rad}} B_1(\Phi_1^{m,{\rm s}}) \theta^{\rm s}_m(\Phi_0,\Phi_{\rm rad}^{m,{\rm s}}) 
 - \frac{\df \Phi_{\rm rad}^{m,{\rm c}}}{\df \Phi^m_{\rm rad}} B_1(\Phi_1^{m,{\rm c}}) 
  + \frac{\df \Phi_{\rm rad}^{m,{\rm cs}}}{\df \Phi^m_{\rm rad}} B_1(\Phi_1^{m,{\rm cs}})  
\bigg]
\nn\\
& + \int \df \Phi_{{\rm rad},a}\ G_a(\Phi_{1,a}) + \int \df \Phi_{{\rm rad},b}\ G_b(\Phi_{1,b}) \,.
\end{align}
As mentioned in \subsubsec{eeFixedOrder}, we choose to partition the phase space by means of
\mbox{$\theta^\Tau_m$-functions} that depend on the jet resolution variable. It is therefore
crucial to evaluate the jet resolution variable and the subtraction in
the same frame. This ensures the proper cancellation of IR
singularities by subtraction counterterms. The preferred frame for the
fixed-order calculations is the partonic center-of-mass frame, since the subtraction is most naturally expressed in terms
of variables defined in that frame.  Also, the
jet resolution variable, \eq{tauB}, is defined in this frame and resummation can be performed in it.  Therefore, our approach is to perform the entire calculation in the partonic center-of-mass frame.

At this point, all the ingredients of \eq{subtraction}  are known and
available in the literature~\cite{Frixione:1995ms,Alioli:2010xd}.
Note that additional complications arise when extending this
construction to higher multiplicities because one must use a map that preserves the value of $\Tau_0$ 
(up to power corrections), just as for $\Tau_2$ in the $e^+e^- \to 3$ jet
case.  For our $pp\to Z/\gamma^* \to \ell^+ \ell^-$ study, this
problem can be avoided since we currently include only up to one
extra parton and, consequently, the $\Tau_0(\Phi_1)$ value is
well defined.  In order to obtain $\Tau_0(\Phi_N)$ with $N>1$ in
general, this issue may be addressed in a similar fashion to what has
been done for $e^+e^-$.

\subsection{Application to Drell-Yan Production}
\label{subsec:plotspp}

We study Drell-Yan production in $pp$ collisions at $\Ecm = 8$ TeV in \geneva, sampling the invariant mass $Q$ of the $\ell^+ \ell^-$ pair around the $Z$ pole between $M_Z - 10 \Gamma_Z$ and $M_Z + 10 \Gamma_Z$, where $M_Z = 91.1876$ GeV is the mass of the $Z$ and its width is $\Gamma_Z = 2.4952$ GeV~\cite{Beringer:1900zz}.  The dominant contribution in this range of $Q$ comes from $Z$ exchange, although the photon does contribute.  Profile scales identical to those used in the $e^+ e^-$ $\Tau_2$ resummation are used, which is justified since the logarithmic structure is the same between the two observables.  The resummation is turned off just above $\Tau_0 \sim M_Z / 2$, and for greater $\Tau_0$, the spectrum reproduces the fixed-order distribution.

\begin{figure*}[t!]
\subfigure[\hspace{1ex} Peak Region]{%
\parbox{0.5\columnwidth}{\includegraphics[scale=0.5]{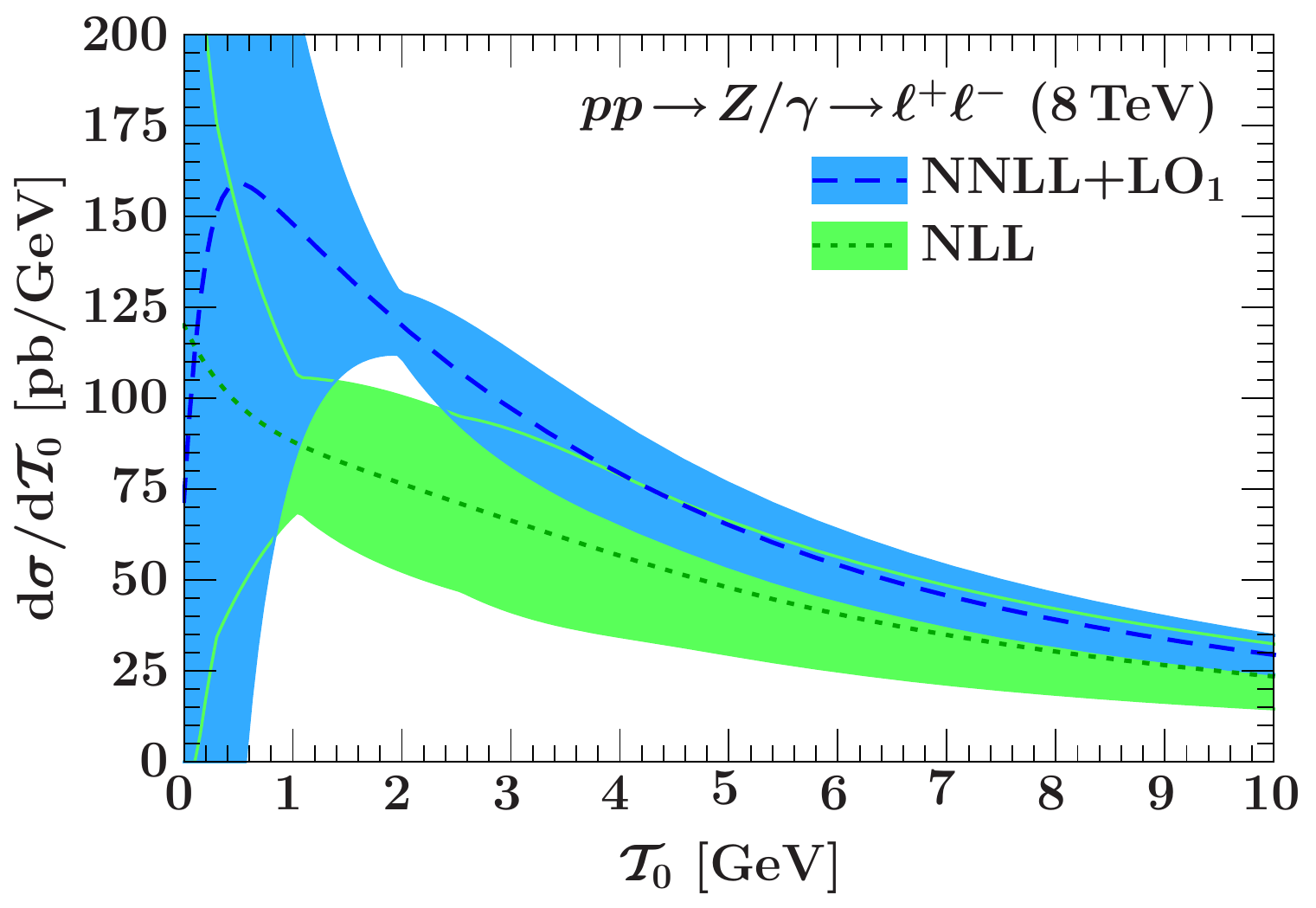}}%
\label{fig:Tau0ConvPeak}}%
\hfill%
\subfigure[\hspace{1ex} Transition Region]{%
\parbox{0.5\columnwidth}{\includegraphics[scale=0.5]{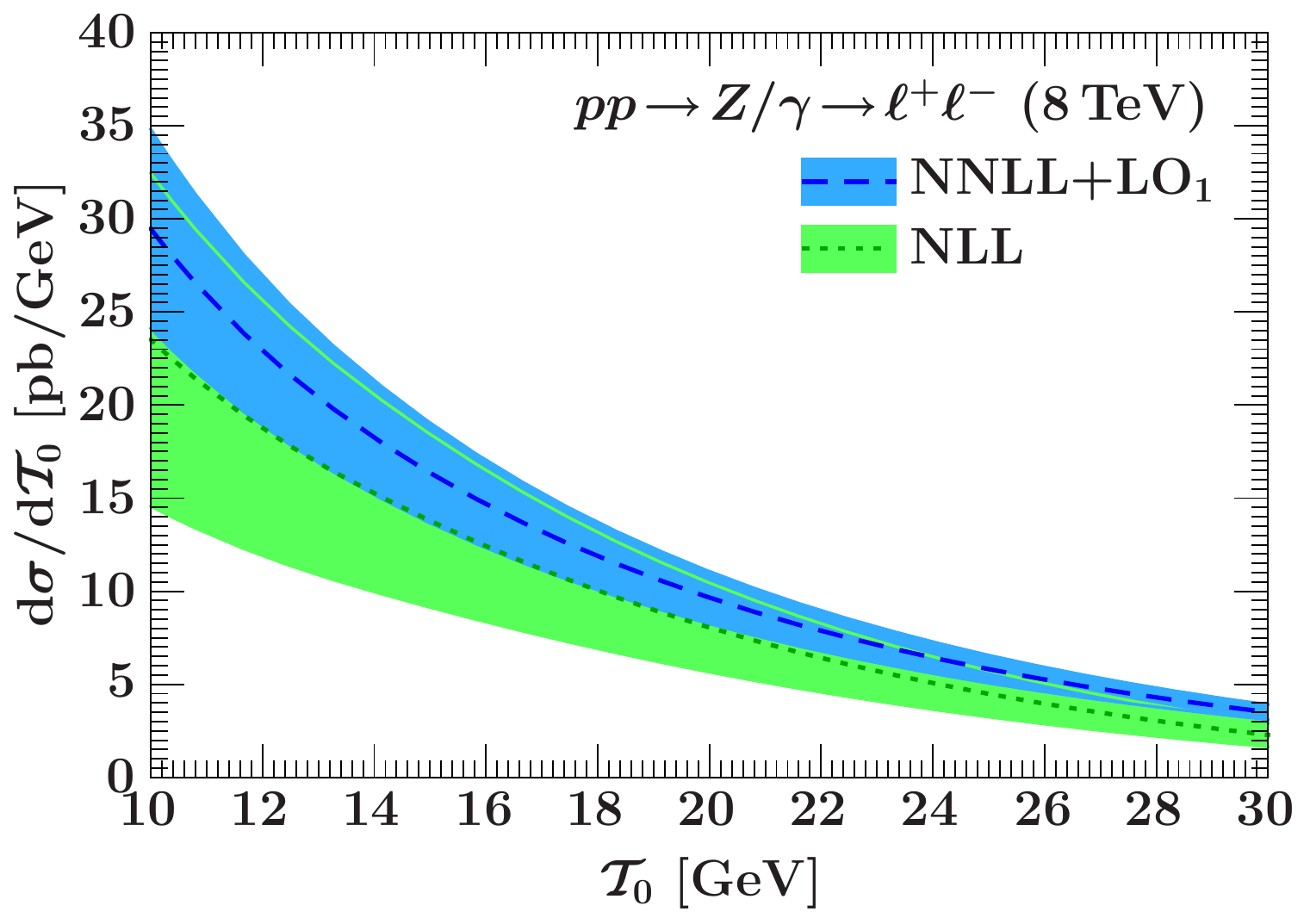}}%
\label{fig:Tau0ConvTrans}}%
\\%
\begin{center}
\subfigure[\hspace{1ex} Tail Region]{%
\parbox{0.5\columnwidth}{\includegraphics[scale=0.5]{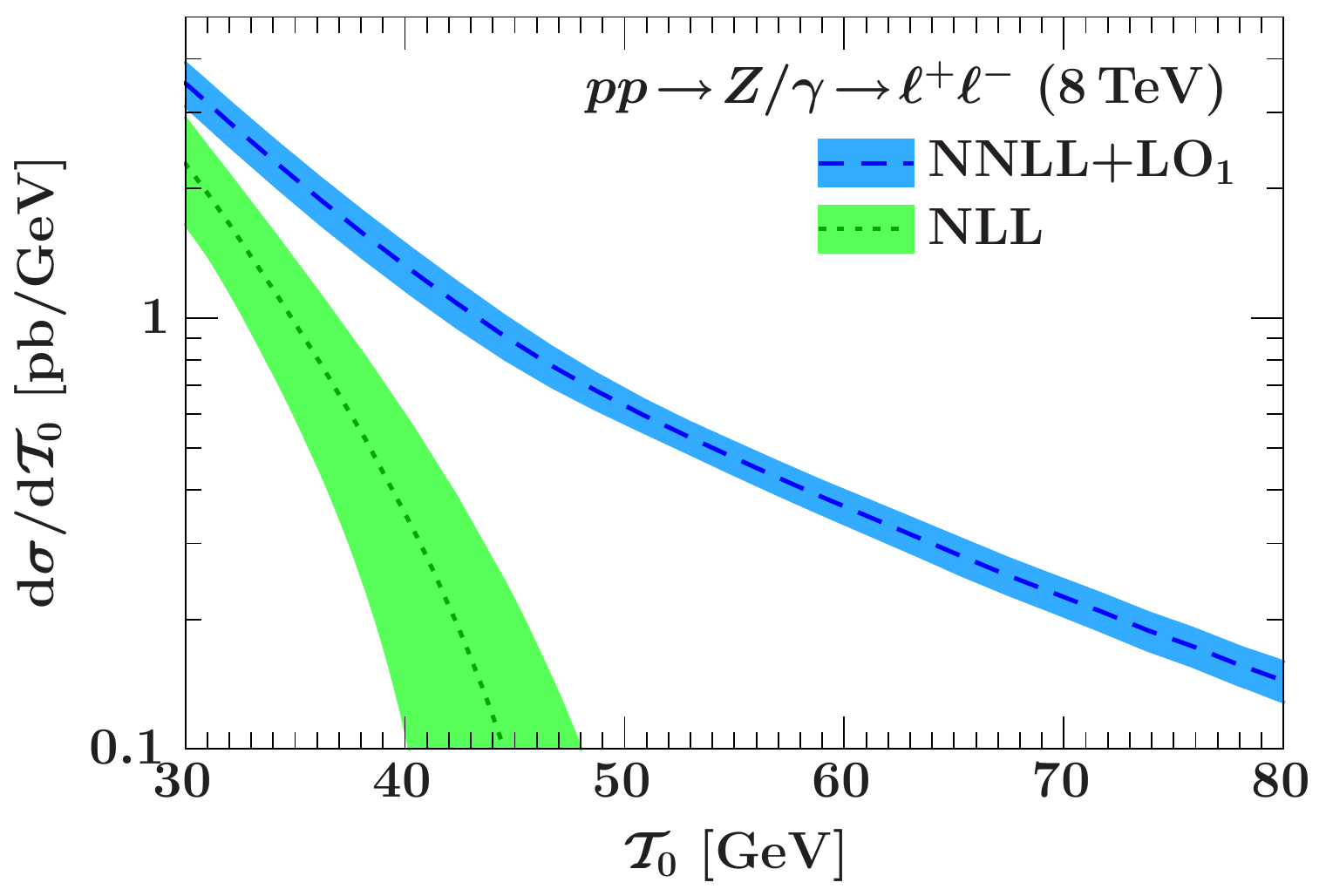}}%
\label{fig:Tau0ConvTail}}%
\end{center}
\vspace{-0.5ex}
\caption{ Analytic resummation of $\Tau_0$ matched to fixed order in the (a) peak, (b) transition, and (c) tail regions. The central value is shown along with the band from scale uncertainties, as discussed in \subsec{plotspp}, at NLL and NNLL$'$+LO$_1$.}
\label{fig:Tau0Conv}
\end{figure*}

\begin{figure*}[t!]
\subfigure[\hspace{1ex} Peak Region]{%
\parbox{0.5\columnwidth}{\includegraphics[scale=0.5]{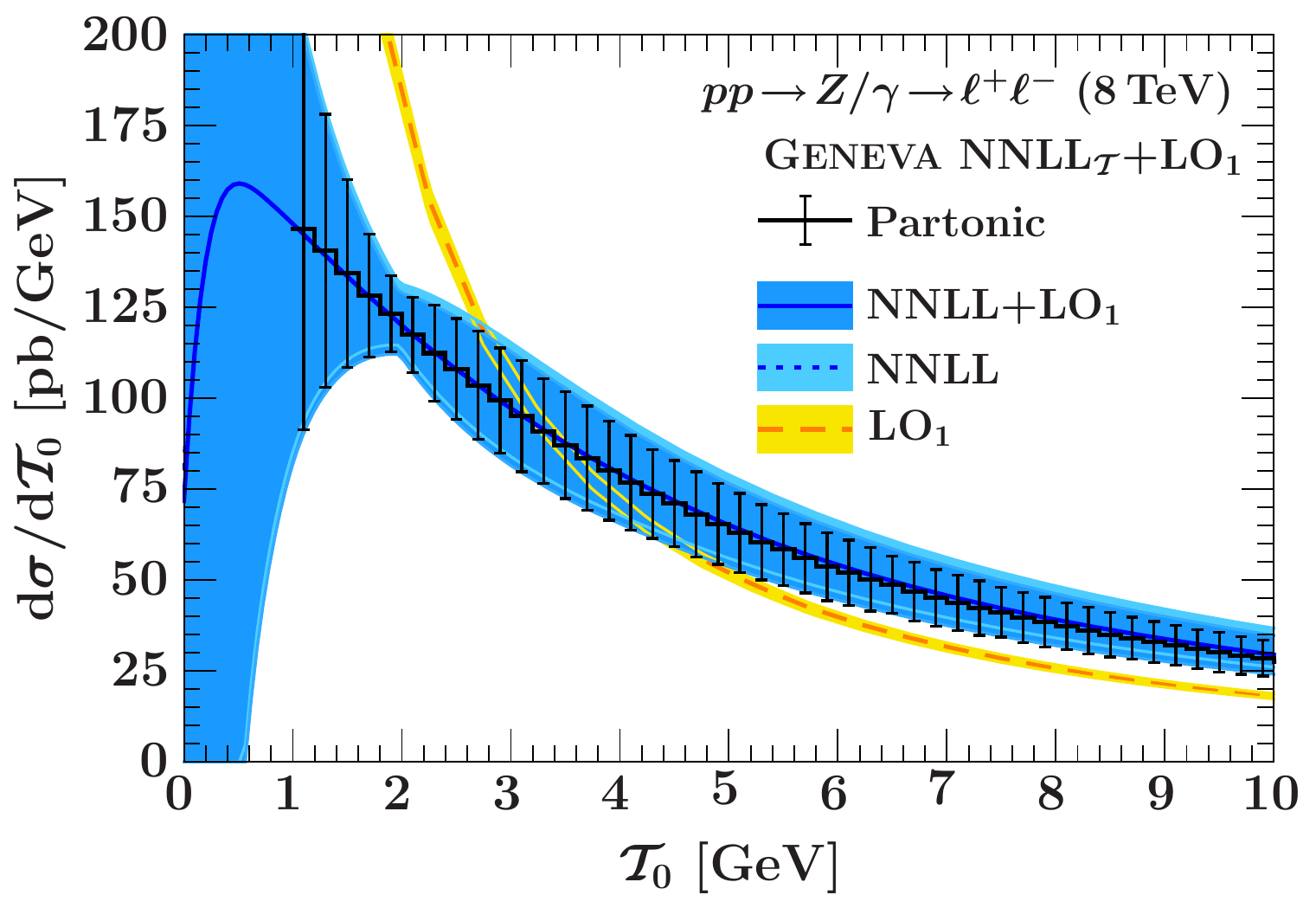}}%
\label{fig:Tau0NNLLPeak}}%
\hfill%
\subfigure[\hspace{1ex} Transition Region]{%
\parbox{0.5\columnwidth}{\includegraphics[scale=0.5]{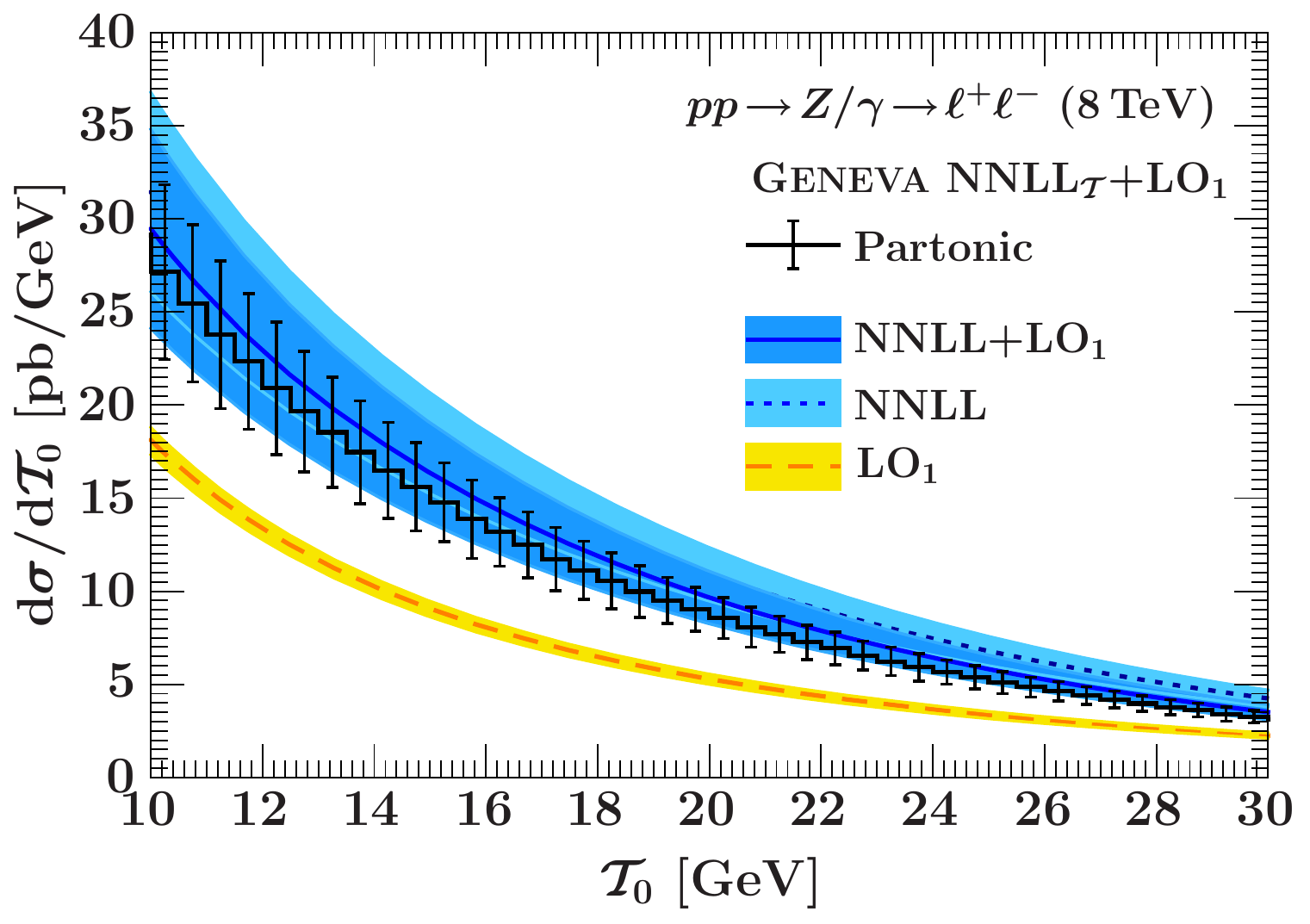}}%
\label{fig:Tau0NNLLTrans}}%
\\%
\begin{center}
\subfigure[\hspace{1ex} Tail Region]{%
\parbox{0.5\columnwidth}{\includegraphics[scale=0.5]{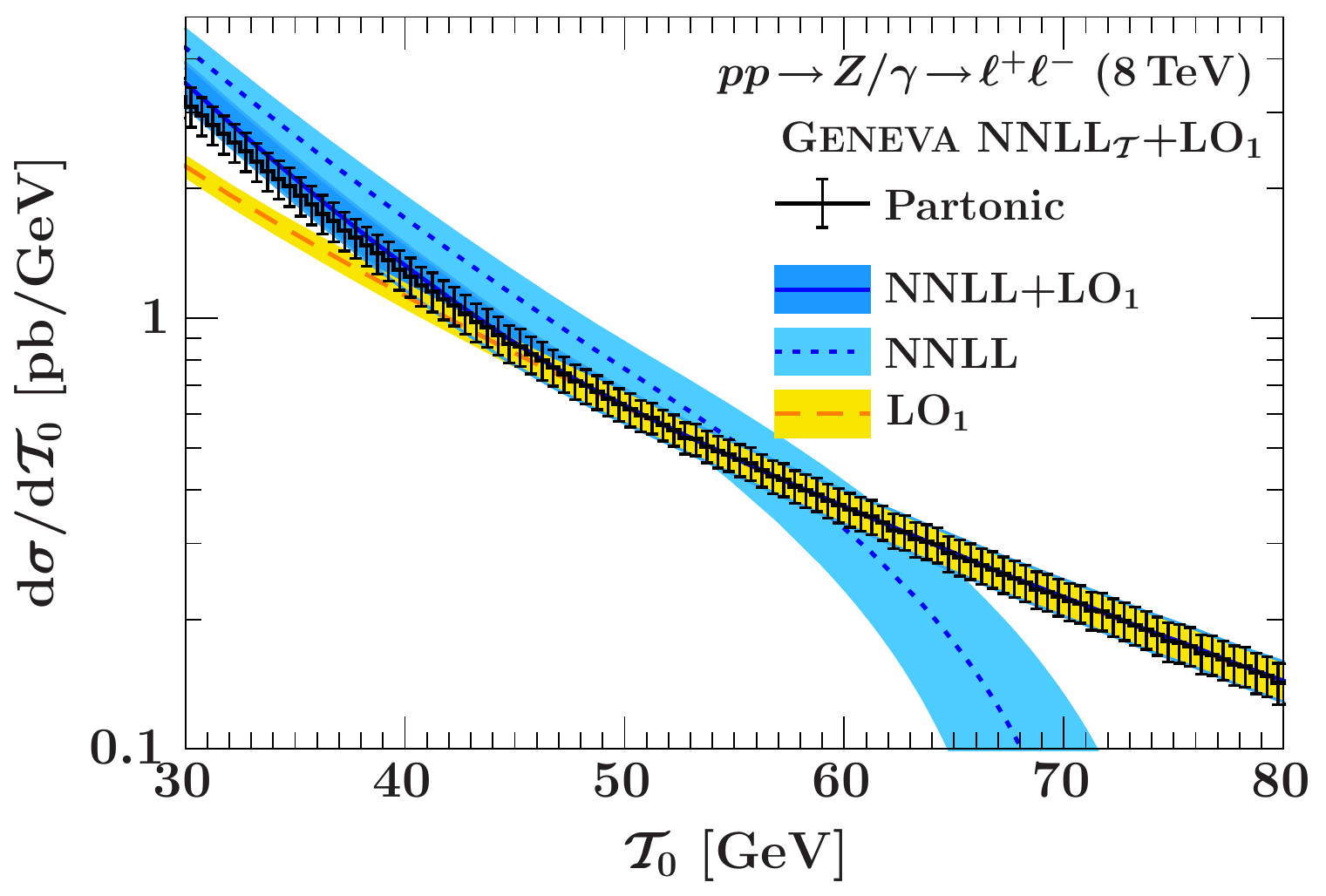}}%
\label{fig:Tau0NNLLTail}}%
\end{center}
\vspace{-0.5ex}
\caption{ The \geneva partonic NNLL+LO$_1$ result is shown compared to the analytic resummation of $\Tau_0$ matched to fixed order at NNLL+LO$_1$  in the (a) peak, (b) transition, and (c) tail regions. Also shown for comparison is the pure resummed result at NNLL  and the fixed-order LO$_1$ contribution. }
\label{fig:Tau0Val}
\end{figure*}

In \fig{Tau0Conv}, we show the analytic beam thrust resummation at NLL and NNLL+LO$_1$ in the peak, transition, and tail regions. In the peak and transition regions, the resummed result converges well. At the end of the transition region and in the tail region, the pure NLL resummed distribution goes to 0 as the resummation is turned off, but the NNLL+LO$_1$ distribution moves into the fixed-order result.

Implementing the Drell-Yan process in \geneva allows us to study the feasibility of the multiplicative matching for the spectrum in \eq{dsigma01} and compare  with the analytic resummed distribution.
We show this comparison in \fig{Tau0Val}, where the analytic curve is evaluated at NNLL+LO$_1$.  Additionally, we show the NNLL and LO$_1$ distributions separately.  Overall, the partonic \geneva distribution is quite close to the NNLL+LO$_1$ distribution, both in terms of the central values and the size of uncertainties.  In the peak region, the spectra match closely and agree well with the pure NNLL resummed distribution.  At the low end of the transition region, the resummed spectra are still in fair agreement while moving to higher $\Tau_0$ values the \geneva partonic prediction and the NNLL+LO$_1$ distributions begin to systematically deviate from the NNLL distribution.  This deviation arises from the LO$_1$ nonsingular terms that are present in the matched spectrum but absent in the pure resummed one.  In the tail region, the \geneva partonic and NNLL+LO$_1$ predictions move into the LO$_1$ spectrum.  After the resummation has been turned off, these spectra match the LO$_1$ precisely in central value and uncertainties, as expected.  Note that in the corresponding comparison in the $e^+e^-$ case, shown in \fig{TauNLLpPeak}-\ref{fig:TauNLLpTail}, the analytic and the partonic \geneva distributions are in closer agreement because the resummed components of the multiplicative matching in \eq{mastermult} include the nonsingular terms at LO$_3$, which are known analytically.  These are not included in the Drell-Yan case, and so the difference between the analytic and \geneva distributions in \fig{Tau0Val} is more sensitive to the subleading corrections that the nonsingular terms generate.

As in the $e^+e^-$ case, the uncertainty bands for the resummed curves and \geneva predictions in \figs{Tau0Conv}{Tau0Val} are obtained by adding the fixed-order and resummation uncertainties in quadrature.  In the peak and transition regions of the distribution, the resummation uncertainties dominate, while the fixed-order uncertainty dominates as the resummation is being turned off in the tail region.  Comparing the uncertainty of the resummed distributions with that of the LO$_1$ distribution, which is much smaller, one can see that the fixed-order uncertainty is an underestimate of the missing higher-order terms. The reason for this is twofold: the missing large logarithmic corrections at higher orders, whose associated uncertainties are instead included in the resummed distribution, and the partial cancellation between the renormalization and factorization scale dependence, whose variations are correlated in the results we show.

In both the $e^+e^-$ and Drell-Yan processes, the partonic \geneva spectrum is determined by \eq{mastermultgen}, which for an event multiplies the fully exclusive fixed-order cross section by the ratio of the resummed cross section for the jet resolution variable divided by the fixed-order expansion of that resummation.  Compared to the $e^+ e^-$ case, where each subprocess contributing to the cross section is trivially proportional, in Drell-Yan, the convolution with the PDFs requires treating every possible  $q\bar{q}$ initial state separately, in both the fixed-order and the resummed cross sections.  In \geneva, the flavor sum is performed in the Monte Carlo sense, since every event has a definite flavor for the initial-state quarks and the correct flavor-summed cross section is obtained after a sum over all events.  This means that the separate factors in \eq{mastermultgen} are evaluated for an individual flavor, and the entire expression is summed over flavors.  In the analytic resummation, since the matching between the resummed and fixed-order cross sections is additive, there is instead only one way to perform the sum over flavors.

A version of the master formula where both the resummed and the resummed-expanded are separately flavor summed before entering \eq{dsigma01} would be equally valid. We have checked, however, that this is a very minor effect and is not the main contribution to the apparent differences of the \geneva predictions with the analytic resummed cross section.  In fact, the reason for the discrepancy is in the difference of higher-order terms that are included in the \geneva multiplicative approach with respect to the additive matching used in the analytical calculation.  This can also be seen as an indication of the relative freedom in implementing the master formula in \eq{mastermultgen} at a given perturbative accuracy.  As one can evince from figures \ref{fig:TauNNLLpPeak}-\ref{fig:TauNNLLpTail}, should the resummation and fixed-order calculations be evaluated at the next order in perturbative accuracy, the size of the yet-missing terms would decrease, and consequently, the difference between \geneva and the analytic results would be reduced.

The \geneva implementation in this example can be extended to higher accuracy in terms of both fixed-order matrix elements and resummation.  An equivalent accuracy to the $e^+e^-$ results shown in \sec{ee} can be achieved if the fixed-order matrix elements for the 1-jet multiplicity are calculated at NLO, the 2-jet multiplicity are calculated at LO, and the resummation of beam thrust is continued to NNLL$'$.  Although this is beyond the scope of this work, we nonetheless demonstrate that the \geneva framework is capable of merging matrix elements of different jet multiplicities beyond the lowest order.  As in the $e^+e^-$ case, jet multiplicities are defined using a physical jet resolution variable, which allows for a consistent extension of the entire framework to $\ord{\alpha_s}$ perturbative accuracy.

In \sec{ee}, we found that NNLL$'$ resummation of the jet resolution variable $\Tau_2$, when combined with the parton shower, was capable of describing the spectrum in other 2-jet observables with an accuracy that clearly exceeded NLL, the naive accuracy of the parton shower.  The resummation of $\Tau_2$ captures an important set of logarithms that are correlated with other 2-jet observables, and, when combined with the fully exclusive, all-orders description of the parton shower, the accuracy of other observables can be improved beyond NLL.  At a hadron collider, the effective dynamic range of observables is much larger, meaning the correlation between the jet resolution variable and another observable of interest may be small.  In this case, the parton shower may play a greater role in determining the spectrum, and hence the accuracy, of other observables.  We will investigate these features in a future work.

\section{Conclusions}
\label{sec:conclusions}

In this paper, we have shown how to combine higher-order resummation of a jet resolution variable with fully differential next-to-leading-order calculations to extend the perturbative accuracy of cross sections beyond the lowest order for all values of the jet resolution scale.  This framework has been interfaced with a parton shower and hadronization to give the \geneva Monte Carlo program.

Our framework provides both the versatility of fixed-order calculations and the accuracy of higher-order analytic resummation. From the point of view of Monte Carlo generators, the \geneva approach allows the combination of higher-order resummations with higher fixed-order calculations. From the point of view of resummed calculations, it allows one to obtain a fully differential cross section that correctly resums the jet resolution variable to higher logarithmic accuracy. Since this construction maintains the higher perturbative accuracy for all values of the jet resolution scale, it naturally allows NLO calculations of different jet multiplicities to be combined with one another.

The higher logarithmic resummation of the jet resolution scale allows us to use a low cut on the jet resolution scale, much lower than the point where fixed-order perturbation theory breaks down but still above the nonperturbative regime. This is a major difference to other approaches~\cite{Hoeche:2012yf,Gehrmann:2012yg,Frederix:2012ps}, where the jet-merging scale has to be chosen much larger, such that $\alpha_s \ln^2\tau^\cut \ll 1$.

In this paper, we have concentrated on the theoretical construction, which is valid for any number of jets, and for both $e^+e^-$ and hadron colliders. We have shown that one has to carefully choose a jet resolution variable that is resummable to higher logarithmic accuracy.
In our approach, the $N$-jet and $(N+1)$-jet regions are described by the same fully differential calculation without the need of an explicit jet-merging scale. The cut on the jet resolution variable is only needed due to the presence of IR divergences. We have given expressions for both the integrated cross section below the IR cutoff and the differential cross section above that properly combine the higher logarithmic resummation with a higher fixed-order calculation.

This approach has been implemented in the \geneva Monte Carlo. As a first application, we have presented results for $e^+e^-$ collisions. The jet resolution variable for this case was chosen to be 2-jettiness, which is directly related to thrust, and we  combined its NNLL$'$ resummation with the fully differential 3-jet rate at NLO$_3$.  Varying the profile scales and the renormalization scale has allowed us to obtain event-by-event uncertainties. As a final step, we have interfaced our perturbative result with  \pythia 8, which added a parton shower and hadronization to our results.  The parton shower adds additional radiation beyond the highest jet multiplicity in \geneva and has been restricted to only fill out the jets of the exclusive jet cross sections at lower jet multiplicity.  Since the cut on the jet resolution variable could be chosen to be very small, the effect of the perturbative shower (without hadronization) is rather small, and different tunes in \pythia do not affect the resulting distributions significantly. Hadronization has a significant effect, and the difference in final results due to different hadronization parameters is more manifest.

We have shown that \geneva  correctly reproduces the higher-order resummation of the thrust spectrum, even after showering, which serves as a nontrivial validation of our approach. Using $\alpha_s(m_Z) = 0.1135$, as obtained in ref.~\cite{Abbate:2010xh} from fits to the thrust spectrum using higher-order resummation, together with tune 1 of \pythia8, we obtain an excellent description of \Aleph and \Opal data.
The same setup was then used to predict other event shape variables, namely $C$-parameter, heavy jet mass, and jet broadening. In all cases, our results agree remarkably well with the explicit analytical resummations, even though only the thrust resummation  was used as an input. This comparison shows numerically that we achieve a higher resummation accuracy than NLL, which is what one would naively expect to obtain from the parton shower. This is especially remarkable for jet broadening, where the resummation formula has a completely different structure from the thrust resummation. Comparing our results after hadronization to the data, we again find excellent agreement for these other observables.

Finally, we have presented first results toward an implementation for hadron colliders in \geneva. Choosing the Drell-Yan process at the LHC with beam thrust as the jet resolution variable, we combined the resummation of beam thrust at NNLL with the leading-order matrix element for the emission of an extra jet. The results from \geneva agree well with analytical results, which shows the applicability of the framework to hadron colliders.

As we have shown, our theoretical framework to combine higher-order resummation with fixed-order matrix elements and parton shower Monte Carlos is very general, and there are many avenues to pursue in the future. Obvious next steps for $e^+ e^-$ collisions are to include NLO calculations for 4 jets, which would require including the logarithmic resummation for 3-jettiness~\cite{Bauer:2011uc} as well as additional tree-level matrix elements. In addition, one can consider the resummation for other jet resolution variables. For hadronic collisions, next steps are to include the resummation and NLO calculations for higher jet multiplicities, as well as adding parton showering and hadronization using the different available programs.

\begin{acknowledgments}
We thank Jesse Thaler, Nicholas Dunn, Erik Strand, and Simon Berman for collaboration
during early stages of this work.

The primary support for SA, CWB, CB, CKV, and SZ comes from a Department of Energy Early Career Award with Funding Opportunity No. DE-PS02-09ER09-26. 
AH is supported by the US Department of Energy under Grant No. DE-FGO2-96ER40956.
FT is supported by the DFG under Emmy-Noether Grant No. TA 867/1-1.
CKV and JRW are partially supported by the National Science Foundation under Grant Nos. NSF-PHY-0705682, NSF-PHY-0969510 (LHC Theory Initiative, Jonathan Bagger, PI).
Additional support comes from the Director, Office of Science, Office of High Energy Physics of the U.S. Department of Energy under Contract No. DE-AC02-05CH11231.
This research used resources of the National Energy Research
Scientific Computing Center, which is supported by the Office of
Science of the U.S. Department of Energy under Contract No.
DE-AC02-05CH11231.

\end{acknowledgments}

\bibliographystyle{../jhep}
\bibliography{../geneva}

\end{document}